\documentclass[acmtog,nonacm,screen]{acmart}

\acmConference{}{}{}
\acmYear{}
\acmBooktitle{}
\acmISBN{}
\acmDOI{}
\acmPrice{}

\renewcommand\footnotetextcopyrightpermission[1]{}

\title{Mechanics Simulation with Implicit Neural Representations of Complex Geometries}

\author{Samundra Karki}
\affiliation{\institution{Iowa State University} \city{Ames} \state{Iowa} \country{USA}}
\email{samundra@iastate.edu}

\author{Ming-Chen Hsu}
\affiliation{\institution{Iowa State University} \city{Ames} \state{Iowa} \country{USA}}
\email{mhsu@iastate.edu}

\author{Adarsh Krishnamurthy}
\authornote{Corresponding author}
\affiliation{\institution{Iowa State University} \city{Ames} \state{Iowa} \country{USA}}
\email{adarsh@iastate.edu}

\author{Baskar Ganapathysubramanian}
\authornote{Corresponding author}
\affiliation{\institution{Iowa State University} \city{Ames} \state{Iowa} \country{USA}}
\email{baskarg@iastate.edu}

\usepackage{amsmath,amsfonts}
\usepackage{graphicx}
\usepackage{subcaption}
\usepackage{multirow}
\usepackage{booktabs}
\usepackage{algorithm}
\usepackage{algpseudocode}
\usepackage{array}
\usepackage{enumitem}
\usepackage{xcolor}
\usepackage{wrapfig}
\usepackage{tikz, pgfplots}
\usepackage{nicefrac}
\pgfplotsset{compat=1.17}

\newcommand{\cref}[2]{\hyperref[#2]{#1~\ref*{#2}}} 
\newcommand{\colref}[3]{\hyperref[#2]{#1~\ref*{#2}{#3}}} 
\newcommand{\figref}[1]{\cref{Figure}{#1}} 
\newcommand{\secref}[1]{\cref{Section}{#1}} 
\newcommand{\eqnref}[1]{\cref{Equation}{#1}} 
\newcommand{\tabref}[1]{\cref{Table}{#1}} 
\newcommand{\algoref}[1]{\cref{Algorithm}{#1}} 
\begin{document}

\begin{abstract}
Implicit Neural Representations (INRs), characterized by neural network-encoded signed distance fields, provide a powerful means to represent complex geometries continuously and efficiently. While successful in computer vision and generative modeling, integrating INRs into computational analysis workflows, such as finite element simulations, remains underdeveloped. In this work, we propose a  computational framework that seamlessly combines INRs with the Shifted Boundary Method (SBM) for high-fidelity linear elasticity simulations without explicit geometry transformations. By directly querying the neural implicit geometry, we obtain the surrogate boundaries and distance vectors essential for SBM, effectively eliminating the meshing step. We demonstrate the efficacy and robustness of our approach through elasticity simulations on complex geometries (Stanford Bunny, Eiffel Tower, gyroids) sourced from triangle soups and point clouds. Our method showcases significant computational advantages and accuracy, underscoring its potential in biomedical, geophysical, and advanced manufacturing applications.
\end{abstract}


\begin{teaserfigure}
    \centering
    \includegraphics[trim=0.1in 22in 1.1in 0.1in,clip,width=0.5\linewidth]{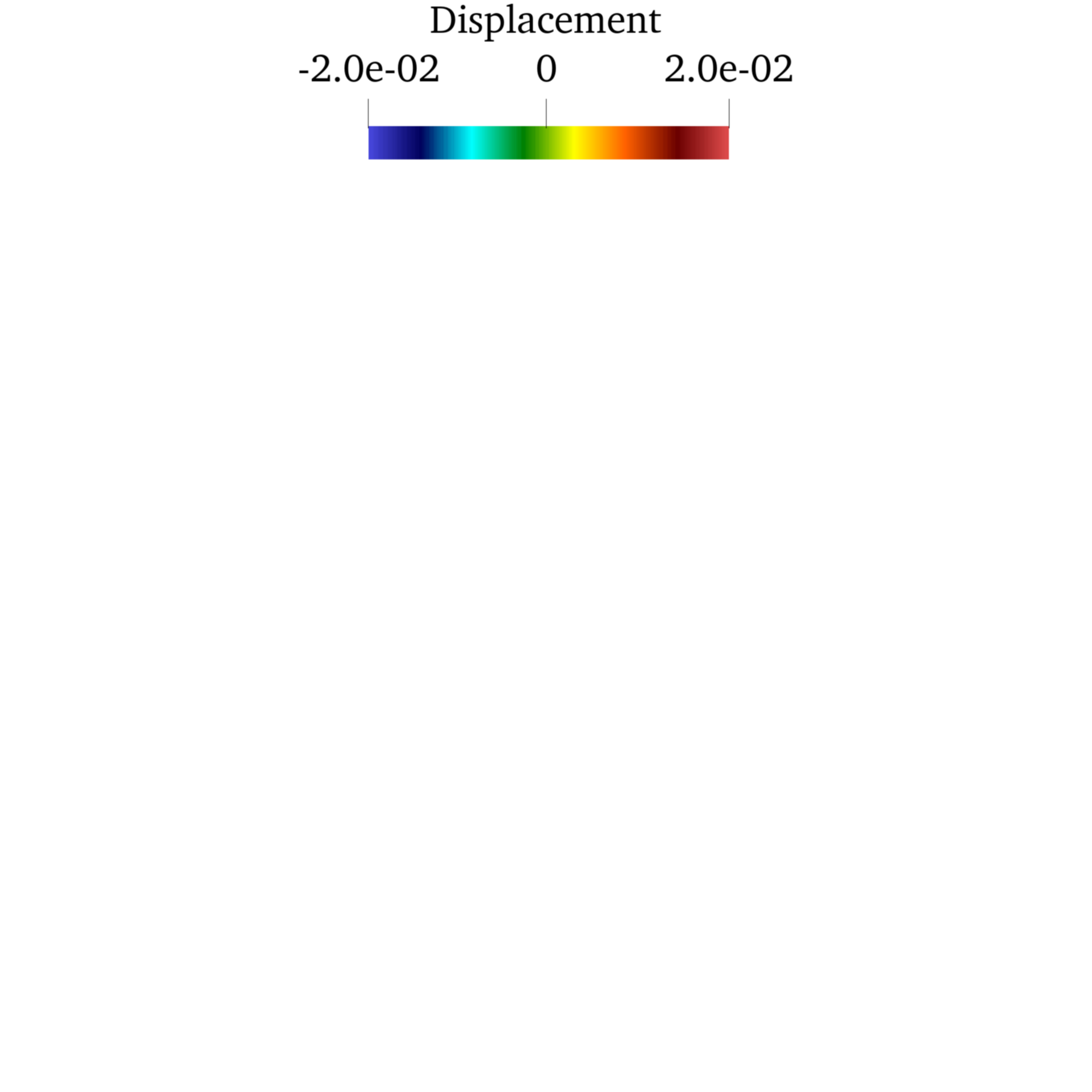}\\
    \begin{subfigure}{0.33\linewidth}
        \centering
        \includegraphics[trim=5.5in 1in 5.5in 2.3in,clip,width=\linewidth]{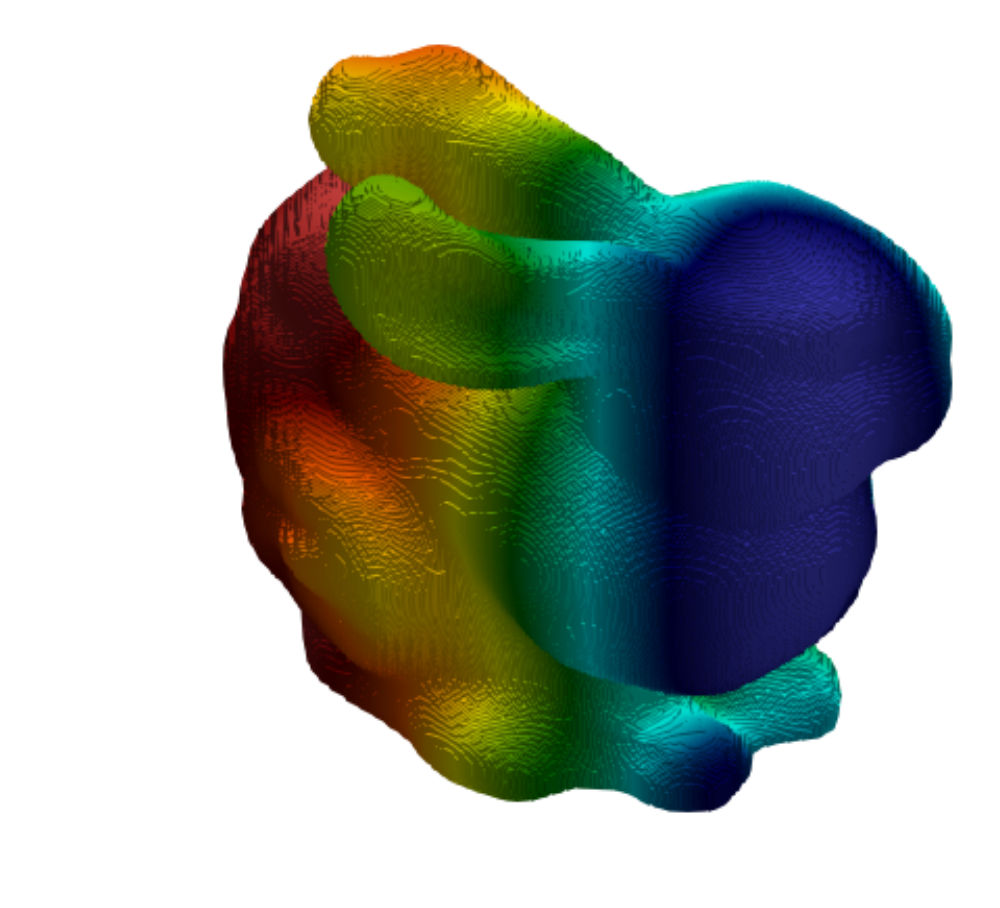}  
    \end{subfigure}
    \begin{subfigure}{0.33\linewidth}
        \centering
        \includegraphics[trim=5.5in 1in 5.5in 2.3in,clip,width=\linewidth]{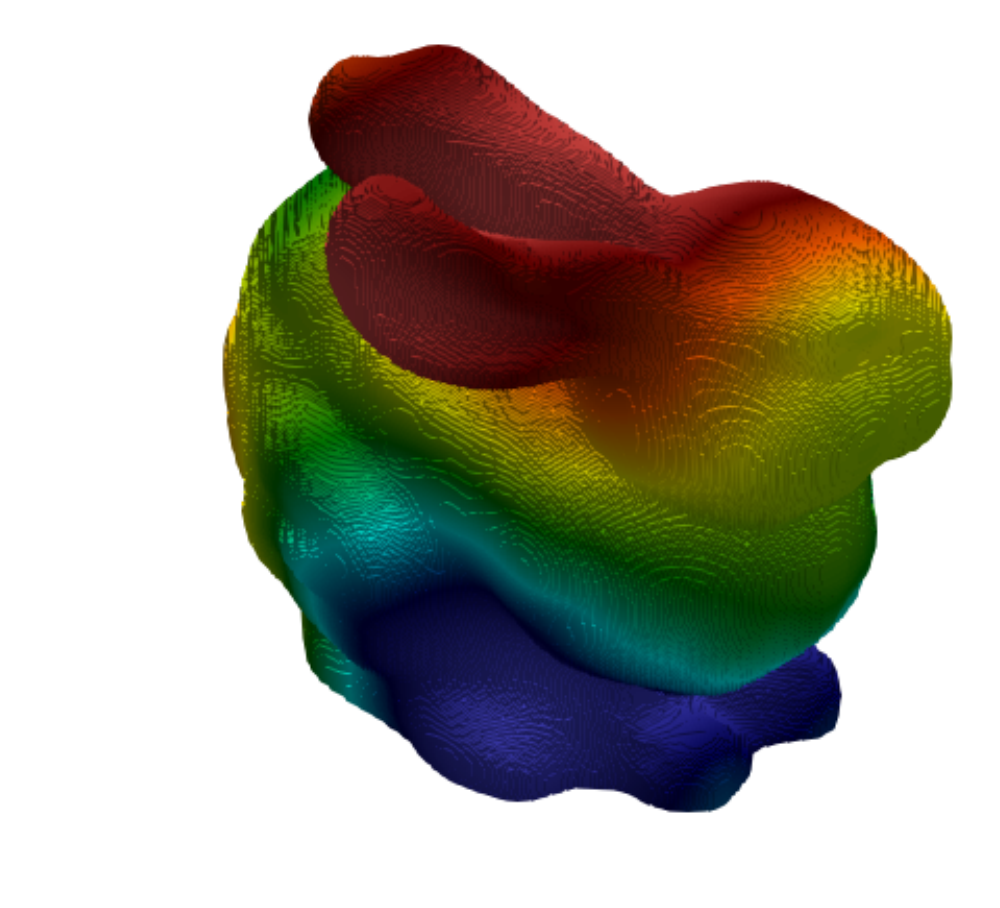}
    \end{subfigure}
    \begin{subfigure}{0.33\linewidth}
        \centering
        \includegraphics[trim=5.5in 1in 5.5in 2.3in,clip,width=\linewidth]{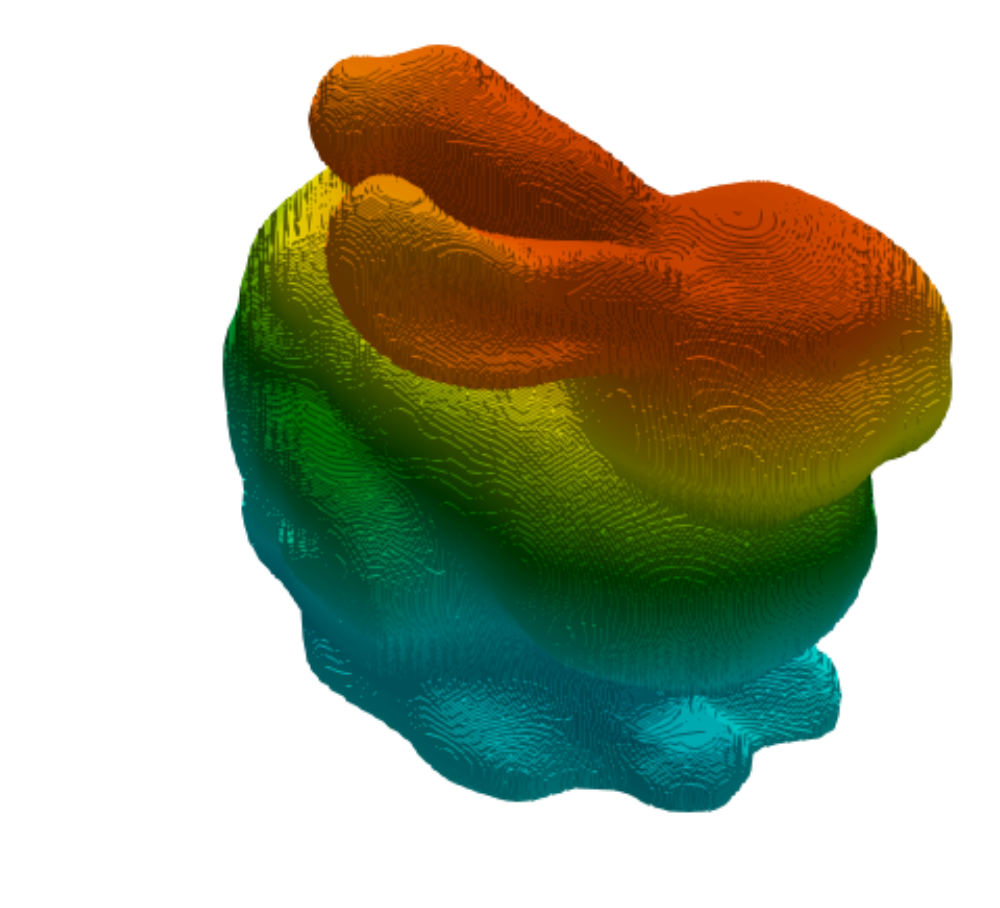}
    \end{subfigure}
    \caption{Left, middle, and right represent the displacement magnitude in $x$-direction ($U_x$), $y$-direction ($U_y$), and $z$-direction ($U_z$) for INR based-bunny. The boundary condition varies sinusoidally for $U_x$ and $U_y$, decays exponentially based on y for $U_z$ }
    \label{fig:bunnyDisplacement}
\end{teaserfigure}

\maketitle

\section{Introduction}

Analyzing complex geometries accurately and efficiently is crucial across various scientific and engineering domains, spanning applications from biomedical engineering and advanced manufacturing to geophysical modeling and digital entertainment. Traditional computational methods, such as finite element analysis (FEA) and finite volume analysis, typically rely on explicit geometric representations like polygonal meshes or boundary representations. While these explicit methods are robust and widely used, they involve substantial computational overhead and discretization errors and often require manual preprocessing, especially when handling intricate, evolving, or dynamically changing geometries~\citep{mchenry2008overview,CHIBA1998145}.

Implicit Neural Representations (INRs) have recently emerg\-ed as a novel paradigm for efficiently encoding complex geometric information. Unlike traditional explicit geometry representations, INRs leverage neural networks to encode shapes implicitly as continuous functions, often as signed distance fields (SDFs). INRs provide compact, smooth, and adaptive representations capable of capturing intricate geometries, topology changes, and fine geometric details without explicit discretization. Such representations have shown remarkable success in computer vision and graphics tasks, such as multi-view reconstruction, generative modeling, and point cloud-based reconstructions~\citep{park2019deepsdf, gropp2020implicit, mescheder2019occupancy, chen2019learning}. However, the direct integration of INRs into established computational simulation pipelines, such as finite element or finite volume analysis, remains limited due to the dependence of these analysis workflows on explicit mesh-based geometric inputs.

Addressing this challenge, this work introduces an innovative computational framework that directly couples INRs with the Shifted Boundary Method (SBM). SBM is a numerical method uniquely suitable for implicit geometric representations because it imposes boundary conditions on grid-aligned surrogate boundaries, thus entirely circumventing the explicit mesh generation step~\citep{main2018shifted1,main2018shifted2}. Readers are referred to the following for history of development and deployment of SBM in recent work~\citep{hsu2016direct,burman2015cutfem,burman2012fictitious,saurabh2021industrial,colomes2021weighted,chou2023diffusion,huang2020simulation,karatzas2020reduced,atallah2021shifted,yang2024optimal,yang2024simulating}. By leveraging the inherent properties of INRs, such as continuous differentiability and resolution independence, our framework directly queries the neural implicit geometry to generate surrogate boundaries and distance vectors essential for SBM-based simulations.

In this paper, we present a detailed description and illustration of our integrated INR-SBM framework for linear PDEs with application to linear elasticity. We demonstrate the efficacy of our method along with efficiency through comprehensive simulations involving linear elasticity spanning canonical benchmarks in \secref{sec:validation_2d} and \secref{sec:validation_3d} as well as complex geometries derived from diverse sources, such as point clouds image-based reconstructions and generative AI models (gyroids, Stanford Bunny, and Eiffel tower models). Our results showcase significant computational advantages, including reduced manual preprocessing, minimized discretization errors, streamlined workflows, and the ability to adaptively refine simulations based on geometric complexity. By enabling direct utilization of implicit neural geometric representations, the proposed approach enhances the precision, flexibility, and efficiency of geometric and physical modeling tasks, highlighting its potential for transformative impacts across numerous scientific and engineering disciplines.

The main contributions of this work are:
\begin{itemize}[topsep=0pt,itemsep=0pt]

    \item Development and demonstration of an integrated computational framework that directly couples INRs with the Shifted Boundary Method (SBM), completely removing the need for explicit mesh generation. We use the implicit neural geometries to dynamically provide surrogate boundaries and distance vectors required for boundary condition enforcement. 
    
    \item Validation of the method through a canonical 2D benchmark, along with demonstration of efficacy and efficiency for the 3D case.
    
    \item Demonstration of the framework's versatility across diverse and complex geometries derived from various real-world and synthetic data sources.
\end{itemize}

We believe our approach can provide significant computational efficiency improvements and enhanced modeling capabilities, including automated adaptive refinement and streamlined handling of intricate geometric complexities, highlighting potential transformative impacts across numerous scientific and engineering applications.

The rest of the paper is arranged as follows. In \secref{Sec:Math}, we provide the mathematical preliminaries on the shifted boundary method and signed distance field required to understand this work. We present a method to generate analysis suitable INR to demonstrate the training complexity involved for INR along with that to get INR with favorable ground truth described in \secref{section:generating_INR_polygon}. We show the formulation for linear elasticity along with the shifted boundary conditions in \secref{Sec:Elasticity}. In \secref{Sec:Results}, we present analysis suitability analysis for different complexities, followed by validation in 2D and 3D cases. Then, we demonstrate the method's capability by simulations performed across INR obtained from diverse sources, along with different complexities. Finally, we outline some future directions and conclude in \secref{Sec:Conclusions}.

\section{Mathematical Preliminaries}
\label{Sec:Math}

We first introduce our approach combining INRs with SBM via a robust and adaptive octree-based meshing strategy~\citep{saurabh2021industrial}. Octree-based meshing efficiently represents complex geometries using Cartesian-aligned, hierarchical grids that adaptively refine near boundaries to resolve intricate geometric features accurately~\citep{saurabh2021scalable}. Subsequently, we describe the true and surrogate boundaries, which are essential concepts underlying SBM. We then introduce mapping techniques for computing distance vectors. Finally, we detail the use of INRs, particularly Signed Distance Fields, discussing methods for generating INRs from various data sources, including point clouds, images, and generative AI models.

\subsection{True and Surrogate Boundaries}
\begin{figure}[t!]
    \centering
    \begin{subfigure}[t]{0.48\linewidth}
        \centering
        \begin{tikzpicture}[scale=0.18] 
            \draw[step=1cm, gray!30] (-10,-10) grid (10,10);
            \draw[blue] (0,0) circle (5cm);
            \draw[red, thick] 
                (-3,-4) -- (-4,-4) -- (-4,-3) -- (-5,-3) -- (-5,3) -- (-4,3) -- (-4,4) -- (-3,4) --
                (-3,5) -- (3,5) -- (3,4) -- (4,4) -- (4,3) -- (5,3) -- 
                (5,-3) -- (4,-3) -- (4,-4) -- (3,-4) -- (3,-5) -- (-3,-5) -- cycle;
            \node[blue] at (3,3) {$\Gamma$};
            \node[blue] at (0,0)
            {$\Omega$};
            \node[red] at (-7,-6)
            {$\tilde{\Omega}$};
            \node[red] at (-5,-4) {$\tilde{\Gamma}$};
            \node at (-9.5,-9.5) {$\mathcal{O}$};
        \end{tikzpicture}
        \caption{ The true domain $\Omega$, the surrogate domain $\tilde{\Omega} \subset \Omega$, the true boundary $\Gamma$, and the surrogate boundary $\tilde{\Gamma}$.}
        \label{fig:SBM_Definition}
    \end{subfigure}
    \hspace{0.05\linewidth}
    \begin{subfigure}[t]{0.4\linewidth}
        \centering
        \begin{tikzpicture}[scale=0.5]
                \draw [line width = 0.5mm,blue] plot[smooth] coordinates {(1,-0.5) (2.25,2.5) (0.75,6)};
                \draw[line width = 0.5mm,red] (0,0.5) -- (0,5);
                \draw[->,line width = 0.25mm,red] (0,2.5) -- (1,2.5);
                \node[text width=0.5cm] at (0.7,2.1) {\small${\color{red}{\tilde{n}}}$};

                \node[text width=0.5cm] at (0.5,5.5) {\small${\color{red}}$};
                \node[text width=0.5cm] at (1.75,5.5) {\small${\color{blue}}$};
                \node[text width=0.5cm] at (1.55,2.7) {\small$d$};
                \node[text width=0.5cm] at (3,2.9) {\color{blue}\small$n$};
                \node[text width=0.5cm] at (3,2.1) {\color{blue}\small$\tau$};
                \draw[->,line width = 0.25mm,-latex] (0,2.5) -- (2.12,3.1);
                \draw[->,line width = 0.25mm,-latex,blue] (2.12,3.1) -- (2.95,3.3);
                \draw[->,line width = 0.25mm,-latex,blue] (2.12,3.1) -- (2.45, 2.1);
        \end{tikzpicture}
        \caption{The distance vector $\mathbf{d}$, the true normal $\mathbf{n}$, the true tangent $\boldsymbol{\tau}$, and the surrogate normal $\tilde{\mathbf{n}}$.
}
        \label{fig:dist_vec}
    \end{subfigure}
    \caption{The domain $\Omega$ is a square grid, with a circle at the center, featuring the true boundary \textcolor{blue}{$\Gamma$} and the surrogate boundary \textcolor{red}{$\tilde{\Gamma}$}.}
    \label{fig:SBM}
\end{figure}
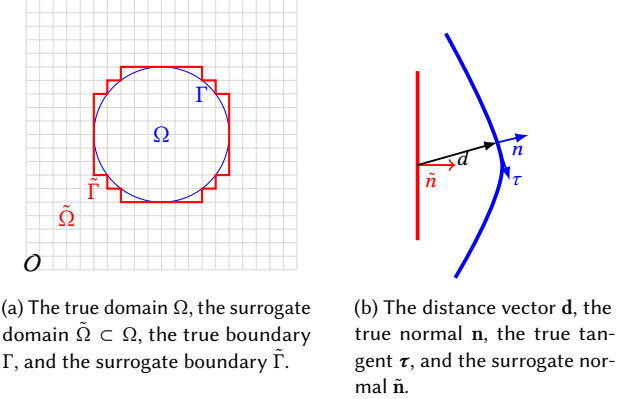 
\figref{fig:SBM_Definition} illustrates a closed region $\mathcal{O}$, representing a complete quadtree, or equivalently, a Cartesian grid consisting of $T_h(\mathcal{O})$ cartesian decompositions of $\mathcal{O}$. Then, the domain of our interest, $\Omega$, is embedded inside $\mathcal{O}$  where $\text{clos}(\Omega) \subseteq \mathcal{O}$ (with $\text{clos}(\Omega)$ denoting the closure of $\Omega$), and with a (true) boundary, $\Gamma$. The union of $T_h$ forms the complete tree but we are only interested in $T \in T_h(\mathcal{O})$ that have a non-empty intersection with the domain of interest $\Omega$.

We define the family of grids as: 
\begin{equation} 
\tilde{T}_h := \left\{ T \in T_h(\mathcal{O}) : \text{meas}(T \cap \Omega) > 0 \right\}
\end{equation}

\textbf{Note:} A key feature for any abstract geometry representation is ease and effectiveness of computing $T \cap \Omega$.  
\begin{align*}
        T \cap \Omega \neq \phi \implies \exists  v \in ~\text{vertices (T)}  \ni v~\text{is inside}~\Omega 
\end{align*}
We can define the operation ($v~\text{is inside}~\Omega$) which returns a classification flag (F) as a function.
\begin{align}
    f:v \rightarrow F
    \label{eq:classifier}
\end{align}
A suitable geometric representation of $\Omega$ efficiently returns this flag. For instance, in a triangulated (polygonal) surface mesh, denoted by $T_{\Delta}$, this function is implemented in terms of an in/out test, which requires $\mathbf{O}(cardinality(T_{\Delta}))$ for every function call. 

Now, we can define the surrogate domain: 
\begin{equation} 
\tilde{\Omega}_h := \text{int} \left( \bigcup_{T \in \tilde{T}_h} T \right)
\end{equation}

This gives us the surrogate domain, $\tilde{\Omega}_h$, with surrogate boundary $\tilde{\Gamma}_h := \partial \tilde{\Omega}_h$ and outward-oriented unit normal vector $\tilde{n}$ to $\tilde{\Gamma}$ as shown in \figref{fig:dist_vec}. The set of grids $T_h/\tilde{T}_h$ does not contribute towards the analysis of the domain $\Omega$, and these grids (in octree-based terminology leaves) are pruned, which significantly improves memory overhead \citep{saurabh2021scalable}. Although the $T_h$ are presented as regularly uniform grids, each $T \in T_h$ can be further sub-divided based on some criteria $F(.)$. This ensures the adaptive refinement of octree-based grids. To preserve the well-posedness of the tree and keep the refinement process gradual, 2:1 balancing is enforced~\citep{saurabh2021scalable}.

\subsection{Distance Vector Computation}

The mapping sketched in \figref{fig:dist_vec} is defined as follows:
\begin{equation}
M_h : \tilde{\Gamma}_h \to \Gamma,
\end{equation}
\begin{equation}
\tilde{x} \mapsto x,
\end{equation}
where $M_h$ maps any point $\tilde{x} \in \tilde{\Gamma}_h$ on the surrogate boundary to a point $x = M_h(\tilde{x})$ on the physical boundary $\Gamma$. In this study, $M_h$ is defined as the closest-point projection of $\tilde{x}$ onto $\Gamma$. 

Using this mapping, a distance vector function $d_{M_h}$ can be expressed as:
\begin{equation}
d_{M_h}(\tilde{x}) = x - \tilde{x} = [M - I](\tilde{x}),
\end{equation}
where $M$ is the mapping operator, and $I$ is the identity operator. For simplicity, we denote $d = d_{M_h}$ and further decompose it as:
\begin{equation}
d = \| d \| \nu,
\end{equation}
where $\| d \|$ is the magnitude of the distance vector, and $\nu$ is a unit vector indicating the direction of the distance. \\ \\
In light of above discussion, to perform SBM analysis our geometric representation must meet two criteria of providing (i)~a distance vector or correspondingly the map $M$; and (ii)~the function defined in \eqnref{eq:classifier} to perform classification of octree vertices to obtain $\tilde{\Omega}$.

\subsection{Implicit Neural Representation}
\label{sec:INR}

A signed distance field (SDF), a type of implicit representation,  is a scalar field that represents the shortest distance from any point in space to the surface of a given shape. The signed distance field (SDF) for a surface $\mathbf{\Gamma}$ is defined by \eqref{sdf definition}.
\begin{equation}
f(\mathbf{x}) = 
\begin{cases} 
\phantom{-}\min\limits_{\mathbf{y} \in \Gamma} \|\mathbf{x} - \mathbf{y}\| & \text{if } \mathbf{x} \quad \text{inside} \quad \boldsymbol{\Omega} \\
-\min\limits_{\mathbf{y} \in \Gamma} \|\mathbf{x} - \mathbf{y}\| & \text{if }  \mathbf{x} \quad \text{outside} \quad \boldsymbol{\Omega} \\
0 & \text{if } \mathbf{x} \in \boldsymbol{\Gamma}
\end{cases}
\label{sdf definition}
\end{equation}
By virtue of definition in \eqnref{sdf definition}, SDF meets the second criteria and acts as the classifier.

INRs provide a powerful approach to encoding complex geometries in a continuous and memory-efficient manner. By using neural networks to represent surfaces implicitly, we can capture intricate shapes and topologies that traditional explicit methods often struggle to model accurately. When the function in \eqnref{sdf definition} is represented by a neural network with parameters $\theta$, such a form of representation is called INR. Such representation, if it represents the signed distance field of geometry, follows the Eikonal equation, and the gradient of the field near the surface scaled by the signed distance value gives the distance vector $\mathbf{d}$ given by \eqref{eikonal}. The distance vector $\mathbf{d}$ is the vector pointing towards the closest point in the true surface $\Gamma$.
\begin{equation}
    ||\nabla_x f_{\theta} (x)||=1,
    \mathbf{d}=-f_{\theta}(x) \cdot \nabla_x f_{\theta} (x)
    \label{eikonal}
\end{equation}


We outline three primary methods for generating INR: from point clouds, images, and through generative AI models.

\subsubsection{Point Clouds}
Point cloud represents the surface of 3D geometry as points, and several methods exist to obtain point clouds for real-world geometries like LiDAR, photogrammetry, and structured light scanning. A variety of work has been focused on generating INRs from point cloud data \citep{gropp2020implicit,jignasu2024stitch,liu2019learning,ben2022digs,atzmon2020sal}. These works play in the paradigm of several architectural elements of the neural network and take advantage of the eikonal constraint and geometric features of the signed distance field to obtain INRs. 

\subsubsection{Images}
Multi-view 3D reconstruction from images is one of the most studied tasks in terms of 3D reconstruction \citep{wang2021neus,niemeyer2020differentiable}. Different methods exist utilizing differentiable rendering frameworks to obtain INRs from Images. All these frameworks are based on volume rendering method using INRs to render images. The difference between rendered images and actual images guides the training of INRs.

\subsubsection{Generative AI Models}
Diffusion-based models are gaining prominence in generative tasks \citep{ho2020denoising} and inspired from \citet{rombach2022high} work in latent diffusion \citet{erkocc2023hyperdiffusion} have developed diffusion over neural network weights of INRs to generate shapes. In their work, they have demonstrated the ability of the diffusion-based model to generate INRs for complex shapes. They successfully present a generation of complex shapes like planes, chairs, and cats. 

While raw geometric representations (images or point clouds) can be converted into INRs, they are not ideal sources for conducting ground truth comparisons of signed distance fields (SDFs) and associated distance vectors. In particular, obtaining accurate interior/exterior classifications and corresponding distance vectors from images or point clouds is nontrivial and often unreliable. These modalities lack the structural fidelity and geometric richness required to act as a precise ground truth for comparison.

To address this, we adopt triangle soup representations as the ground truth for our study. Triangle soups are widely used in computational geometry and simulation pipelines, particularly they are frequently used in SBM analysis to represent complex geometries. Unlike raw images or point clouds, triangle soups allow for well-defined and robust algorithms to compute both In/Out tests (ray tracing) and exact distance vectors. Our training procedure focuses on learning an INR from a given triangle soup, leveraging its geometric consistency to evaluate the accuracy of the learned representation for SBM-based analysis. The process of computing distance vectors from triangle soups is detailed in the work of \citet{yang2024optimal}. In \secref{section:generating_INR_polygon}, we devise sampling strategy and loss function to obtain INR which respects the requirement of SBM to have correct distance vectors close to the boundary.

\section{Generating Analysis Suitable INR from Triangle Soups}
\label{section:generating_INR_polygon}
This workflow outlines the process of converting a triangular soup into an INR. To fulfill the requirements for simulating using SBM, the training process leverages a carefully designed loss function and a strategic sampling approach. The loss function is structured to penalize errors in distance vector magnitude and direction, particularly in regions within the narrow-band $\delta$ (a region near the boundary where we want distance vectors to be very accurate), while also promoting the correct classification of grid points. The sampling strategy prioritizes regions within the narrow band, ensuring that the model learns to represent critical geometric features accurately. This approach ensures that the INR remains compatible with the SBM, resulting in a highly accurate representation that adapts to complex boundary geometries. We discuss below some of the critical aspects of the NN training process for generating the INR.

\paragraph{Geometry Re-scaling:} We define a cubic domain \\  
$\boldsymbol{\Omega} = [-1, -1, -1] \times [1, 1, 1]$. All the geometries are rescaled such that the volume occupied by $\Omega^-$ and $\Omega^+$ are in close tolerance for training purposes.

\paragraph{Hybrid Sampling:} We use a hybrid sampling method, where points are sampled uniformly on the surface \( \Gamma \), in the narrow band defined by a specified width \( \delta \),and as well as uniformly in cube $\Omega$. \( P_S \) be the set of points sampled uniformly on the true boundary \( \Gamma \):
\[
P_S = \{ \mathbf{x} \in \Gamma \mid \exists \, \mathbf{u} \in [0, 1]^2, \, \mathbf{x} = \text{Sample}(\mathbf{u}) \}
\]
\( P_{NB} \) be the set of points sampled uniformly within the narrow band around the true boundary \( \Gamma \):
\[
P_{NB} = \{ \mathbf{x} \in \mathbb{R}^n \mid \text{dist}(\mathbf{x}, S) \leq \delta \text{ and } \exists \, \mathbf{u} \in [0, 1]^m, \mathbf{x} = \text{Sample}(\mathbf{u}) \}
\]
\( P_{U} \) be the set of points sampled uniformly in the overall sampling space \( \Omega \):
\[
P_{U} = \{ \mathbf{x} \in \Omega \mid \exists \, \mathbf{u} \in [0, 1]^n, \, \mathbf{x} = \text{Sample}(\mathbf{u}) \}
\]
The total sampled points \( P \) is then defined as the union of these three sampling techniques:
\[
P = P_S \cup P_{NB} \cup P_{U}
\]
The number of points taken from each set can be controlled to balance the sampling strategy based on the requirements of the problem. The ablation study of impact of varying, $\boldsymbol{n (P_{S})}$,$\boldsymbol{n (P_{U})}$, and $\boldsymbol{n (P_{NB})}$ is presented in \secref{appendix:Neural Implicit}.

\paragraph{Loss Function:} The loss function is defined in \eqnref{equation:loss_function}, which takes the location of point $\mathbf{x}$, prediction $f_{\theta}(x)$, true distance $\mathbf{s}$, and true normal $\hat{n}$. The loss function uses clamped loss, which ensures that the distance near the narrow band, given by the width \(\delta\), is given more priority. This approach helps to stabilize the training process by focusing on the values of \(\mathbf{s}\) that are within a certain proximity to the true boundary $\mathbf{\Gamma}$. Similarly, the eikonal constraint and the property of INR as described in \eqref{eikonal} which is termed as Geometric Regularization Loss in \citet{gropp2020implicit} is applied wherever $|\mathbf{s}|<\omega$, where $\omega$ is a region close to the true boundary $\mathbf{\Gamma}$.
\begin{equation}
\begin{aligned}
    &L(f_\theta(\mathbf{x}), \mathbf{x}, \mathbf{s}, \hat{\mathbf{n}}) =  \int_{\Omega} \left( \text{clamp}(\mathbf{s}, \delta) - \text{clamp}(f_\theta(\mathbf{x}), \delta) \right)^2 \, d\mathbf{\Omega} \\
    & + \begin{cases}
        \lambda_g \int_{\Omega} \left( \left\| \nabla_{\mathbf{x}} f_\theta(\mathbf{x}) \right\| - 1 \right)^2 \, d\mathbf{\Omega} 
         \\ +~~\tau \int_{\Omega} \left( \frac{\nabla_{\mathbf{x}} f_\theta(\mathbf{x})}{\left\| \nabla_{\mathbf{x}} f_\theta(\mathbf{x}) \right\|} \cdot \hat{\mathbf{n}}(\mathbf{x}) - 1 \right)^2 \, d\mathbf{\Omega} & \text{if } |\mathbf{s}| < \omega \\
        0 & \text{otherwise}
    \end{cases}
\end{aligned}
\label{equation:loss_function}
\end{equation}
Here, $\lambda_g$ and $\tau$ are the Lagrange multipliers (hyperparameters) for the eikonal constraint and the normal similarity constraint, respectively. 

\paragraph{Network Architecture:} The foundational architecture used in INR is Multi-layer perceptron. The work here uses ImplicitNet (eight hidden layers with 512 neurons each with one skip-in layer). The Implicit Net was proposed in \citet{park2019deepsdf} as an Auto-Decoder network and used by~\citep{wang2021neus,sitzmann2020implicit,gropp2020implicit} and has been shown to perform well.

\paragraph{Evaluation Metric:} Evaluating the network for the particular task at hand is very crucial. The network should perform well in near boundary regions, but how close to the boundary is a hyper-parameter, which would depend on the computational discretization size we want for our computational analysis. In this analysis, we obtain a 3D grid of size $1024^3$ in the bounding box and select the points near the boundary region given by a threshold($\delta$). Then, Normalized Mean Squared Error is obtained to analyze the performance of the network as given by:
\begin{equation}
    \text{NMSE}_\delta = \frac{\frac{1}{N}\sum_{i=1}^{N} (s_i - f_\theta(x_i))^2}{\Delta},
    \label{nmse}
\end{equation}
where \( |y_i| < \delta \), and \( \Delta \) is the characteristic dimension.

\algoref{Algorithm 1 Implicit Network Training} presents the full training pipeline where a polygonal soup is taken as input and converted into respective INR. Hybrid Sampling and the loss function are described to train the architecture as explained to obtain appropriate INR of a particular shape. In this work, $target(p)$ and $normal(\hat{n})$ are obtained using libgl package~\citep{libigl}. The INRs are evaluated on Normalized Mean Square error as given in \eqnref{nmse}. \

\begin{algorithm}[t!]
  \footnotesize
    \caption{\textsc{ImplicitNetworkTraining:} Obtain Implicit Network from a Polygonal Soup}
    \label{Algorithm 1 Implicit Network Training}
    \begin{algorithmic}[1]
\Require Polygonal soup $P$, Bounding box $B$, Number of samples $N$, Implicit Neural Network $f_\theta$, Loss function $L(f_\theta, s)$
\Ensure Trained neural network $f_\theta$
\State Initialize a list $\mathcal{P}_U$ for sampled points within the bounding box

\For{$i = 1$ to $N$}
    \State $p_i \gets \text{sample point uniformly from } B$
    \State $\mathcal{P}_U \gets \mathcal{P}_U \cup p_i$ \Comment{Store uniform samples in bounding box}
\EndFor

\State Initialize a list $\mathcal{P}_S$ for sampled surface points
\For{each polygon $\triangle \in P$}
    \State Sample barycentric coordinates $(u, v, w)$ where $u + v + w = 1$ and $u, v, w \geq 0$
    \State Project point $q$ onto the surface of polygon $\triangle$ using barycentric coordinates
    \State $\mathcal{P}_S \gets \mathcal{P}_S \cup q$ \Comment{Store surface samples}
\EndFor

\State Define a narrow band around the surface with distance threshold $\delta$
\For{each point $q \in \mathcal{P}_S$}
    \State Sample points $q_{\text{band}}$ around $q$ within distance $\delta$
    \State $\mathcal{P}_{\text{NB}} \gets \mathcal{P}_{\text{NB}} \cup q_{\text{band}}$ \Comment{Store points in narrow band}
\EndFor

\State Combine all sampled points: $\mathcal{P} = \mathcal{P}_S \cup \mathcal{P}_U \cup \mathcal{P}_{\text{NB}}$
\For{each point $p \in \mathcal{P}$}
    \State Pass $p$ to the implicit neural network $f_\theta(p)$
    \State Compute the loss $L(f_\theta, \text{target}(p))$ \Comment{Calculate loss for each point}
\EndFor

\While{not converged}
    \State Update parameters $\theta \gets \theta - \eta \nabla_\theta L(f_\theta, \text{targets})$ \Comment{Optimize network parameters}
\EndWhile

\State \Return Trained implicit network $f_\theta$
\end{algorithmic}
\end{algorithm}

\subsection{Ablation Studies on INR Generation for Triangle Soup}
\label{appendix:Neural Implicit}

The sphere is one of the most computationally analyzed geometry.  We use icosphere in this case to approximate the sphere and obtain INR. The experiments, until specifically specified, take 100K uniformly sampled points, 25K points in narrowband with width $\delta = 0.001$, and 25K points in the surface.

In this study, we perform following comparisons:
\begin{enumerate}
    \item Comparison of the Implicit Net with different loss functions along the modified Implicit Loss as given in \eqnref{equation:loss_function}.
    \item Ablation study of loss function in \eqnref{equation:loss_function}.
    \item Ablation study of the sampling strategy.
\end{enumerate}

\begin{figure}[t!]
    \centering
    \includegraphics[width=0.2\linewidth]{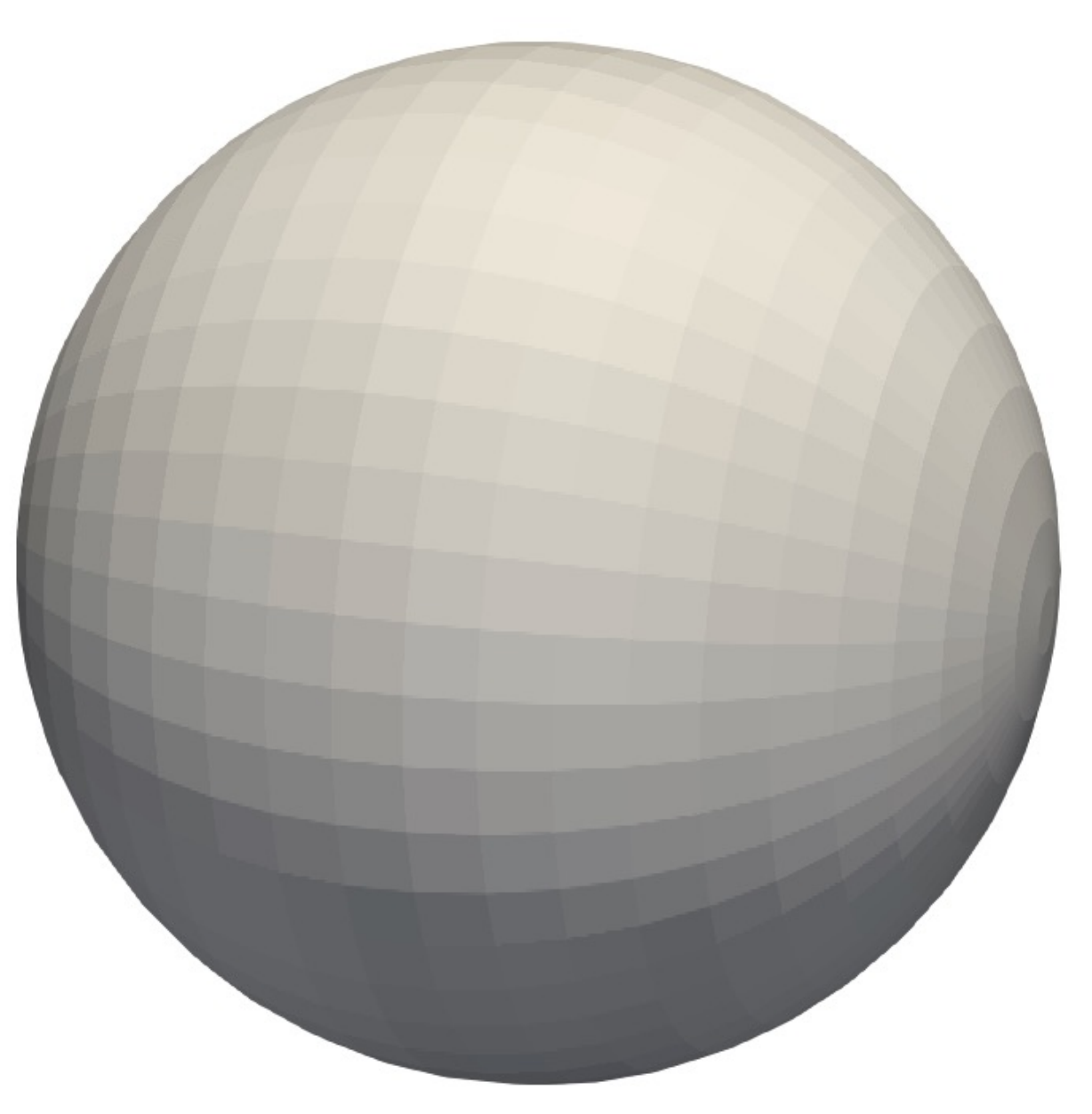}
    \caption{Ico-Sphere}
    \label{fig:ico_sphere}
\end{figure}

\begin{table*}[t!]
    \centering
    \caption{Comparison of Different Loss Functions with Fully Connected Network}
    \small
    \setlength{\extrarowheight}{2pt}
    \begin{tabular}{|c|l|c|c|}
        \hline
        \textbf{Case} & \textbf{Loss Function} & \textbf{Expression} & $NMSE_{0.1}$ \\ \hline
        a & $L_1$ Clamped Loss~\citep{park2019deepsdf} & $L_{1,Clamped}(\mathbf{s}, f_\theta(\mathbf{x})) = \frac{1}{N}\sum_{i=1}^{N} \left| \text{clamp}(\mathbf{s}_i, \delta) - \text{clamp}(f_\theta(\mathbf{x}), \delta) \right|$ & 0.0566 \\ \hline
        b & $L_2$ Clamped Loss & $L_{2,Clamped}(\mathbf{s}, f_\theta(\mathbf{x})) = \frac{1}{N}\sum_{i=1}^{N} \left( \text{clamp}(\mathbf{s}_i, \delta) - \text{clamp}(f_\theta(\mathbf{x}), \delta) \right)^2$ & 0.0407 \\ \hline
        c & $L_2$ Smooth Loss & $L_{2,Smooth}(\mathbf{s}, f_\theta(\mathbf{x})) = \frac{1}{N}\sum_{i=1}^{N} (1+\alpha ^{\left | \mathbf{s}_i \right|})(\mathbf{s}_i - f_\theta(\mathbf{x}))^2$ & 0.01895 \\ \hline
    \end{tabular}
    \label{tab:loss_functions}
\end{table*}

\tabref{tab:implicitnet_loss_functions} makes comparison between loss functions used and defined in \tabref{tab:loss_functions} along with loss function from \eqnref{equation:loss_function}. \eqnref{equation:loss_function} is hybrid of better performing $L_2 Clamped Loss$ from \tabref{tab:loss_functions} and geometric regularization as proposed in \citet{gropp2020implicit}. Similarly, the model with loss function as \eqnref{equation:loss_function} has the least value of the error. This inspires us to perform the ablation study to analyze the effect of removing different terms from the loss function, as shown in \tabref{tab:ablation_of_loss}. The Eikonal loss and Normal loss are defined below which are components of the original loss function.
\[\text{Eikonal Loss:} \quad \lambda_g \int_{\Omega} \left( \left\| \nabla_{\mathbf{x}} f_\theta(\mathbf{x}) \right\| - 1 \right)^2 \, d\Omega\]
\[\text{Normal Loss:} \quad \tau \int_{\Omega} \left( \frac{\nabla_{\mathbf{x}} f_\theta(\mathbf{x})}{\left\| \nabla_{\mathbf{x}} f_\theta(\mathbf{x}) \right\|} \cdot \hat{\mathbf{n}}(\mathbf{x}) - 1 \right)^2 \, d\Omega\]

\begin{table}[t!]
    \centering
    \caption{Comparison of Different Loss Functions with Implicit Net}
    \label{tab:implicitnet_loss_functions}
    \small
    \setlength{\extrarowheight}{2pt}
    \begin{tabular}{|c|c|c|}
        \hline
        \textbf{Case} & \textbf{Loss Function} & $\mathbf{NMSE_{2^{-10}}}$ \\ \hline
        a & $L_1$ Clamped Loss~\citep{park2019deepsdf} & $2.17 \times 10^{-7}$ \\ \hline
        b & $L_2$ Clamped Loss & $1.96 \times 10^{-7}$ \\ \hline
        c & $L_2$ Smooth Loss & $2.08 \times 10^{-7}$ \\ \hline
        d & \textbf{IGR Loss (\eqnref{equation:loss_function})} & $\mathbf{1.54 \times 10^{-7}}$ \\ \hline
    \end{tabular}
\end{table}

\begin{table}[t!]
    \centering
    \caption{Ablation study of IGR Loss Functions based on \( NMSE_{2^{-10}} \)}
    \label{tab:ablation}
    \small
    \begin{tabular}{|c|c|}
        \hline
        \textbf{Loss Function} & $\mathbf{NMSE_{2^{-10}}}$ \\ \hline
        \textbf{IGR Loss (\eqnref{equation:loss_function})} & $\mathbf{1.54 \times 10^{-7}}$ \\ \hline
        w/o Eikonal loss & $1.97 \times 10^{-7}$ \\ \hline
         w/o Eikonal and Normal loss& $1.96 \times 10^{-7}$ \\ \hline
    \end{tabular}
    \label{tab:ablation_of_loss}
\end{table}

\begin{table*}[t!]
    \caption{Comparison of sampling strategies based on NMSE at resolution \(2^{-10}\).}
    \centering
    \begin{tabular}{|c|c|c|c|}
        \hline
        \textbf{Approach} & $n(\mathcal{P})$ & $n(\mathcal{P_{U}})$,$n(\mathcal{P_{S}})$, $n(\mathcal{P_{NB}})$ & $\mathbf{NMSE_{2^{-10}}}$ \\ 
        \hline
        \textbf{Only Uniform} & 150K & 150K, 0, 0 & $2.0 \times 10^{-7}$ \\
        \hline
        \textbf{Only Narrow Band} & 150K & 0, 0, 150K & $2.25 \times 10^{-7}$ \\
        \hline
        \textbf{Only Surface} & 150K & 0, 150K, 0 & $2.4 \times 10^{-7}$ \\
        \hline
        \textbf{Hybrid} & 150K & 90K, 28K, 32K & $\mathbf{3.81 \times 10^{-8}}$ \\
        \hline
    \end{tabular}
    \label{tab:hybrid_sampling}
\end{table*}

It can be clearly seen from \tabref{tab:implicitnet_loss_functions} that the loss function proposed in \eqnref{equation:loss_function} performs well in comparison to other intuitive loss functions. 
All of the above experiments were performed with consideration to the same sampling set. In \tabref{tab:hybrid_sampling}, the loss function and the network architecture are fixed, and a comparison is made between different sampling strategies with an equal number of points but sampled differently The number of points in hybrid sampling is obtained through Bayesian Optimization which has the least value of the Normalized Mean Square Error.

\section{Linear Elasticity}
\label{Sec:Elasticity}
Consider the region $\Omega $ with boundary $\Gamma = \Gamma_D \cup \Gamma_N$ (as shown in \figref{fig:SBM}). 
The strong form partial differential equation for linear elasticity which relates the internal stress $\sigma_{ij}$ with body force $f_i$ is
\begin{align}
\partial_i \sigma_{ij} = -f_i.
\end{align}
The material constitutive relation relates the stress induced, $\sigma_{ij}$, with strain, $\epsilon_{kl}$ as given by:
\begin{align}
    \sigma_{ij}= \mathbf{\mathcal{C}}_{ijkl} \epsilon_{kl} 
\end{align}
The strain is defined as the symmetric gradient of the displacement field, $u_i$:
\begin{align}
    \epsilon_{kl} = \frac{(\partial_k u_l + \partial_l u_k)}{2}
\end{align}
The system has boundary conditions on the displacement field given by the following
\begin{align}
    u_i=g_i \quad on \quad \Gamma_D\\
    \sigma_{ij}n_j = h_i \quad on \quad \Gamma_N
\end{align}

Using finite element method, the Galerkin formulation with test function $w_i$ can be written as 
\begin{align}
   (w_i,\partial_j \sigma_{ij})_{\Omega}= (w_i,-f_i)_{\Omega}.
\end{align}
By applying the Gauss-Divergence theorem, the weak form can be written  as:
\begin{align}
   (\partial_i w_j,\sigma_{ij})_{\Omega}= (w_i,f_i)_{\Omega} + \langle w_i,\sigma_{ij}n_j \rangle_{\Gamma}.
\end{align}
Now, $\Gamma = \Gamma_D \cup \Gamma_N$ , we choose $w_i=0$ on $\Gamma_D$ and   $\sigma_{ij}n_j = h_i$ on  $\Gamma_N$. So, the above equation reduces to:
\begin{align}
   (\partial_i w_j,\sigma_{ij})_{\Omega}= (w_i,f_i)_{\Omega} + \langle w_i,h_i\rangle_{\Gamma_N}
\end{align}

\subsection{Shifted Dirihclet Boundary Conditions}

Let us consider a surrogate Dirichlet boundary $\tilde{\Gamma}_D$ positioned near the actual Dirichlet boundary $\Gamma_D$. By utilizing the distance measure between these boundaries, we can express the velocity vector through its Taylor expansion:
\begin{equation}
\tilde{u}_i + \partial_j\tilde{u}_i d_i+ R_D(\tilde{u}_i, d_i)(\tilde{x}) = u_i(M_h(\tilde{x})), \quad \text{on } \tilde{\Gamma}_{D,h},
\end{equation}
where the remainder term $R_D(u, d)$ exhibits the property that $\| R_D(u, d) \| = o(\| d \|^2)$ as $\| d \| \to 0$. Also
$\quad u_i(M_h(\tilde{x}))=g_i$. Then,
\begin{equation}
    u_i-\tilde{g_i}=\tilde{u}_i + \partial_j\tilde{u}_i d_i - g_i
    \label{taylor}
\end{equation}
where, 
\[S_{D,h}v_i:=v_i + \partial_jv_i d_i\]
$S_{D,h}$ is defined as a shift operator which is shifting the boundary condition from true boundary to surrogate boundary.
\subsection{Linear Elasticity on Surrogate Domain}
Instead of solving the original problem we solve the problem in an extended domain as described earlier. Then, the Galerkin formulation in $\tilde{\Omega}$ is:
\begin{align}
   (\partial_i w_j,\sigma_{ij})_{\tilde{\Omega}}= (w_i,f_i)_{\tilde{\Omega}} + \langle w_i,\sigma_{ij}\tilde{n_j} \rangle_{\tilde{\Gamma}}
\end{align}
Now, the assumptions that were made for $\Gamma$ don't hold for $\tilde{\Gamma}$. We apply Nitsche's method assuming that we have $u=\tilde{g}$ on $\tilde{\Gamma}_D$. Similarly, assume $ \sigma_{ij}\tilde{n}_j = \tilde{h}_i $ on $\tilde{\Gamma}_N$. Then, the weak form is as follows:
\begin{equation}
   \begin{aligned}
      (\partial_i w_j,\sigma_{ij})_{\tilde{\Omega}} &= (w_i,f_i)_{\tilde{\Omega}} +
      \langle w_i,\tilde{h}_i\rangle_{\tilde{\Gamma}_N} \\
      &\quad +  \underbrace{\langle w_i,\sigma_{ij} \tilde{n}_j \rangle_{\tilde{\Gamma}_D}}_{\text{Consistency}} \quad
      \underbrace{-\langle \mathbf{\mathcal{C}}_{ijkl} \partial_k w_l \tilde{n}_j,u_i-\tilde{g}_i \rangle_{\tilde{\Gamma}_D}}_{\text{Adjoint Consistency}} \\
      &\quad \underbrace{- \langle w_i, \gamma h^{-1}(u_i - \tilde{g}_i)\rangle_{\tilde{\Gamma}_D}}_{\text{Penalty Term}}
   \end{aligned}
   \label{eq:basic_nitsche1}
\end{equation}
Then, using the taylor series exapnsion from \eqnref{taylor}. We get,
\begin{equation}
   \begin{aligned}
      (\partial_i w_j,\sigma_{ij})_{\tilde{\Omega}} &= (w_i,f_i)_{\tilde{\Omega}} +
      \langle w_i,\tilde{h}_i\rangle_{\tilde{\Gamma}_N} \\
      &\quad +  \underbrace{\langle w_i,\sigma_{ij} \tilde{n}_j \rangle_{\tilde{\Gamma}_D}}_{\text{Consistency}} \quad
      \underbrace{-\langle \mathbf{\mathcal{C}}_{ijkl} \partial_k w_l \tilde{n}_j,S_{D,h}u_i-g_i \rangle_{\tilde{\Gamma}_D}}_{\text{Adjoint Consistency}} \\
      &\quad \underbrace{- \langle w_i, \gamma h^{-1}(S_{D,h} u_i - g_i)\rangle_{\tilde{\Gamma}_D}}_{\text{Penalty Term}}
   \end{aligned}
   \label{eq:basic_nitsche2}
\end{equation}
\textbf{Remarks:} The consistency, adjoint consistency, and penalty term are applied on the $\tilde{\Gamma}_D$. The method is the application of Nitsche's method, which is popularly used in the Finite Element Method to enforce the Dirichlet boundary condition weakly. The Neumann boundary condition requires additional treatment, which is not discussed here, and readers are referred to~\citet{atallah2021shifted}.

\subsection{Integrating INR with SBM}
\label{sec:Implementation}

In this work, the INR is inferred to obtain the incomplete octrees. The implicit neural network is employed to selectively refine or discard octree elements, guided by a function \( F() \), which determines the level of refinement needed for a particular location.The algorithm begins by initializing a complete octree \( O \), and the implicit network \( f_{\theta} \) is applied to each octree element. The function \( F() \) encodes refinement criteria and is used to decide whether a given element should be refined further or pruned. The objective is to construct an incomplete octree \( O_{\text{incomplete}} \) that retains only the essential octants, thereby optimizing both storage and computational requirements. During the traversal of the complete octree, each octant \( S \in \mathcal{O} \) is evaluated by the implicit network. If the network infers values greater than 0 (not inside the geometry) and the octant satisfies the refinement criteria imposed by \( F() \), the octant is retained. Otherwise, it is pruned. Once this process is complete, the remaining octants form the incomplete octree \( \mathcal{O}_{\text{incomplete}} \), which is refined accordingly based on the remaining leaf nodes. 

\textbf{Remarks:} The algorithms presented here are for all generalized cases. We are only interested where $\text{meas}(T \cap \Omega) > 0 $ which is equivalent to $\lambda_{criteria}=1$.

\begin{algorithm}[t!]
  \footnotesize
    \caption{\textsc{ImplicitOctreeGeneration:} Obtain incomplete octree using implicit network}
    \label{Algorithm ImplicitOctreeGeneration}
    \begin{algorithmic}[1]
\Require Complete octree $\mathcal{O}$, Implicit network $f_\theta$, Function $F()$
\Ensure Incomplete octree $\mathcal{O}_{\text{incomplete}}$
\State Initialize empty set $T$ for storing octree leaf nodes

\State \textbf{Step 1: Apply implicit network to prune octree}
\For{each octant $S \in \mathcal{O}$}
    \If{$f_\theta(S) \geq 0$} \Comment{Use implicit network to determine active octants}
        \If{level of $S$ is acceptable based on $F()$}
            \State $T.\texttt{push}(S)$ \Comment{Store the selected octants in $T$}
        \EndIf
    \EndIf
\EndFor

\State \textbf{Step 2: Generate incomplete octree}
\State $\mathcal{O}_{\text{incomplete}} \gets$ Refine and prune $\mathcal{O}$ based on the leaf nodes in $T$

\State \Return $\mathcal{O}_{\text{incomplete}}$
\end{algorithmic}
\end{algorithm}

\algoref{Algorithm ImplicitOctreeGeneration} is for generating incomplete octree based on \citet{saurabh2021scalable}. The algorithm traverses through each element and classifies each Gauss point as outside (if $f_\theta(gp)\geq0$)  or inside (if $f_\theta(gp)<0$). The count is saved for inside the Gauss point, and $\lambda$ and $\lambda_{criteria}$ are used for classifying whether an element is classified as "FalseIntercepted," "Exterior," "Interior," or "TrueIntercepted." Next, the algorithm presented in \algoref{alg:boundary} based on \citet{yang2024optimal}, which takes the marker \( M \), is used to generate the optimal surrogate boundary. \algoref{Algorithm: DistanceFunctionCalculationUsingImplicitNetworkOneGP} outlines the procedure for computing the distance vector for the Gauss points located at the surrogate boundary by calculating the gradient of $f_{\theta}$ as presented in  \eqnref{eikonal}. The gradient is computed numerically by using two stencils on each axis using the central difference method. To optimize the process, a mapping mechanism is employed, ensuring that each gradient computation is performed only once.

\begin{algorithm}[t!]
  \footnotesize
    \caption{\textsc{IdentifySurrogateBoundary:} Surrogate Boundary Identification Using Neural Implicit Network}
    \label{alg:boundary}
    \begin{algorithmic}[1]
\Require Octree mesh $\mathcal{O}$, threshold factor $\lambda$, implicit network $f_\theta$
\Ensure Surrogate boundary $\tilde{\Gamma}$, Element marker $M$
\State Initialize marker $M \gets []$
\For{each element $e \in \mathcal{O}$} \Comment{Loop over all elements in the octree mesh}
    \State Initialize $count \gets 0$
    \For{each Gauss point $gp \in \text{GaussPoints}(e)$} \Comment{Loop over Gauss points in element $e$}
        \If{$f_\theta(gp) < 0$} \Comment{Classify Gauss point as interior based on implicit network}
            \State $count \gets count + 1$ \Comment{Increment count for Interior Gauss points}
        \EndIf
    \EndFor
    \State $\lambda_c \gets \frac{\text{count}}{\text{num\_gp}}$ \Comment{Compute fraction of Interior Gauss points}
    \If{$\lambda_c \geq \lambda$} 
        \State $M[e] \gets \text{FalseIntercepted}$ \Comment{Mark element as FalseIntercepted}
    \ElsIf{$count == 0$} 
        \State $M[e] \gets \text{Exterior}$ \Comment{Mark element as Exterior}
    \ElsIf{$count == \text{num\_gp}$} 
        \State $M[e] \gets \text{Interior}$ \Comment{Mark element as Interior}
    \Else
        \State $M[e] \gets \text{TrueIntercepted}$ \Comment{Mark element as TrueIntercepted}
    \EndIf
\EndFor
\State Extract the surrogate boundary $\tilde{\Gamma}$ based on marker $M$ as outlined in \citet{yang2024optimal}
\State \Return $\tilde{\Gamma}$, $M$
\end{algorithmic}
\end{algorithm}

\begin{algorithm}[t!]
  \footnotesize
    \caption{\textsc{ComputeDistanceVector:} Distance Vector Calculation using Neural Implicit Network}
    \label{Algorithm: DistanceFunctionCalculationUsingImplicitNetworkOneGP}
    \begin{algorithmic}[1]
\Require Gauss point position on the surrogate boundary ($Q$), Implicit network $f_\theta$, Mapping $M$
\Ensure Distance vector $(\boldsymbol{d}_{gp})$ for Gauss point $Q$
\If{$Q$ exists in $M$}
    \State Retrieve $(\boldsymbol{d}_{gp})$ from $M(Q)$ \Comment{Retrieve precomputed distance vector if available}
\Else
    \State Compute $\nabla f_\theta(Q)$ \Comment{Calculate gradient of implicit network at $Q$}
    \State Compute signed distance vector $\mathbf{d}_{gp} = \left( \frac{\nabla f_\theta(Q)}{\|\nabla f_\theta(Q)\|} \right) \times f_\theta(Q)$ \Comment{Determine distance via implicit network}
    \State Store $\mathbf{d}_{gp}$ in $M(Q)$ \Comment{Save mapping from $Q$ to $\mathbf{d}_{gp}$ for future reference}
\EndIf
\State \Return Distance vector $(\boldsymbol{d}_{gp})$
\end{algorithmic}
\end{algorithm}

\begin{figure}[t!]
    \centering
    \begin{tikzpicture}[scale=.6]
    \draw[thick, ->] (0.5, 0) -- (6, 0) node[right] {};
    \draw[thick, ->] (0, 0.5) -- (0, 4) node[above, rotate=90] {};

    \draw[thick, green!70!black] (0.5, 2) -- (5.5, 2); 
    \draw[thick, blue] (0.5, 0.5) -- (5, 3.5);          

    \node[rotate=90] at (-0.5, 2.5) {No. of operations};
    \node at (3, -0.5) {Triangles};
    \begin{scope}[shift={(4.5, 3.5)}] 
        \draw[thick, blue] (0.2, 0.9) -- (1, 0.9) node[right, black] {\small Traversal in $\Delta s$};
        \draw[thick, green!70!black] (0.2, 0.4) -- (1, 0.4) node[right, black] {\small Neural Inference};
    \end{scope}
    \end{tikzpicture}
    \caption{Neural inference required for computing the distance vector occurs in constant operations. It depends upon the number of neural network layers with other hardware and software constraints. For Triangle Soup, distance vector computation needs traversals across all the triangles. The number of required operations increases with an increase in the number of triangles.}
    \label{fig:neural_inference_against_traversal_in_triangles}
\end{figure}
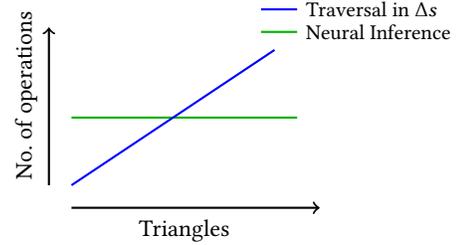

All the algorithms are based on function queries (forward pass across the neural network or Neural Inference) in place of exhaustive traversal through a polygonal mesh~\citep{atallah2021shifted,yang2024optimal}. \figref{fig:neural_inference_against_traversal_in_triangles} represents the linearly increasing number of operations for computing distance vectors through Triangle Soup. For each triangle, the shortest distance needs to be computed. As outlined earlier, computing distance vectors through INR only requires Inference calls, which are independent of the geometry. The calls will take a constant number of operations dependent upon the size of the Multi-Layer Perceptron (number of matrix-matrix multiplications or matrix-vector multiplications).

\section{Results}
\label{Sec:Results}

This section begins with \secref{section:Val_Implicit} where the analysis suitability of an Implicit Neural Representation is quantified by comparing it to the ground truth (in this case is taken to be a polygonal mesh). The comparison is performed across geometries of different complexities. Next, we present the validation of the method with convergence analysis on a ring in \secref{sec:validation_2d}. Then, \secref{sec:validation_3d} presents the comparison between the displacement magnitude of INR of ico-sphere and its corresponding ground truth to establish the accuracy in a 3D case, along with a comparison of assembly time and octree mesh generation time between ico-sphere (with increasing number of triangles) and its INR to detail the efficiency associated with the method. The rest of the sections include simulations with a given INR (generated either from the method described in \algoref{Algorithm 1 Implicit Network Training} or from state of art method) using our framework. Wherever details regarding the generation of INR aren't provided, we use the algorithm described in \algoref{Algorithm 1 Implicit Network Training}. Wherever explicitly not mentioned, INRs are placed inside cubic domain of $[-1,-1,-1]\times[ 1, 1, 1]$ (or $[-1,-1]\times[ 1, 1]$ square in 2D), after which we throw out the grids not contributing towards the surrogate domain $\tilde{\Omega}$ (referred as carving out).

\subsection{Accuracy of INRs for SBM}
\label{section:Val_Implicit}

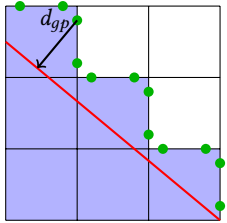
\begin{figure}[b!]
    \centering
    \begin{tikzpicture}[scale=0.95]
        \fill[blue!30] (0,0) rectangle (1,3)
                       (1,0) rectangle (2,2)
                       (2,0) rectangle (3,1);
        
        \draw[thin] (0,0) grid (3,3);
        
        \draw[thick, red] (0,2.5) -- (3,0);
        
        \foreach \x/\y in {0.2/3,0.8/3,1/2.8,1/2.2,1.2/2,1.8/2,2/1.2,2/1.8,2.2/1,2.8/1,3/0.2,3/0.8}{
            \fill[green!70!black] (\x, \y) circle (2pt);
        }
        
        \draw[->, thick] (1,2.8) -- (0.443,2.131);
        \node at (0.7, 2.8) {$d_{gp}$};
        
    \end{tikzpicture}
    \caption{Distance Vector, $\mathbf{d_{GP}}$, corresponding to the gauss points (marked in green) at the surrogate boundary. Distance Vector points in the shortest distance to the true boundary (line in red).}
    \label{fig:evaluating the distance vector}
\end{figure}

INRs of different shapes, as classified as complex and simple, as presented in  \tabref{tab:complexity_objects}, are obtained using \algoref{Algorithm 1 Implicit Network Training}. he end goal of this work is to use INRs for analysis using SBM. To ensure accuracy in this framework, it is essential to compute correct distance vectors at the Gauss points located on the surrogate boundary as presented in  \figref{fig:evaluating the distance vector}.

\begin{figure*}[t!]
    \centering
    \begin{subfigure}[t]{0.2\linewidth}
        \centering
        \includegraphics[width=\linewidth, trim=1.5in 1.5in 1.5in 1.5in, clip]{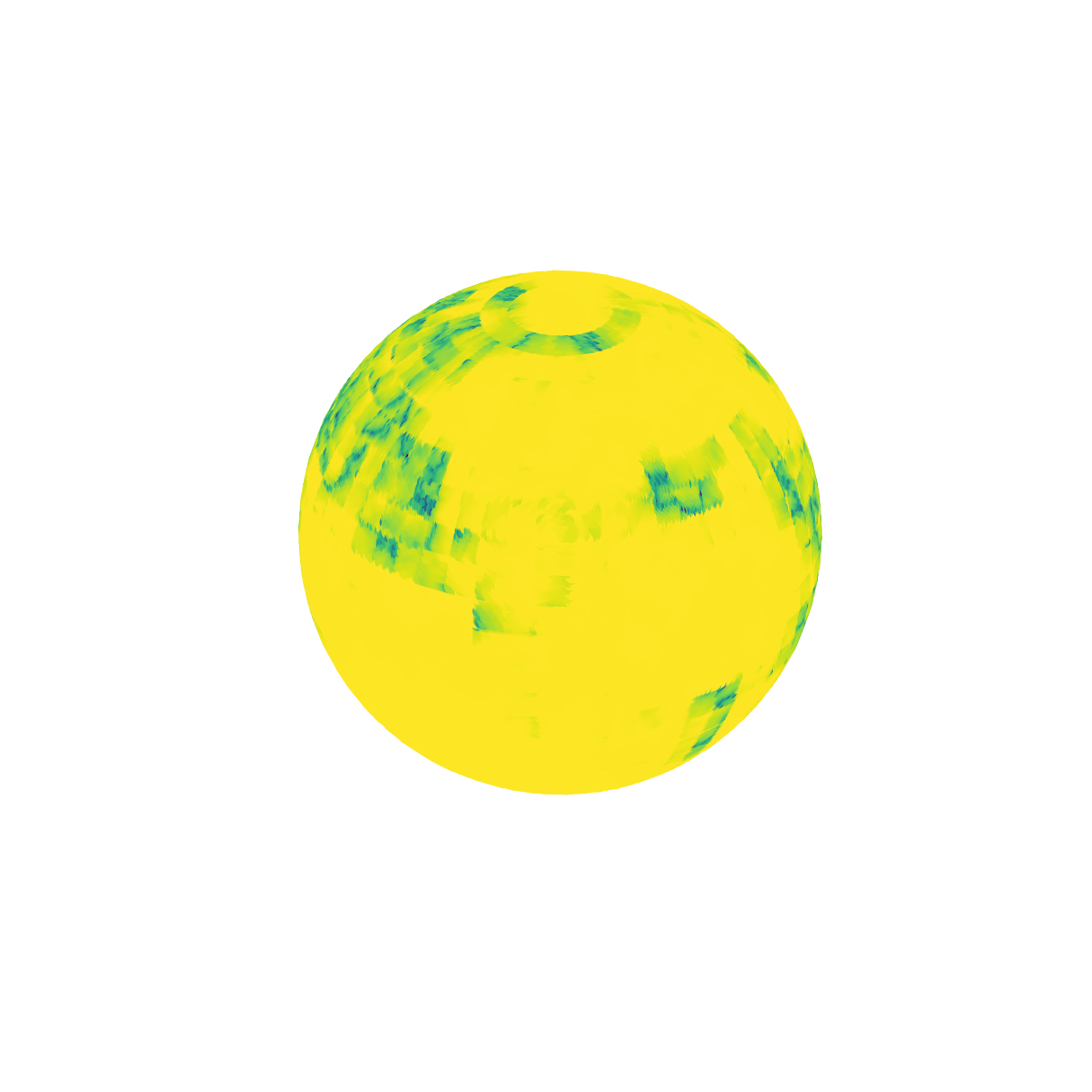}
        \caption{Sphere}
    \end{subfigure}
    \begin{subfigure}[t]{0.2\linewidth}
        \centering
        \includegraphics[width=\linewidth, trim=1.5in 1.5in 1.5in 1.5in, clip]{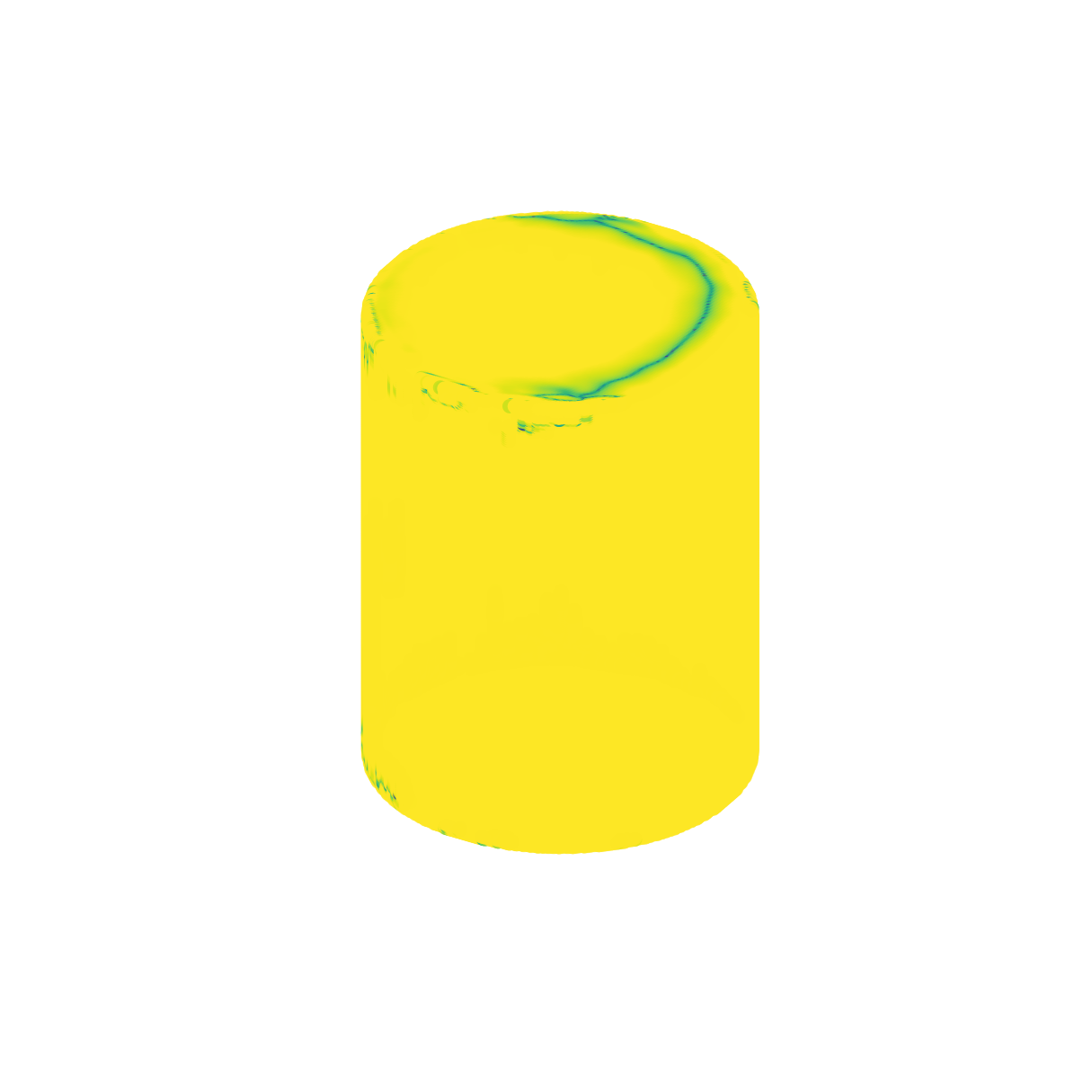}
        \caption{Cylinder}
    \end{subfigure}
    \begin{subfigure}[t]{0.2\linewidth}
        \centering
        \includegraphics[width=\linewidth, trim=2.3in 2.8in 2.3in 1.0in, clip]{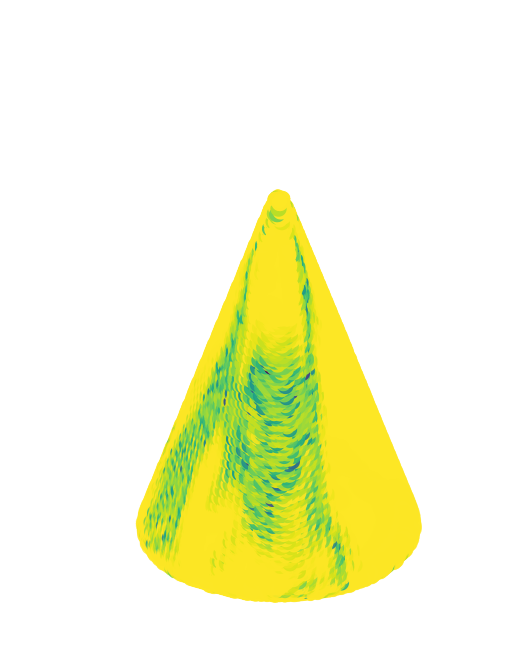}
        \caption{Cone}
    \end{subfigure}
    \begin{subfigure}[t]{0.2\linewidth}
        \centering
        \includegraphics[width=\linewidth, trim=1.5in 2.0in 1.5in 1.5in, clip]{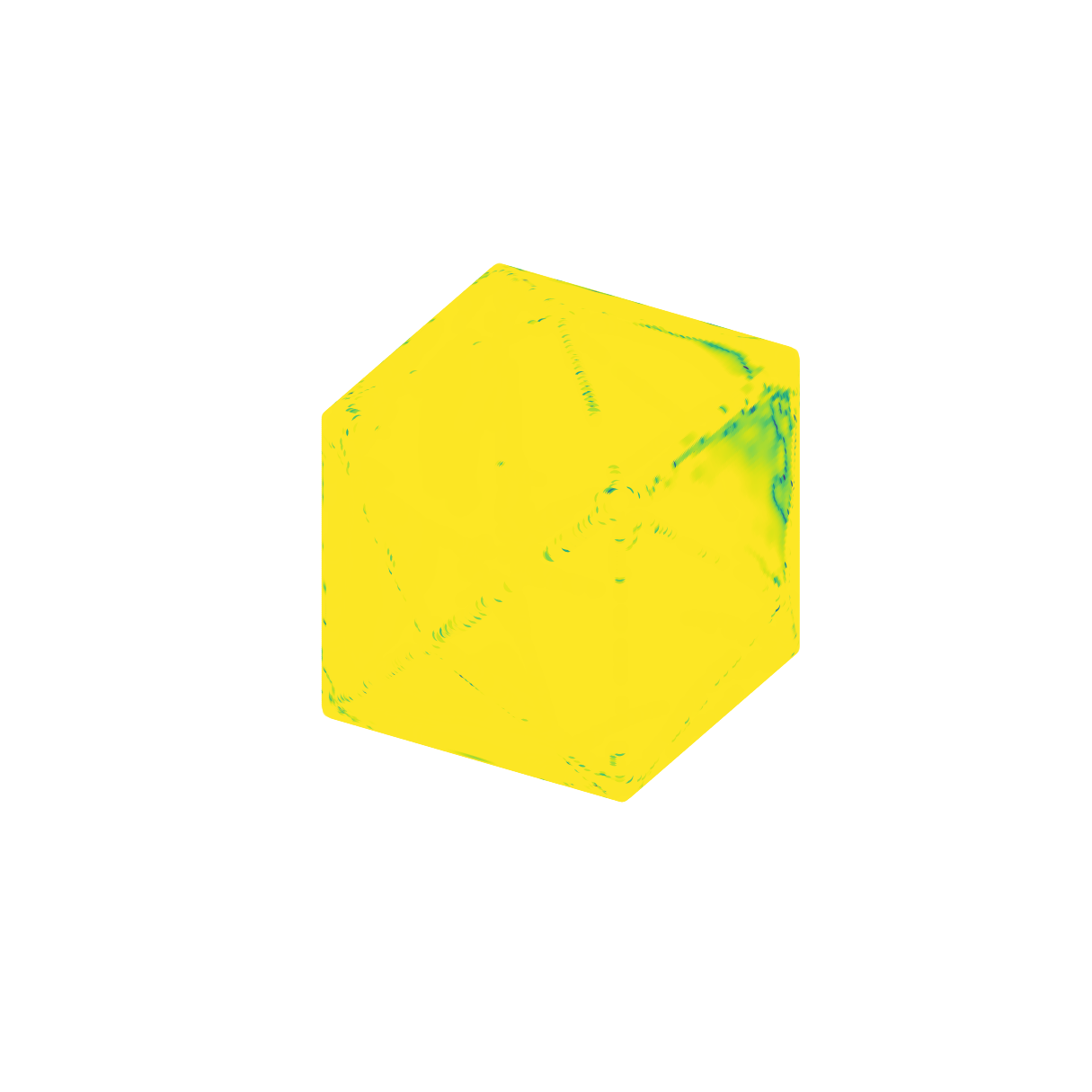}
        \caption{Tetrakis}
    \end{subfigure}
    \begin{subfigure}[t]{0.2\linewidth}
        \centering
        \includegraphics[width=\linewidth, trim=1.5in 2.0in 1.5in 1.5in, clip]{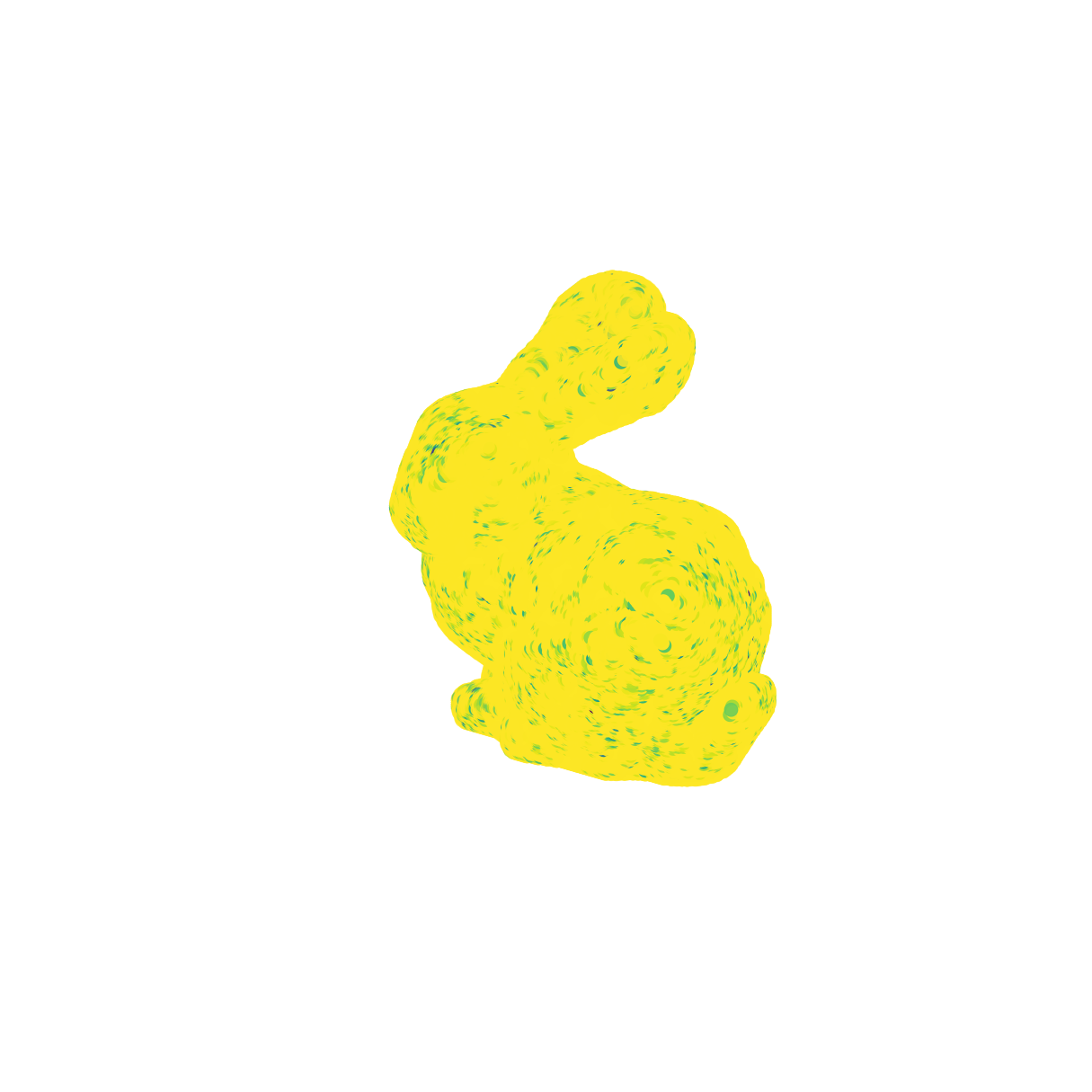}
        \caption{Bunny}
    \end{subfigure}
    \begin{subfigure}[t]{0.2\linewidth}
        \centering
        \includegraphics[width=\linewidth, trim=2.0in 2.3in 2.0in 2.0in, clip]{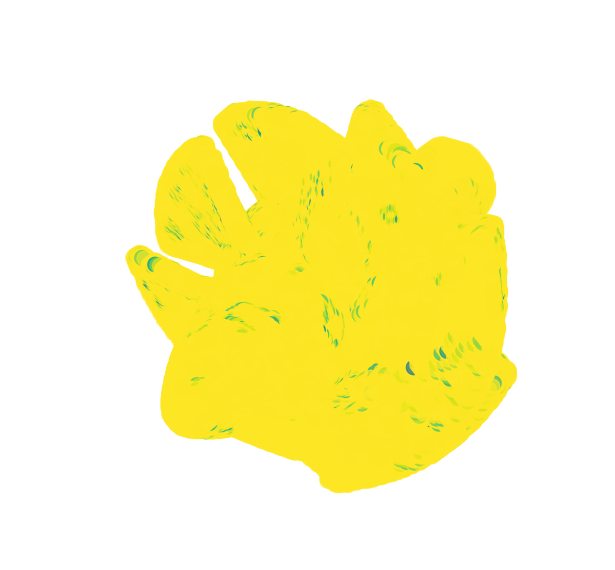}
        \caption{Turbine}
    \end{subfigure}
    \hspace{0.02\linewidth}
    \begin{subfigure}[t]{0.4\linewidth}
        \centering
        \includegraphics[width=\textwidth,trim={0.0in 0.5in 0.0in 0.0in}, clip]{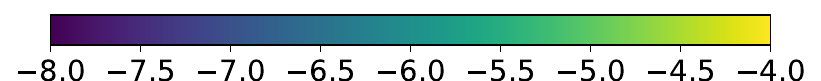}
    \end{subfigure}
    \caption{Plot of $log_{10}(|f_\theta(x_{gp})-s(x_{gp})|)$ for refinement $h=\nicefrac{\Delta}{2^8}$. The error mostly is in order of $10^{-4}$ for all the geometries. The plot shows the spatial variation of error in the magnitude of the distance vectors.}
    \label{fig:comparison_gp_distv}
\end{figure*}

\begin{figure*}[t!]
    \centering
    \begin{subfigure}[t]{0.2\linewidth}
        \centering
        \includegraphics[width=\linewidth, trim=1.5in 1.5in 1.5in 1.5in, clip]{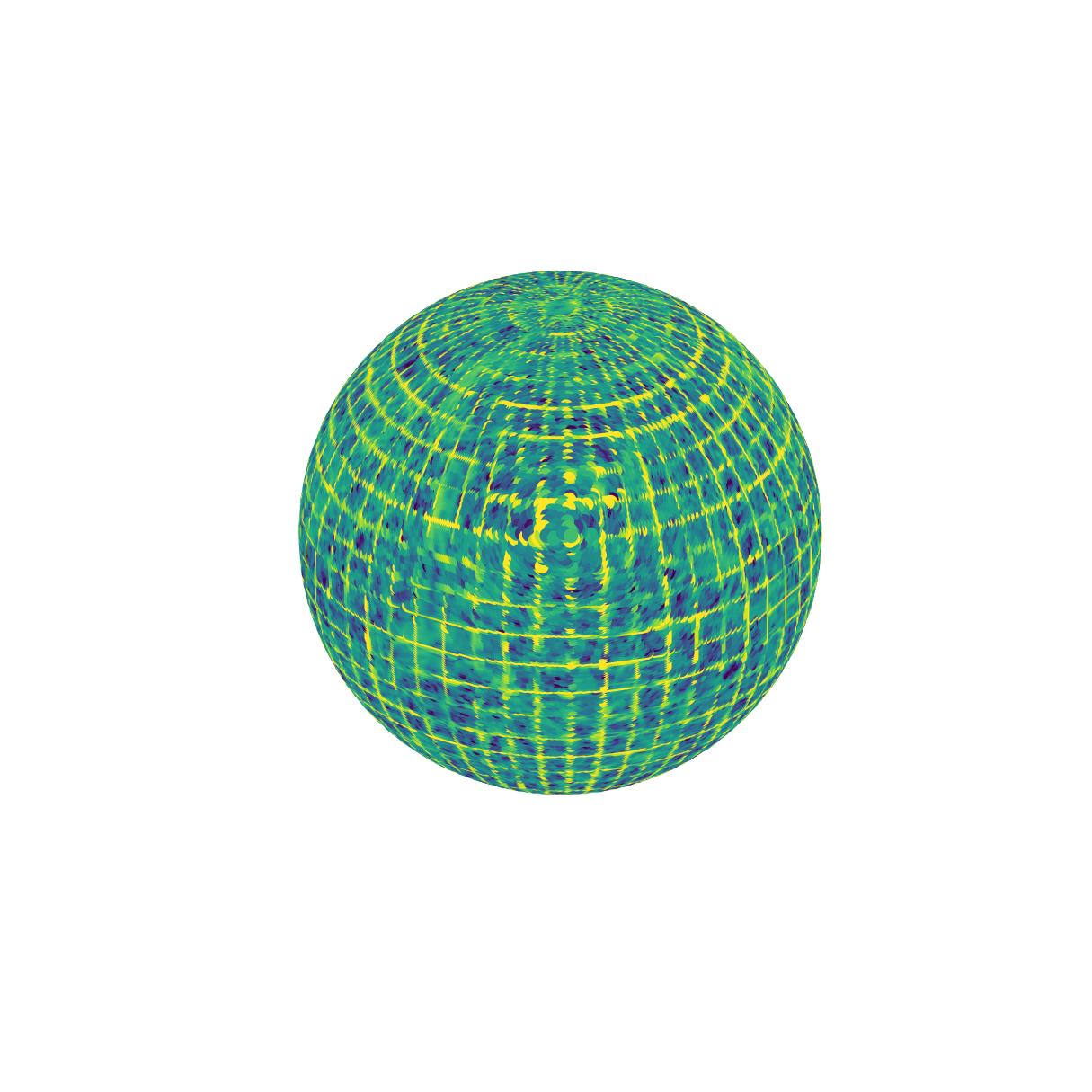}
        \caption{Sphere}
    \end{subfigure}
    \begin{subfigure}[t]{0.2\linewidth}
        \centering
        \includegraphics[width=\linewidth, trim=1.5in 1.5in 1.5in 1.5in, clip]{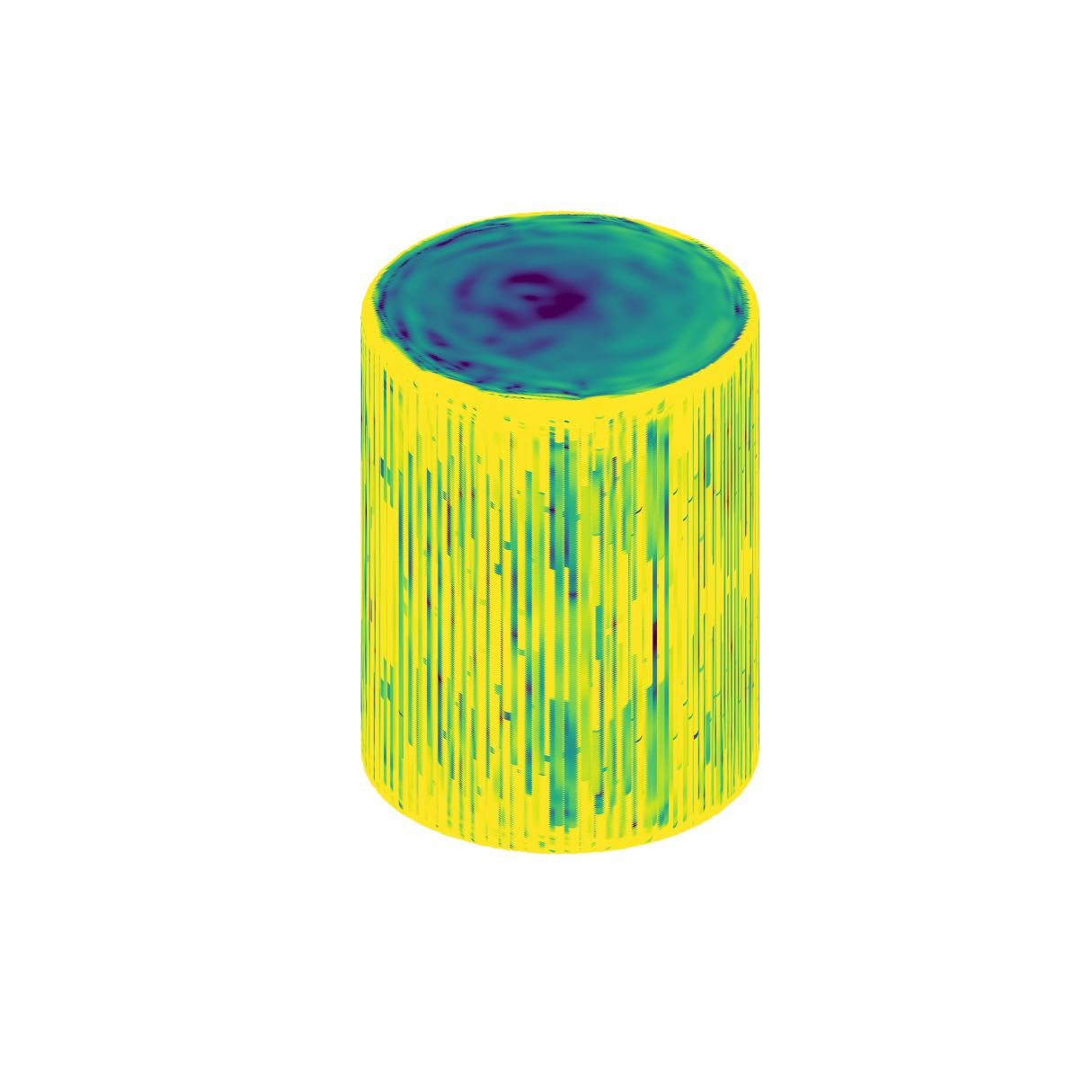}
        \caption{Cylinder}
    \end{subfigure}
    \begin{subfigure}[t]{0.2\linewidth}
        \centering
        \includegraphics[width=\linewidth, trim=2.3in 2.8in 2.3in 1.0in, clip]{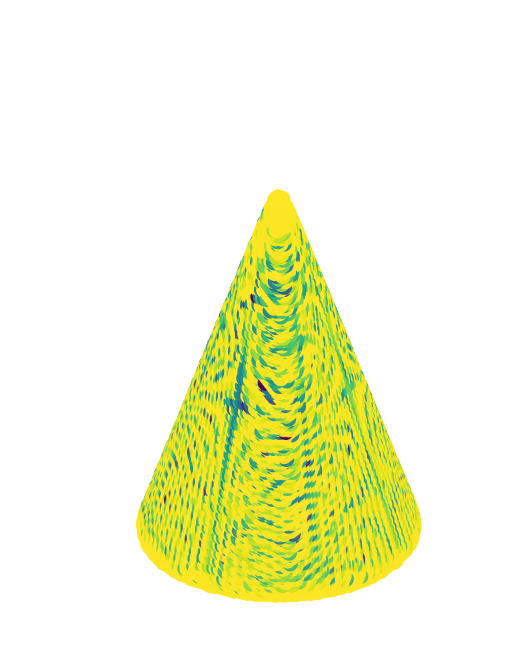}
        \caption{Cone}
    \end{subfigure}
    \begin{subfigure}[t]{0.2\linewidth}
        \centering
        \includegraphics[width=\linewidth, trim=1.5in 2.0in 1.5in 1.5in, clip]{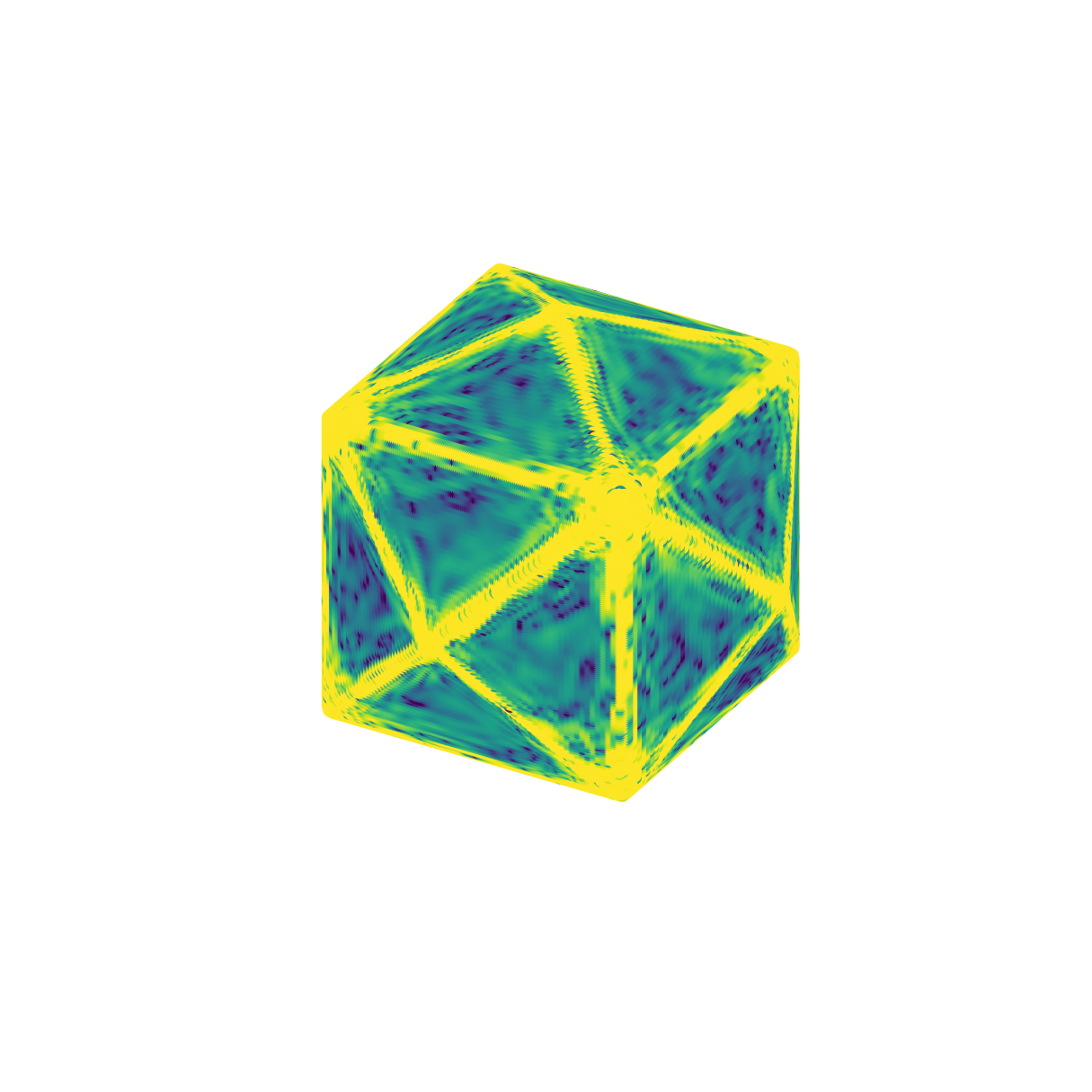}
        \caption{Tetrakis}
    \end{subfigure}
    \begin{subfigure}[t]{0.2\linewidth}
        \centering
        \includegraphics[width=\linewidth, trim=1.5in 2.0in 1.5in 1.5in, clip]{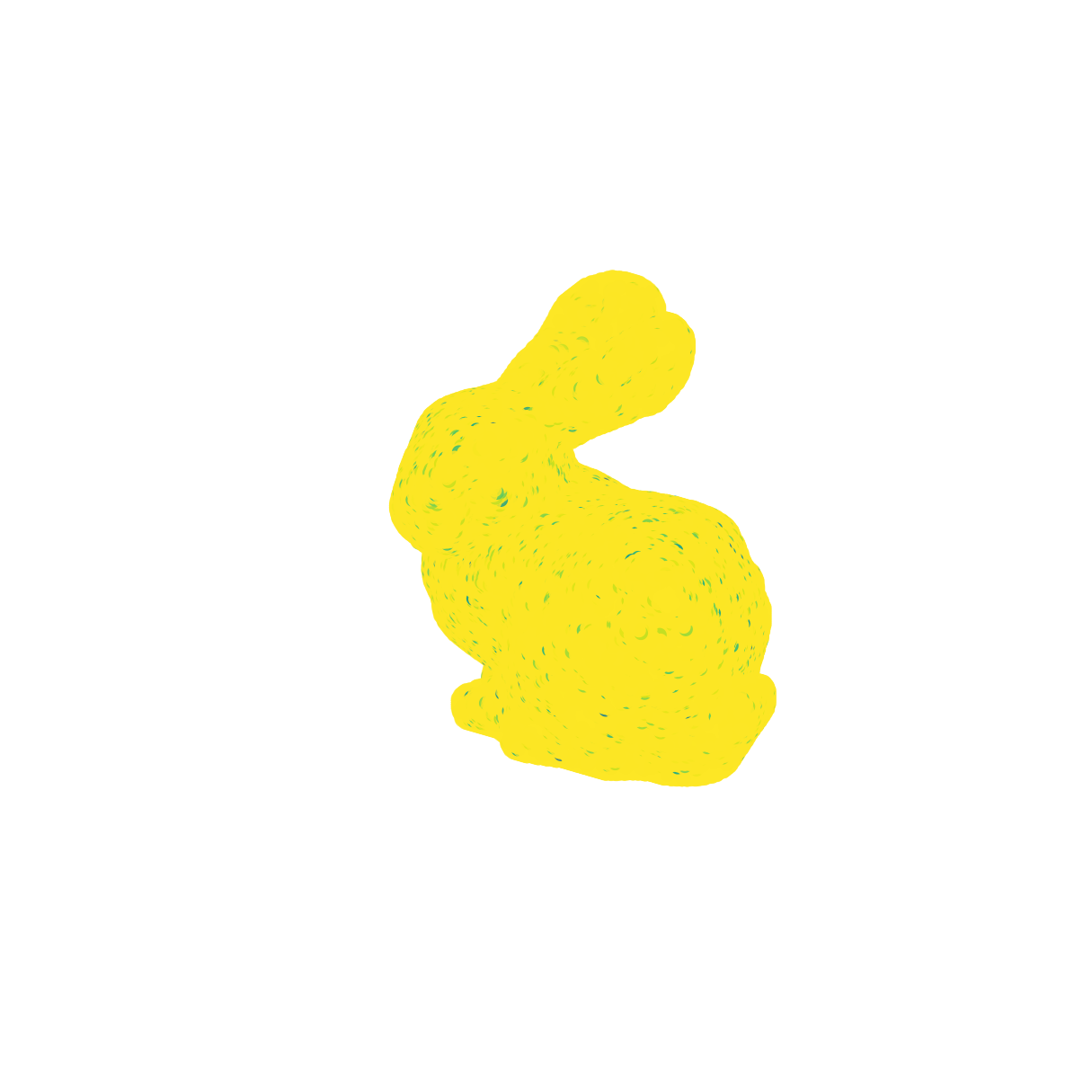}
        \caption{Bunny}
    \end{subfigure}
    \begin{subfigure}[t]{0.2\linewidth}
        \centering
        \includegraphics[width=\linewidth, trim=2.0in 2.3in 2.0in 2.0in, clip]{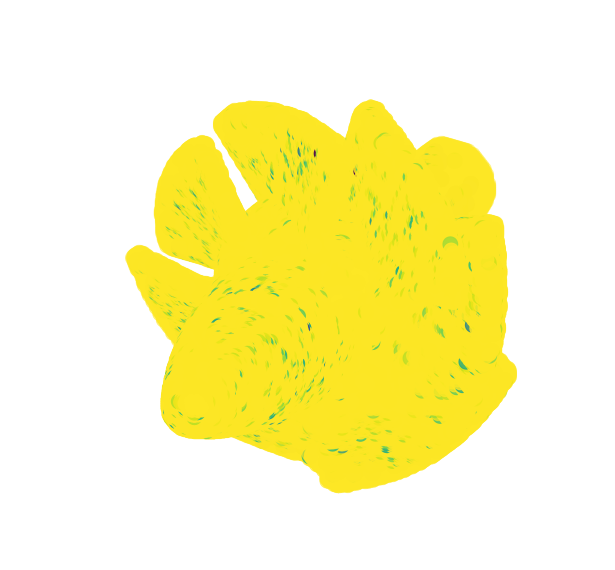}
        \caption{Turbine}
    \end{subfigure}
    \hspace{0.02\linewidth}
    \begin{subfigure}[t]{0.4\linewidth}
        \centering
        \includegraphics[width=\textwidth,trim={0.0in 0.5in 0.0in 0.0in}, clip]{Figure_compressed/geom_colorbar.pdf}
    \end{subfigure}
    \caption{Plot of $log_{10}(1 - \langle d_{gp}^{\text{true}} \cdot d_{gp}^{f_\theta} \rangle)$ for refinement $h=\nicefrac{\Delta}{2^8}$. The error shows the magnitude of the misalignment of the distance vector. As is more evident in tetrakis, the error is very high in the edges where there is a sharp change in curvature. Overall the error is still less than $10^{-4}$.}
    \label{fig:comparison_gp_cs}
\end{figure*}

The analysis of the accuracy of representation is performed by obtaining Gauss points (two per axis) with the surrogate boundary for the level of refinement $h=\frac{\Delta}{2^8}$, where  $\Delta$ is the characteristic length of the bounding box for the INRs. For error computation, the signed distance value is obtained from the implicit representation, $f_\theta(x_{gp})$, and the actual signed distance $s(x_{gp})$ is obtained from libgl library. \figref{fig:evaluating the distance vector} shows the surrogate boundary and the gauss points in the surrogate boundary with the distance vector pointing towards the true boundary, we are basically measuring the correctness of this vector. \figref{fig:comparison_gp_distv} visualizes $log_{10}(|f_\theta(x_{gp})-s(x_{gp}|)$ across different geometries. The accuracy of the direction of distance vector $\mathbf{d_{gp}}$ is computed by obtaining cosine similarly between the distance vector obtained from \algoref{Algorithm: DistanceFunctionCalculationUsingImplicitNetworkOneGP}, $d_{gp}^{f_\theta}$ and the true distance-vector $d_{gp}^{true}$. \figref{fig:comparison_gp_cs} visualizes  $log_{10}(1 - <d_{gp}^{true}\cdot d_{gp}^{f_\theta})>)$ across different geometries. Analysis of the two figures reveals that regions exhibiting significant changes in curvature are particularly prone to error. This is consistent with areas where large deviations in the magnitude of the distance function are observed. In these regions, a corresponding increase in the error of the cosine similarity metric is also evident. This behavior is expected due to the underlying eikonal constraint, which governs the relationship between the gradient of the distance function and the surface geometry. The eikonal equation imposes that the gradient of the distance function maintains a unit norm, and deviations from this condition in regions of high curvature can lead to both larger distance magnitude errors and higher discrepancies in the cosine similarity. Therefore, the correlation between these errors is a natural consequence of the mathematical properties imposed by the eikonal constraint.

\begin{table}[t!]
    \centering
    \caption{Comparison of Normalized Mean Squared Error (NMSE) and Mean Cosine Similarity of Distance Vectors (MCS) for Gauss points at the surrogate boundary. The standard deviation is shown in parentheses.}
    \small
    \label{tab:complexity_objects}
    \begin{tabular}{|c|c|c|c|}
        \hline
        \textbf{Complexity} & \textbf{Object} & $\mathbf{NMSE_{GP}}$ & \textbf{MCS}$_{GP}$ (S.D.) \\
        \hline
        Simple Shape & Sphere & $3.75 \times 10^{-8}$ & 1.000 (0.00044) \\
        & Cone & $2.00 \times 10^{-7}$ & 0.996 (0.045) \\
        & Cylinder & $5.60 \times 10^{-7}$ & 0.999 (0.025) \\
        \hline
        Complex Shape & Bunny & $9.75 \times 10^{-7}$ & 0.995 (0.014) \\
        & Tetrakis & $7.00 \times 10^{-7}$ & 0.997 (0.015) \\
        & Turbine & $3.84 \times 10^{-6}$ & 0.980 (0.130) \\
        \hline
    \end{tabular}
\end{table}

\tabref{tab:complexity_objects} presents a comparison of Normalised Mean Square as defined earlier. The maximum mean squared error is of the order $10^{-6} << h$. Similarly mean cosine similarity of the distance vector with (Standard Deviation) is presented for each geometry. Essentially, the INR gives fairly accurate distance vectors for both complex and simple geometries to perform analysis using SBM.

A plethora of work has been done with a major focus on representational similarity of the INR with ground truth. These INRs are very less looked at from the lens of downstream tasks (especially with a focus on analysis suitability. In our work, the sampling strategy and the loss function are designed to favor the generation of INR with a good distance vector approximation near the boundary. Similarly, we use Normalized Mean Squared error in a narrow band region, which is essentially the region near the surrogate boundary. Though the metric is a good way to make comparisons for  INR, where we have ground truth in the form of Triangle Soup (or any form of representation where it is possible to compute the signed distance values), but for modalities like point cloud, images, and depthmaps the given metric is not trivial. We note that rigorous metric and training pipelines focused on the analysis suitability of INR for other key modalities are not accounted for in our work.

\begin{figure*}[t!]
    \centering
    \begin{subfigure}[t]{0.45\textwidth}
    \centering
    \begin{tikzpicture}
    \begin{loglogaxis}[
        width=0.94\linewidth, 
        height=0.7\linewidth,
        xlabel={\footnotesize Element size, $h$},
        ylabel={\footnotesize $L_2$ Error},
        legend entries={Mesh Convergence, $slope = 2$},
        legend style={at={(0.7,1.0)}, anchor=north, nodes={scale=0.65, transform shape}}, 
        grid=major,
        xmin=0.005, xmax=0.1,
        ymin=1e-6, ymax=1
    ]
        \addplot coordinates {
            (0.0625, 0.00251317)
            (0.03125, 0.00052077)
            (0.015625, 0.00010555)
            (0.0078125, 0.00001713)
        };

        \addplot +[mark=none, red, dashed] [
            domain=0.005:0.1, 
            samples=100
        ] {1*x^2};

    \end{loglogaxis}
    \end{tikzpicture}
    \caption{Mesh convergence using $L_2$ error.}
    \label{fig:L2_convergence}
    \end{subfigure}
    \hfill
    \begin{subfigure}[t]{0.45\textwidth}
    \centering
    \begin{tikzpicture}
    \begin{axis}[
        width=0.94\linewidth, 
        height=0.7\linewidth,
        ybar,
        bar width=10pt,
        xtick=data,
        xlabel={Level},
        ylabel={Normalized $L_2$ Error},
        grid=both
    ]

    \addplot +[fill=blue] coordinates {
        (5, 1.0)
        (6, 0.2072163841)
        (7, 0.04199875058)
        (8, 0.006816092823)
    };

    \end{axis}
    \end{tikzpicture}
    \caption{Normalized $L_2$ error per level.}
    \label{fig:L2_normalized}
    \end{subfigure}

    \caption{Top: Log-log plot showing $L_2$ error convergence with mesh refinement. Right: Normalized $L_2$ error as a bar chart across levels for a ring with Dirichlet boundary condition applied on both inner (using SBM) and outer ring (strongly).}
    \label{fig:mesh_convergence_plots}
\end{figure*}
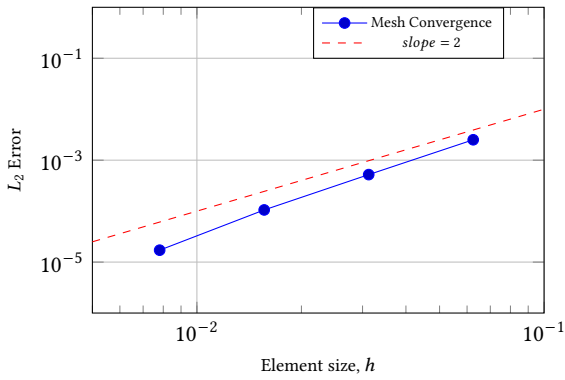
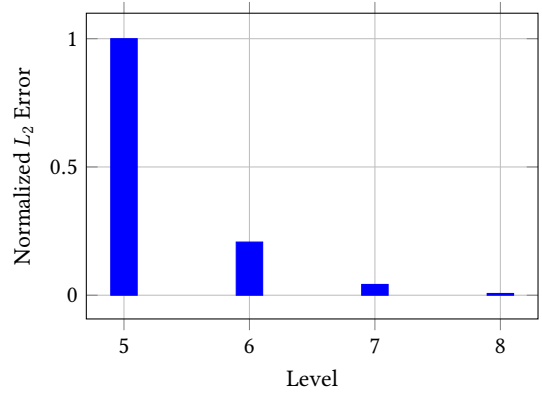

\begin{figure}[t!]
    \centering
    \begin{tikzpicture}
        \node[anchor=south west, inner sep=0] (main) at (-1,0) {
            \includegraphics[trim=200 150 300 100, clip, width=0.8\linewidth]{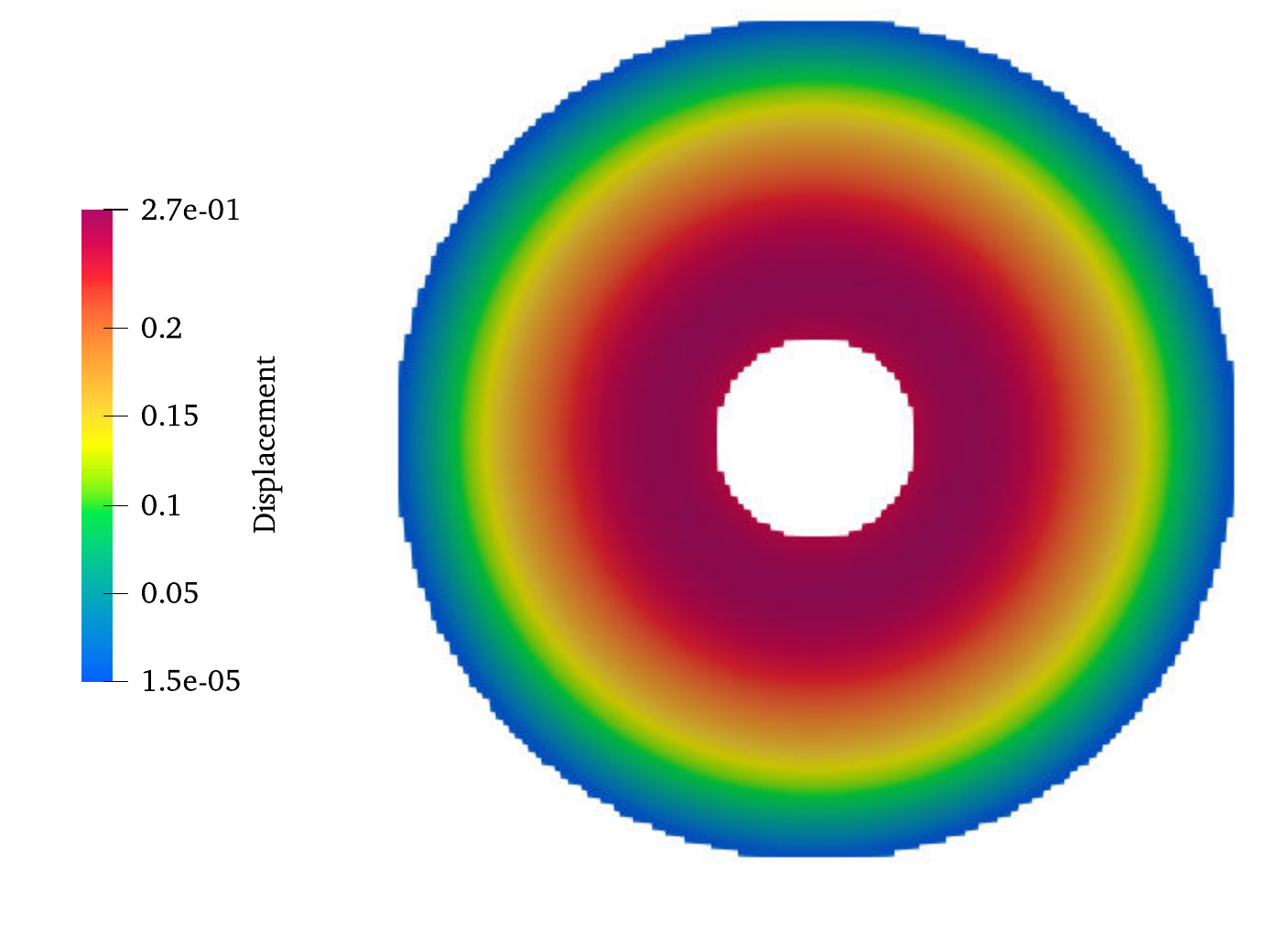}
        };
        \node[anchor=north east, xshift=1.2cm, yshift=1cm] at (main.north east) {
            \fbox{%
                \includegraphics[trim=580 370 300 50, clip, width=0.1\linewidth]{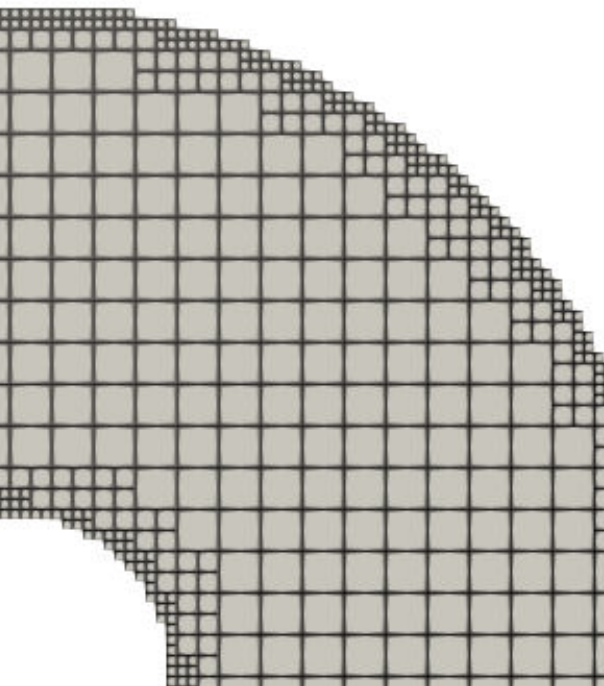}
            }
        };
    \end{tikzpicture}
    \caption{Displacement magnitude of the full ring (background). Top-right inset: region to demonstrate the base and wall refinement }
    \label{fig:ring_displacement_with_inset}
\end{figure}

\subsection{Validation}
\label{sec:validation}

In this section, we validate the proposed framework through two complementary studies. First, we consider a canonical ring domain with a known analytical displacement solution, originally studied by \citet{schillinger2012small}, to assess the accuracy and convergence behavior of our SBM-based numerical formulation. By applying both strong and weak Dirichlet boundary conditions and measuring the $L_2$ error across varying mesh resolutions, we demonstrate second-order convergence consistent with linear basis functions.

\subsubsection{Ring with Analytical Solution}
\label{sec:validation_2d}
We validate our method using a ring domain with an inner radius of \( 0.25 \) and an outer radius of \( 1.0 \), following the benchmark presented in the work by \citet{schillinger2012small}, where a similar test was performed. The analytical solution for the magnitude of the displacement at a radial distance \( r \) is given by
\[
u(r) = -\frac{r \log r}{2 \log 2},
\]
assuming a Young’s modulus \( E = 1.0 \) and Poisson’s ratio \( \nu = 0 \). The ring is centered at the point \( (1,1) \). A strong Dirichlet boundary condition is enforced on the outer boundary of the ring, while a weak Dirichlet boundary condition is imposed on the inner boundary using SBM.

\figref{fig:ring_displacement_with_inset} shows a region representing the refined mesh along with the solution contour for a base level of refinement of 5 and a boundary level of refinement of 7. We vary the base level refinement and obtain the convergence plot as illustrated in \figref{fig:L2_convergence}, where we see the slope of convergence close to 2 as expected for a linear basis function. \figref{fig:L2_normalized} presents the normalized $L_2$ error for the same case. More details on the convergence analysis of the SBM, itself can be found in the literature~\citep{atallah2020second,atallah2021shifted,main2018shifted1,main2018shifted2}.

\subsubsection{Comparison between INR of Ico Sphere and Ico-Sphere}
\label{sec:validation_3d}

To evaluate the accuracy of our proposed method, we consider a simulation on an icosphere geometry represented using an INR. The material is modeled using linear elasticity with Lamé parameters \(\lambda = 1\) and \(\mu = 0.5\). We impose an analytical displacement field as the Dirichlet boundary condition, defined by:
\begin{subequations}
\begin{align}
u_x &= \frac{1}{10} \sin(\pi x)\sin(\pi y)\sin(\pi z), \label{eq:ux_sin} \\
u_y &= \frac{1}{10} \cos(\pi x)\cos(\pi y)\sin(\pi z), \label{eq:uy_cos} \\
u_z &= \frac{1}{10} \cos(\pi x)\sin(\pi y)\cos(\pi z) \label{eq:uz_mix}
\end{align}
\end{subequations}

\begin{figure*}[t!]
    \centering
    \begin{subfigure}{0.33\linewidth}
        \centering
        \includegraphics[trim=3.2in 1in 3.2in 0.1in,clip,width=\linewidth]{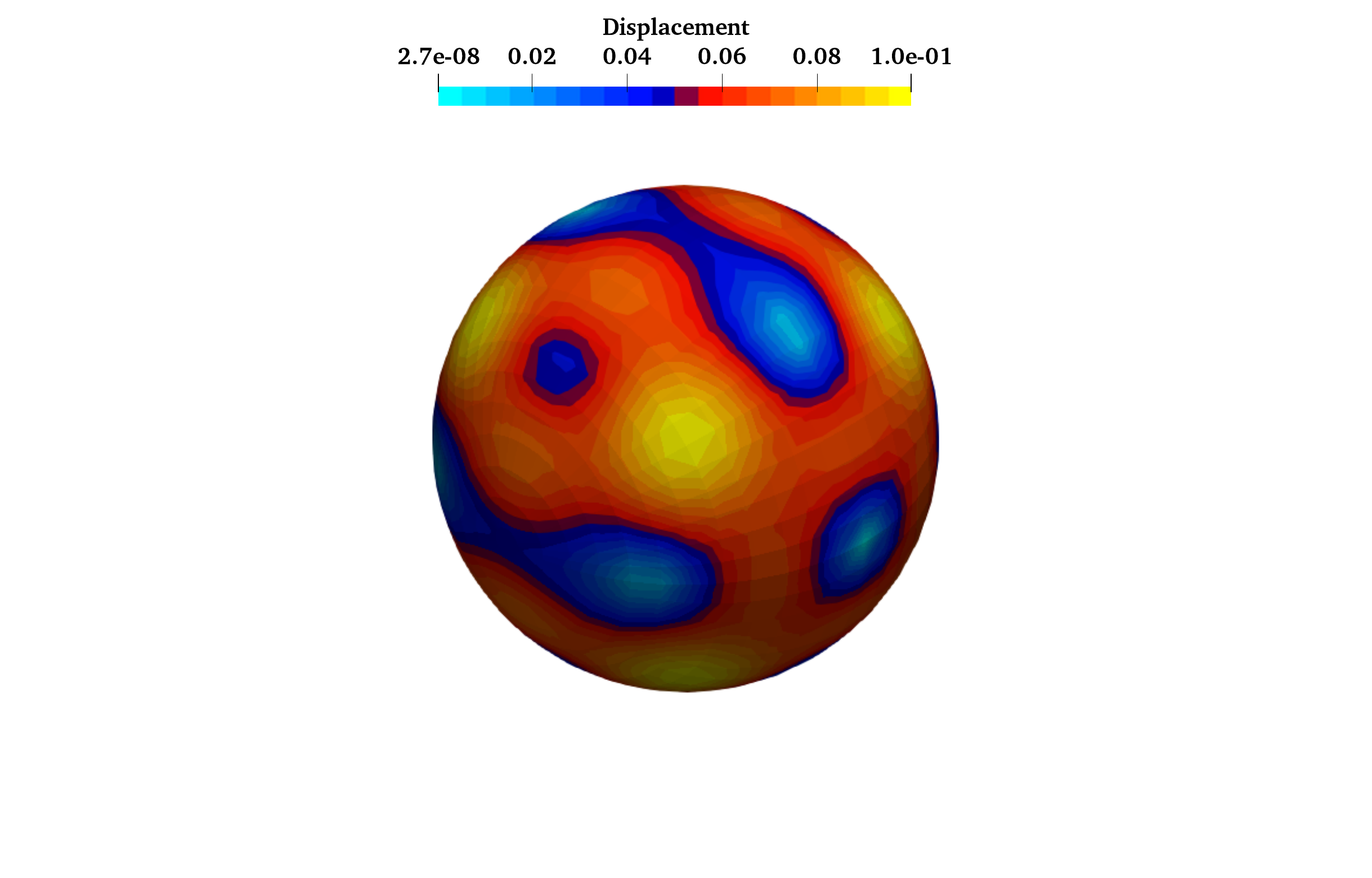} 
        \caption{Solution Displacement Field on the surface ico-sphere based on the prescribed boundary condition.}
        \label{fig:icosphere_solution}
    \end{subfigure}
    \hspace{0.05\linewidth}
    \begin{subfigure}{0.33\linewidth}
        \centering
        \includegraphics[trim=3.2in 1in 3.2in 0.1in,clip,width=\linewidth]{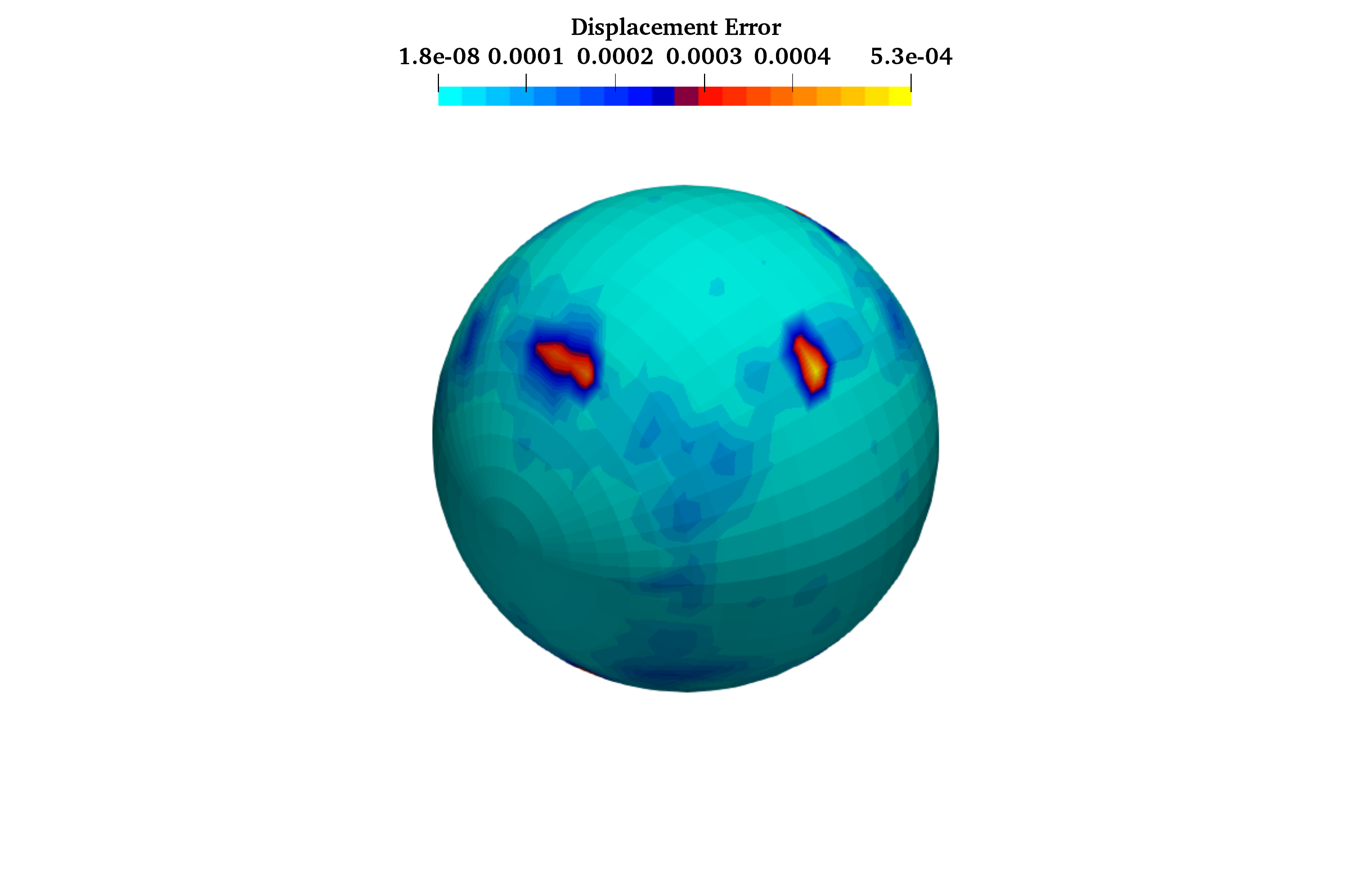}
        \caption{Error in the displacement field using ico-sphere and corresponding INR.}
        \label{fig:icosphere_error}
    \end{subfigure}
    \caption{Solution field and corresponding error in the displacement field between ico-sphere and corresponding INR. The solution obtained from the octree-grids are projected on the actual ico-sphere (ground truth) we are interested in. The error in the solution field represents the error incurred by replacing the ico-sphere with an INR using SBM as the analysis method.}
    \label{fig:icosphere}
\end{figure*}

These conditions are applied via SBM, ensuring consistency with the surrogate boundary representation. The simulation is conducted on an octree mesh with a base refinement level of 4 and a boundary refinement level of 8 to adequately capture boundary features for both representations. INR is trying to mimic the boundaries and for boundary driven problem the error is driven through boundaries. Hence, the computed displacement solution is visualized on the true surface of the icosphere (which in our case is the Ground Truth). \figref{fig:icosphere_solution} illustrates the displacement field, while \figref{fig:icosphere_error} presents the corresponding displacement error. For this case, the surface integral of the error ($\int_{\Gamma}|u_{\theta}-u_{\Delta}|d\Gamma$) is found to be approximately \(2.99 \times 10^{-4}\), validating the accuracy of the INR for suitability of the analysis. Our analysis in \secref{section:Val_Implicit} also presented a similar order of error in the case of an ico-sphere.

\begin{figure}[t!]
    \centering
    \begin{tikzpicture}
    \begin{loglogaxis}[
        width=0.94\linewidth, 
        height=0.7\linewidth,
        xlabel={\footnotesize Number of Triangles},
        ylabel={\footnotesize Wall Time [s]},
        legend entries={
            $\Delta$s Meshing Time, 
            $\Delta$s Assembly Time,
            INR Meshing Time,
            INR Assembly Time
        },
        legend style={at={(0.3,0.98)}, anchor=north, nodes={scale=0.65, transform shape}}, 
        grid=major,
        xmin=1e3, xmax=2e5,
        ymin=1e-1, ymax=1e3
    ]

        \addplot+[mark=*, blue] table[row sep=\\, col sep=comma] {
            Triangles, Meshing
            1520, 2.08262 \\
            2496, 7.44223 \\
            9984, 28.5756 \\
            39936, 113.171 \\
            159744, 454.012 \\
        };

        \addplot+[mark=square*, red] table[row sep=\\, col sep=comma] {
            Triangles, Assembly
            1520, 0.172552 \\
            2496, 0.561698 \\
            9984, 2.09715 \\
            39936, 8.22986 \\
            159744, 32.9037 \\
        };

        \addplot+[domain=1e3:2e5, samples=2, blue, dotted, line width=1pt,mark=none] {28.1971};

        \addplot+[domain=1e3:2e5, samples=2, red, dotted, line width=1pt,mark=none] {2.21228};

    \end{loglogaxis}
    \end{tikzpicture}
    \caption{Log-log plot of meshing and assembly wall times with increasing number of triangles. The experiments were performed on a single core of an Intel Core i9-14900KF processor (24 cores, 32 threads, max 6.1~GHz). Dotted lines indicate the constant reference time incurred by INR-based representation, which does not depend on the number of triangles to be processed.
}
    \label{fig:mesh_vs_time}
\end{figure}
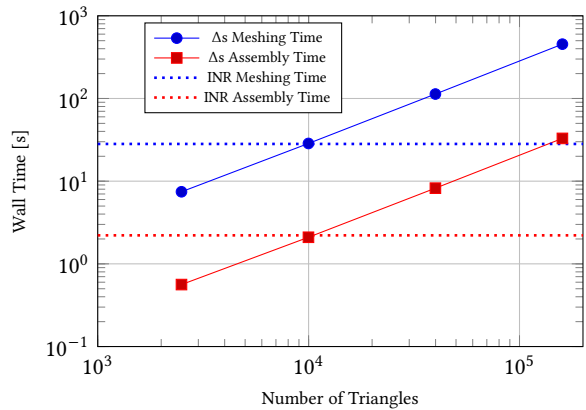

We also compare the computational performance of traditional mesh-based representations with the proposed Implicit Neural Representation (INR) approach by measuring the wall time required for meshing and assembly as a function of the number of surface triangles on an icosphere. As shown in \figref{fig:mesh_vs_time}, both meshing and assembly times for the mesh-based pipeline increase rapidly with the number of triangles, exhibiting approximately linear scaling in log-log scale. In contrast, the INR-based pipeline demonstrates constant wall times for both meshing and assembly, as indicated by the horizontal dotted lines in the plot. This is because INR representations are independent of the explicit surface triangulation, relying instead on neural network evaluations over fixed-resolution sampling. All experiments were executed on a single core of an Intel Core i9-14900KF processor (24 cores, 32 threads, max 6.1~GHz), and the wall times clearly show the scalability advantage of INR over traditional geometric discretizations in scenarios where requirement for number of triangle grows significantly.

\subsection{Bunny Model}

\begin{figure}[b!]
    \centering
    \begin{subfigure}[b]{0.48\linewidth} 
        \centering
        \includegraphics[trim=500 100 400 100,clip,width=0.8\linewidth]{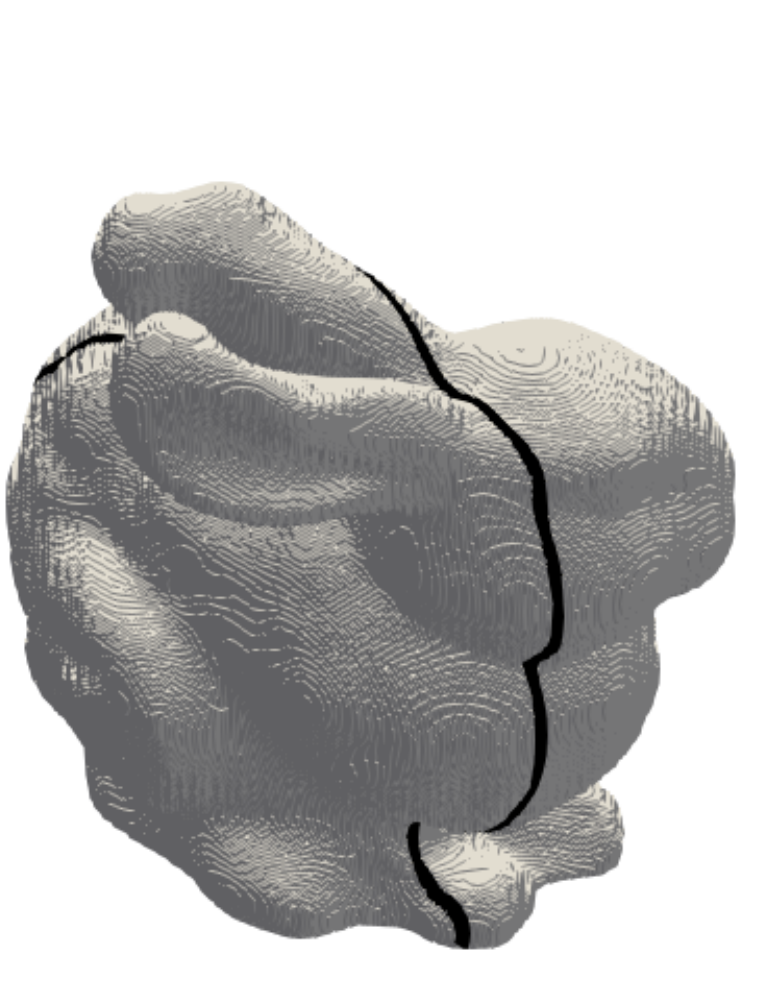}
        \caption{Bunny represented using incomplete Octree. The black line represents the plane used to slice the model.}
        \label{fig:carved_bunny}
    \end{subfigure}
    \hfill
    \begin{subfigure}[b]{0.48\linewidth} 
        \centering
        \includegraphics[trim=50 50 50 100,clip,width=0.95\linewidth]{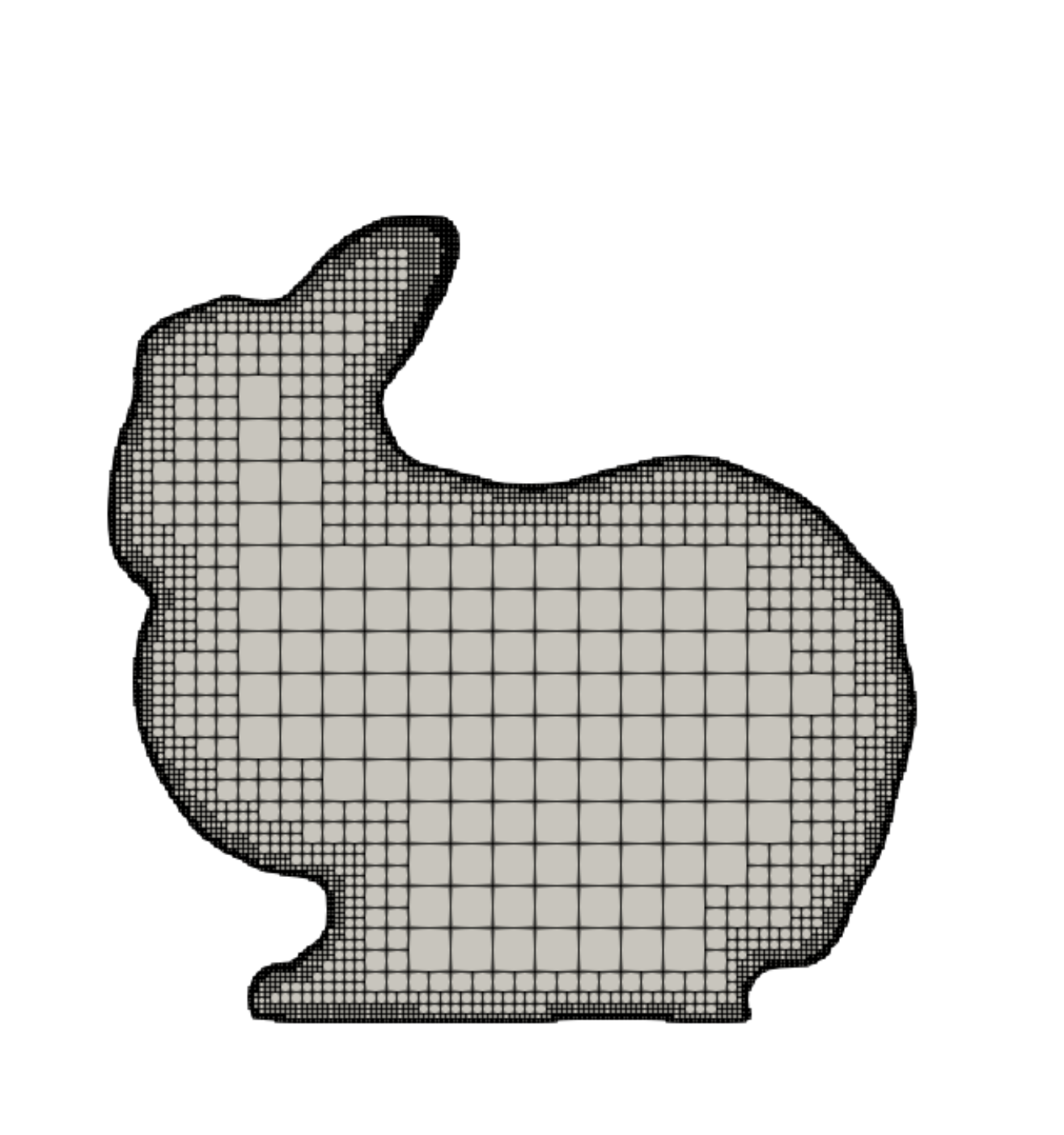}
        \caption{Slice of the Octree grid illustrating adaptive refinement to accurately capture the geometry.}
        \label{fig:bunny_slice}
    \end{subfigure}
    \caption{(a) INR-based Stanford Bunny with a base refinement of level 5. Similarly, boundaries are refined at level 9, as shown by the slice on the right. }
    \label{fig:bunnymesh}
\end{figure}

\figref{fig:bunnymesh} illustrates the Implicit Neural Representation (INR) of the Stanford Bunny~\citep{turk1994zippered}, embedded within an adaptively refined, incomplete octree structure. The INR is constructed using the procedure outlined in \algoref{Algorithm 1 Implicit Network Training}, which enables efficient sampling and learning of the signed distance function (SDF) over the computational domain. This octree-based representation ensures higher refinement near regions of complex geometry while maintaining a coarser discretization elsewhere, striking a balance between accuracy and computational efficiency. To highlight the adaptivity of the octree, \figref{fig:carved_bunny} depicts the Stanford Bunny with a designated slicing plane, and the resulting cross-section is shown in \figref{fig:bunny_slice}. The slice demonstrates the octree’s capability to refine up to level 9 near the boundaries, while retaining a base level of refinement at level 5 in the bulk.

\begin{figure}[t!]
    \centering
    \begin{subfigure}[b]{0.48\linewidth}
        \centering
         \includegraphics[trim=4.0in 20in 4.0in 1.5in,clip,width=0.5\linewidth]{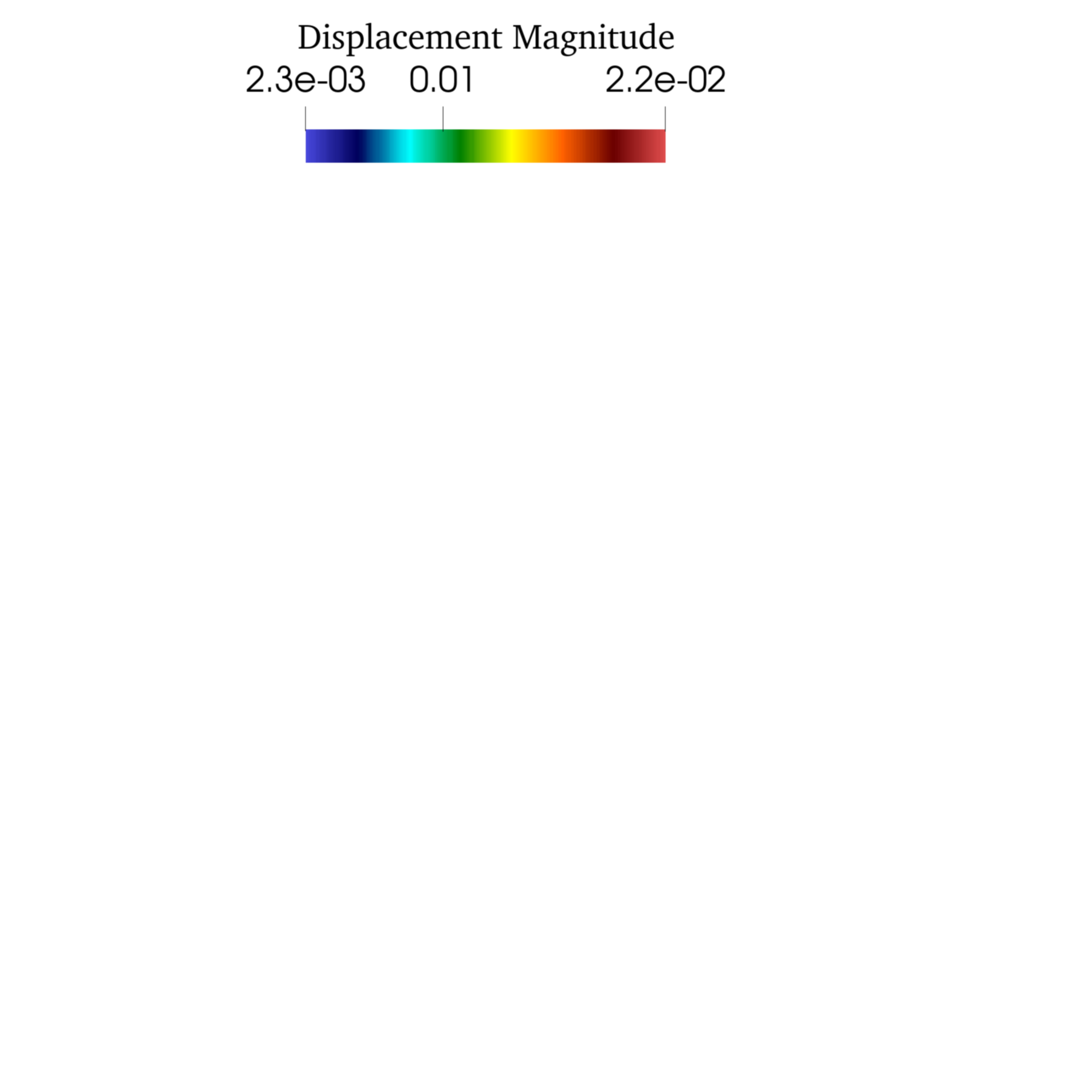}\\
        \includegraphics[trim=300 50 300 100,clip,width=\linewidth]{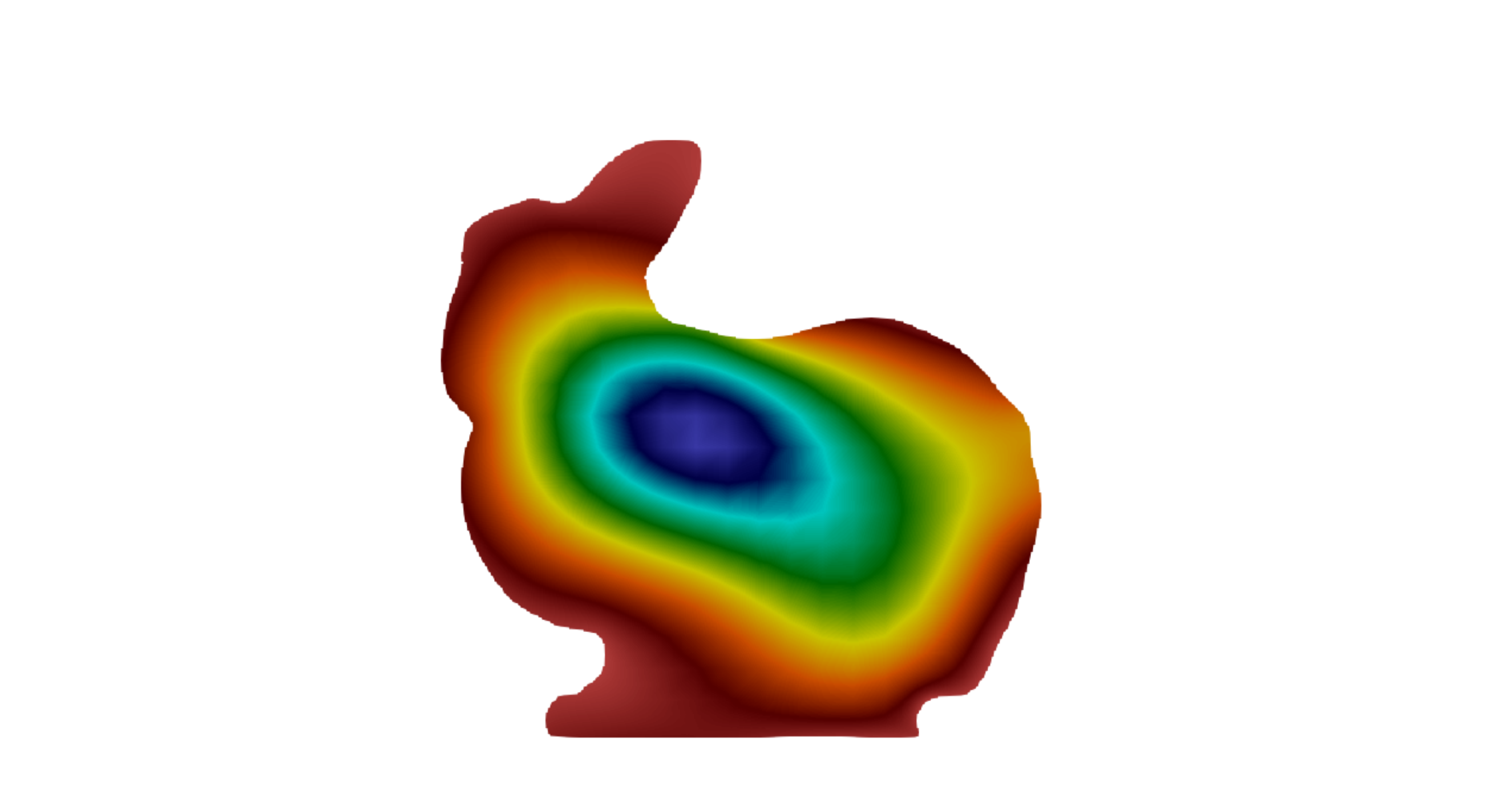}
        \caption{Displacement magnitude}
        \label{fig:bunny_displacement_mag_slice}
    \end{subfigure}%
    \hspace{0.03\linewidth}
    \begin{subfigure}[b]{0.48\linewidth}
        \centering
        \includegraphics[trim=4.0in 19.5in 4.0in 1.5in,clip,width=0.5\linewidth]{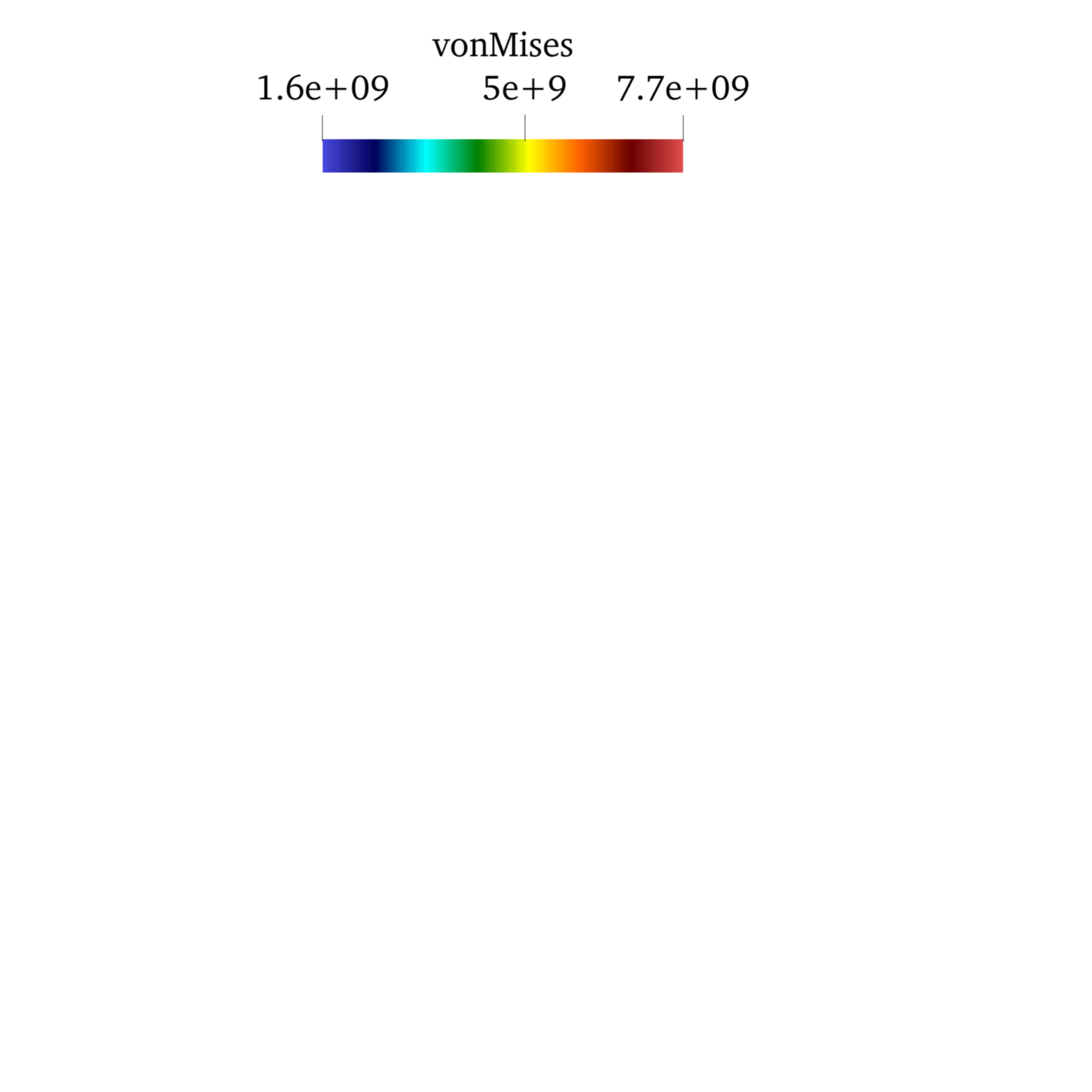}\\
        \includegraphics[trim=300 60 300 100,clip,width=\linewidth]{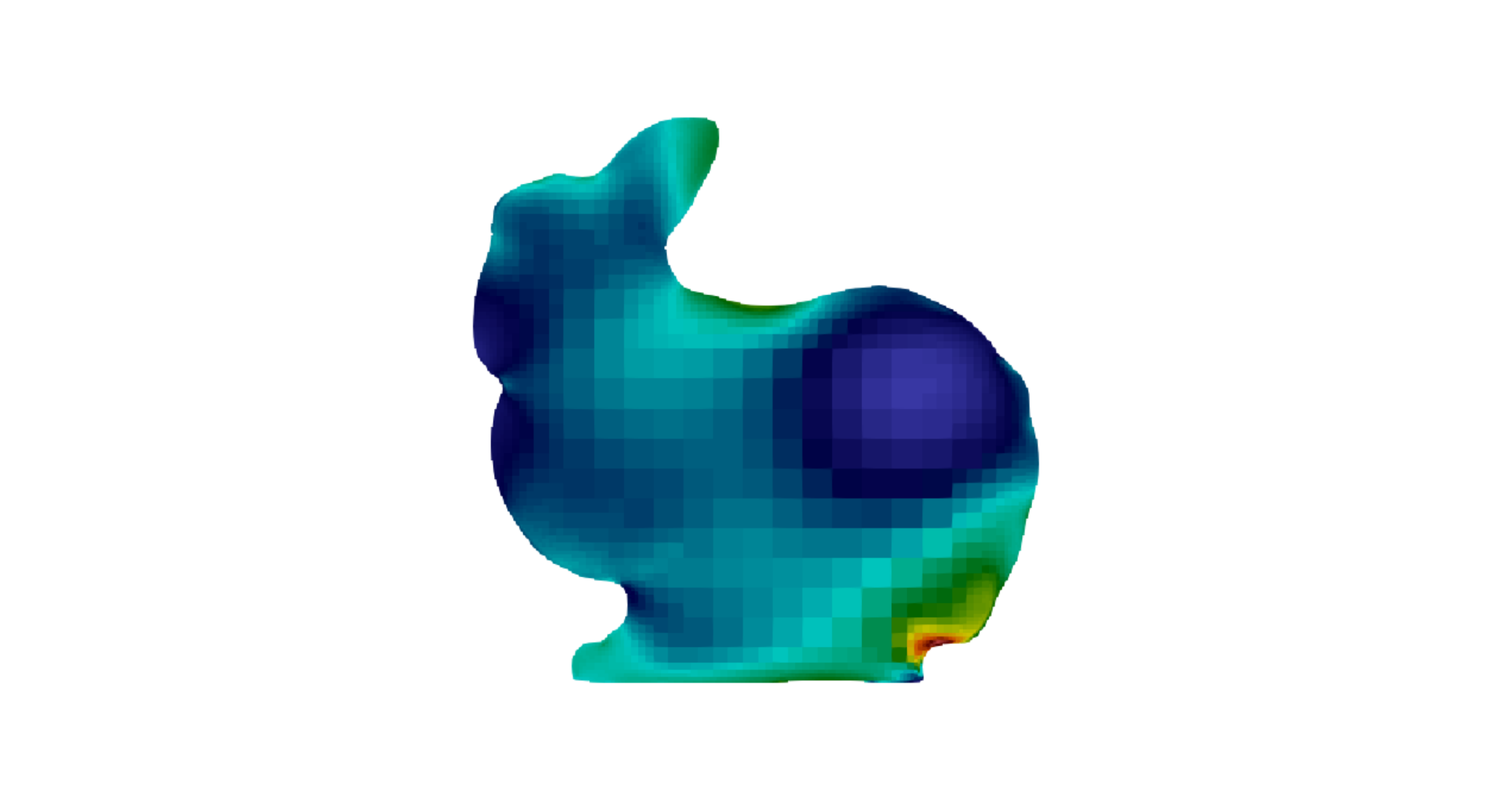}
        \caption{von Mises stress}
        \label{fig:bunny_stress_slice}
    \end{subfigure}
    \caption{(a) Displacement magnitude ($(u_x^2 + u_y^2 + u_z^2)^{1/2}$) along a slice of the Bunny (shown in \figref{fig:bunnymesh}). (b) von Mises stress showing the stress concentration in regions of of sharp curvature and high gradient in the displacement field. }
    \label{fig:bunnyStress}
\end{figure}

For the physical setup, we consider a plane strain condition with Young’s modulus \( E = 7 \times 10^{10} \) and Poisson’s ratio \( \nu = 0.33 \). The displacement boundary condition is defined analytically and directly imposed on the boundary. The body force is computed as the residual of the momentum balance equation when substituting the analytical displacement field. This allows us to use the known displacement field as the exact solution and validate the numerical accuracy of our solver within the INR-defined geometry.

\begin{subequations}
\begin{align}
u_x &= \frac{\sin(\pi x) \cos(\pi y)}{50}, \label{eq:ux_trig} \\
u_y &= \frac{\cos(\pi x) \sin(\pi z)}{50}, \label{eq:uy_trig} \\
u_z &= \frac{e^{-y_{\text{true}}^2} \sin(\pi z_{\text{true}})}{100} \label{eq:uz_trig}
\end{align}
\end{subequations}

The simulation results are shown in \figref{fig:bunnyDisplacement}, where the displacement magnitudes in the $x$-direction (Ux), $y$-direction (Uy), and $z$-direction (Uz) are visualized for the INR-based Bunny. The applied boundary conditions induce sinusoidal variations in Ux and Uy, while Uz exhibits an exponential decay along the $y$-direction.

In \figref{fig:bunnyStress}, we further investigate the results in the slice as in \figref{fig:bunny_slice}. \figref{fig:bunny_displacement_mag_slice} presents the displacement magnitude \(\sqrt{u_x^2 + u_y^2 + u_z^2}\) along the extracted slice, providing insight into the deformation characteristics within the octree structure. \figref{fig:bunny_stress_slice} highlights the von Mises stress distribution, revealing stress concentrations in regions of sharp curvature and high displacement gradients. This behavior is consistent with expected mechanical responses, where localized geometric features and abrupt changes in deformation contribute to stress intensification.

\subsection{Eiffel Tower Model}
\begin{figure}[t!]
    \centering
    \includegraphics[trim=6.0in 0 6.0in 0.9in,clip,width=0.5\linewidth]{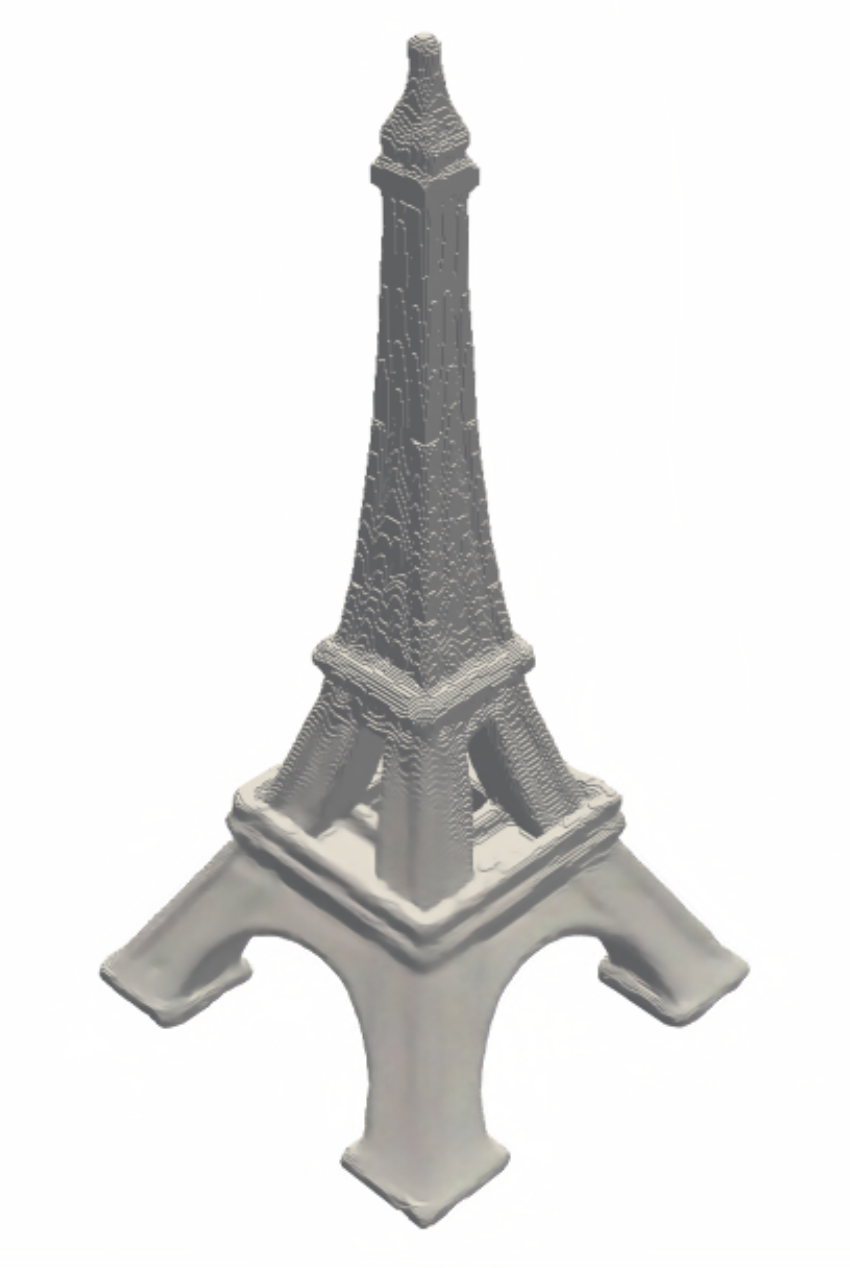} 
    \caption{Visualization of carved-out mesh of Eiffel Tower represented using INR. To capture the tower, we start with a base level of 6 and adaptively refine it to level 10 close to the boundaries.}
    \label{fig:eiffel}
\end{figure}

\figref{fig:eiffel} visualizes the INR-based representation of a carved-out Eiffel Tower model. We begin with a point cloud of a toy Eiffel Tower and employ the method proposed by \citet{jignasu2024stitch} to generate the corresponding INR. We use a point cloud of a toy Eiffel Tower. We refer the reader to \citet{jignasu2024stitch} for details of the generation of INR.
The incomplete octree is obtained with a base refinement level of 6, with adaptive refinement applied near boundaries up to level 10. This approach ensures higher accuracy where geometric details are critical while maintaining computational efficiency in less significant regions. We apply the following Dirichlet boundary condition enforced using SBM everywhere in the boundary:
\begin{subequations}
\begin{align}
u_x &= 0.1 \sin(\pi x) \cos(\pi y), \label{eq:ux} \\
u_y &= 0.05 \sin(\pi y) \sin(\pi z), \label{eq:uy} \\
u_z &= 0 \label{eq:uz}
\end{align}
\end{subequations}

\begin{figure}[t!]
    \centering
    \includegraphics[trim=0.1in 20in 1.1in 1.7in,clip,width=0.8\linewidth]{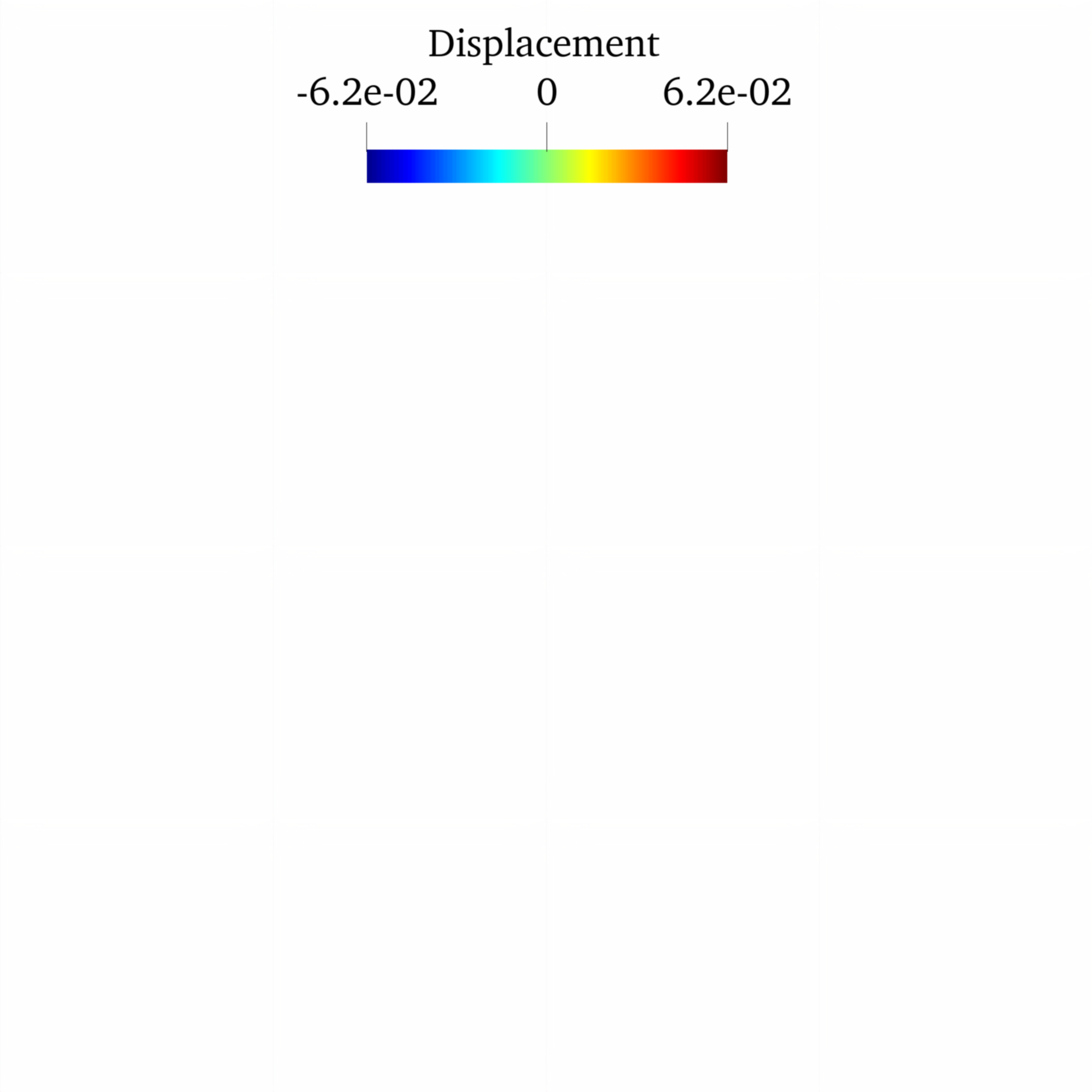}
    \begin{subfigure}[b]{0.47\linewidth}
        \centering
        \includegraphics[trim=6.0in 0 6.0in 1.0in,clip,width=\linewidth]{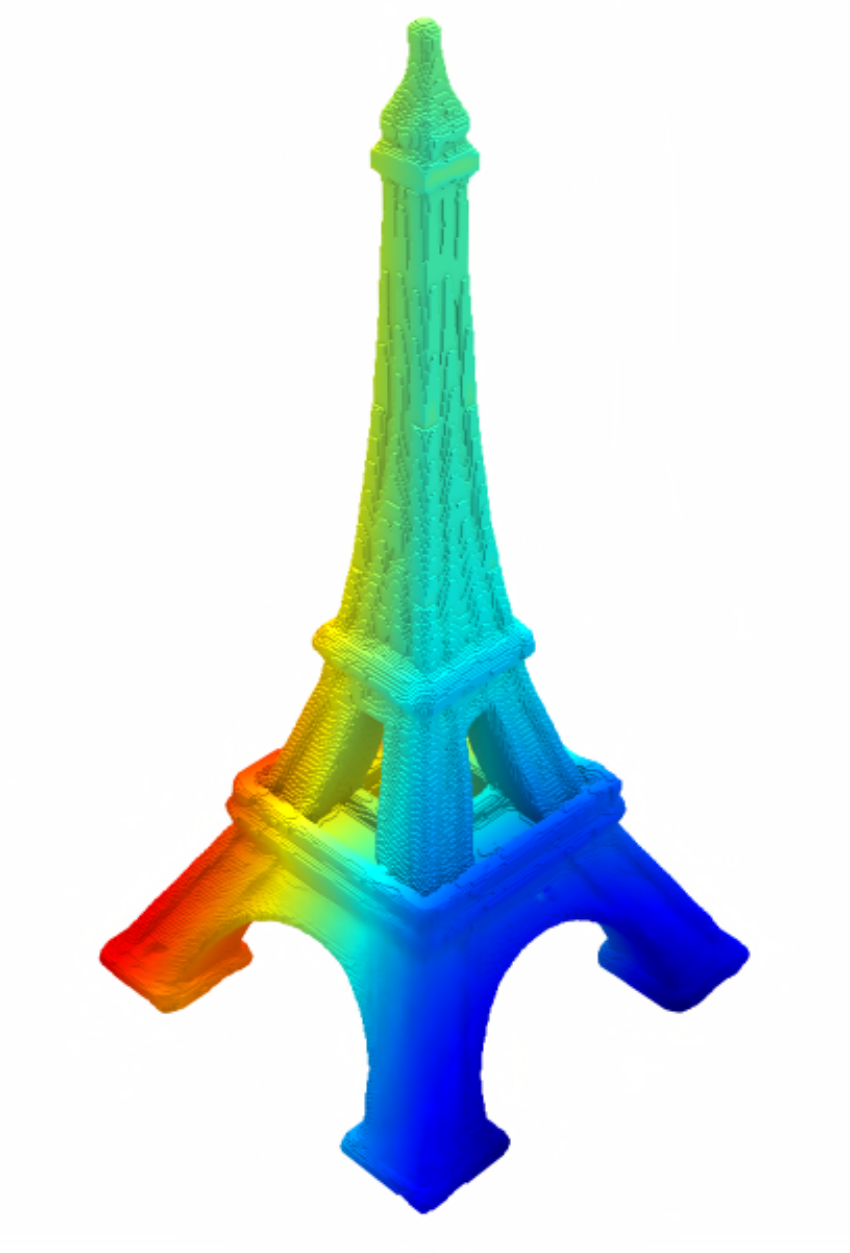}
        \caption{$x$-direction displacement, $U_x$.}
        \label{fig:Eiffelux}
    \end{subfigure}
    \hspace{0.02\linewidth}
    \begin{subfigure}[b]{0.47\linewidth}
        \centering
        \includegraphics[trim=6.0in 0 6.0in 1.0in,clip,width=\linewidth]{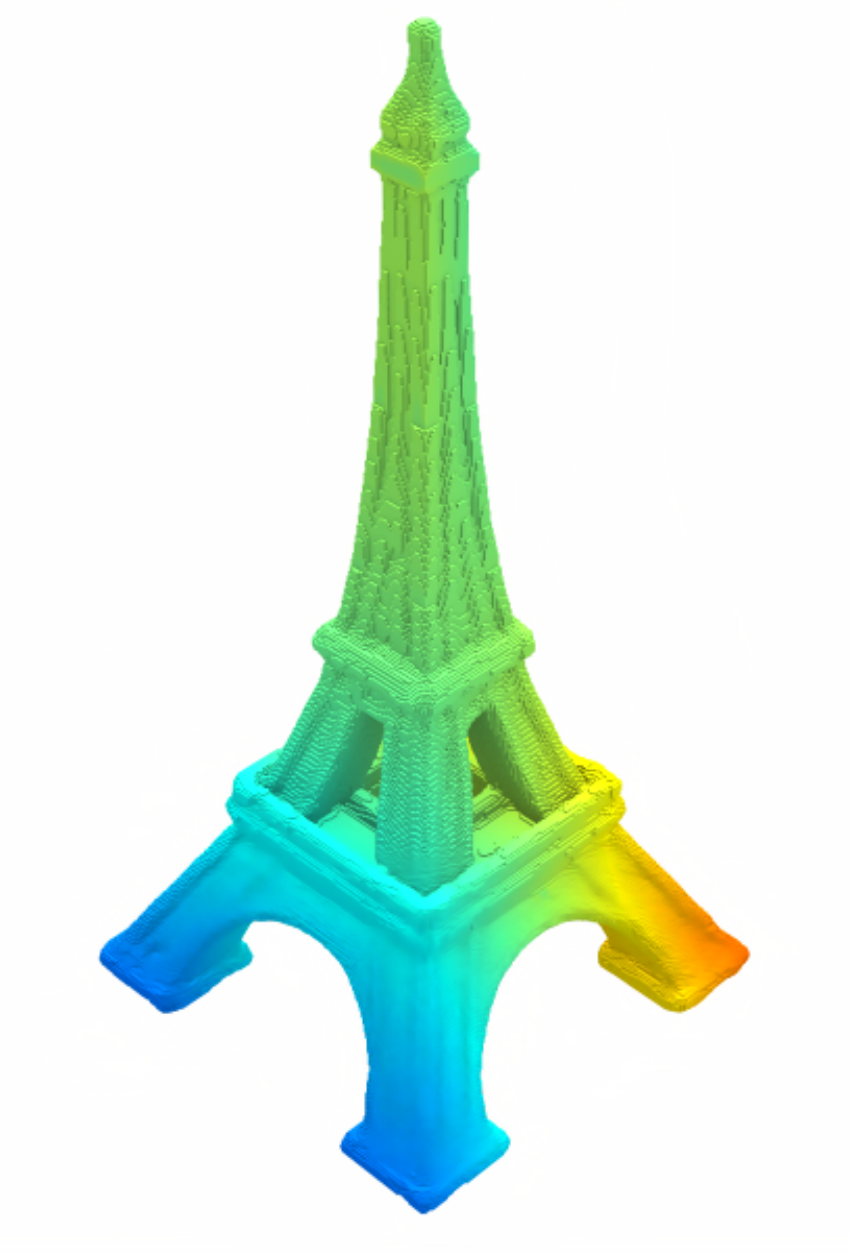}
        \caption{$y$-direction displacement, $U_x$.}
        \label{fig:Eiffeluy}
    \end{subfigure}
    \caption{(a) and (b) show the displacement $U_x$ and $U_y$, which varies sinusoidally as per the prescribed boundary condition. The displacement in the $x$-direction is higher. The tower is fixed in the $z$-direction.}
    \label{fig:EiffelDisp}
\end{figure}

\begin{figure*}[t!]
    \centering
    \begin{subfigure}[!b]{0.99\linewidth}
    \centering
    \includegraphics[trim=0.1in 20in 1.1in 1.3in,clip,width=0.4\linewidth]{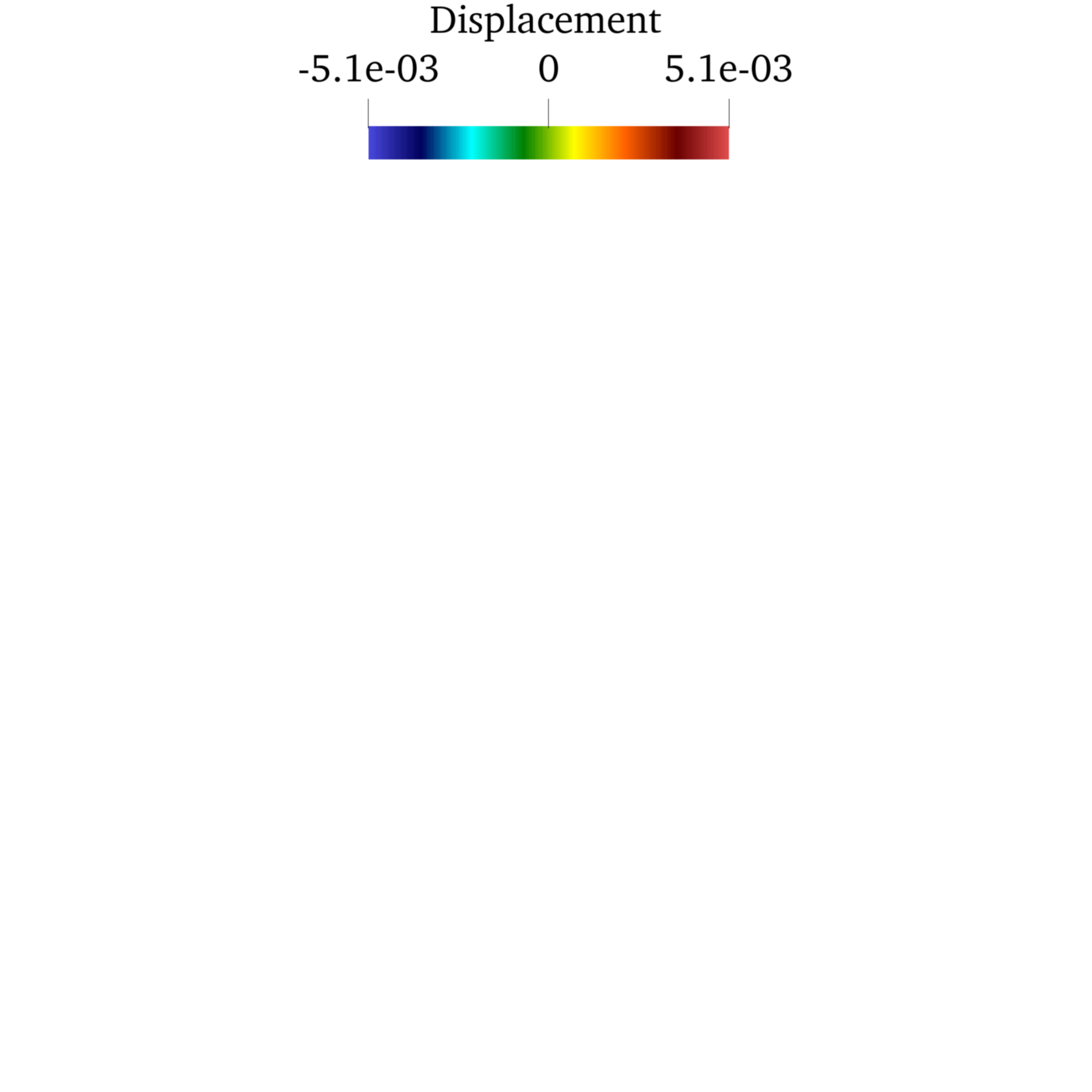}
    \end{subfigure}\\
    \begin{subfigure}[!b]{0.28\linewidth}
        \centering
        \includegraphics[trim=400 100 400 100,clip,width=\linewidth]{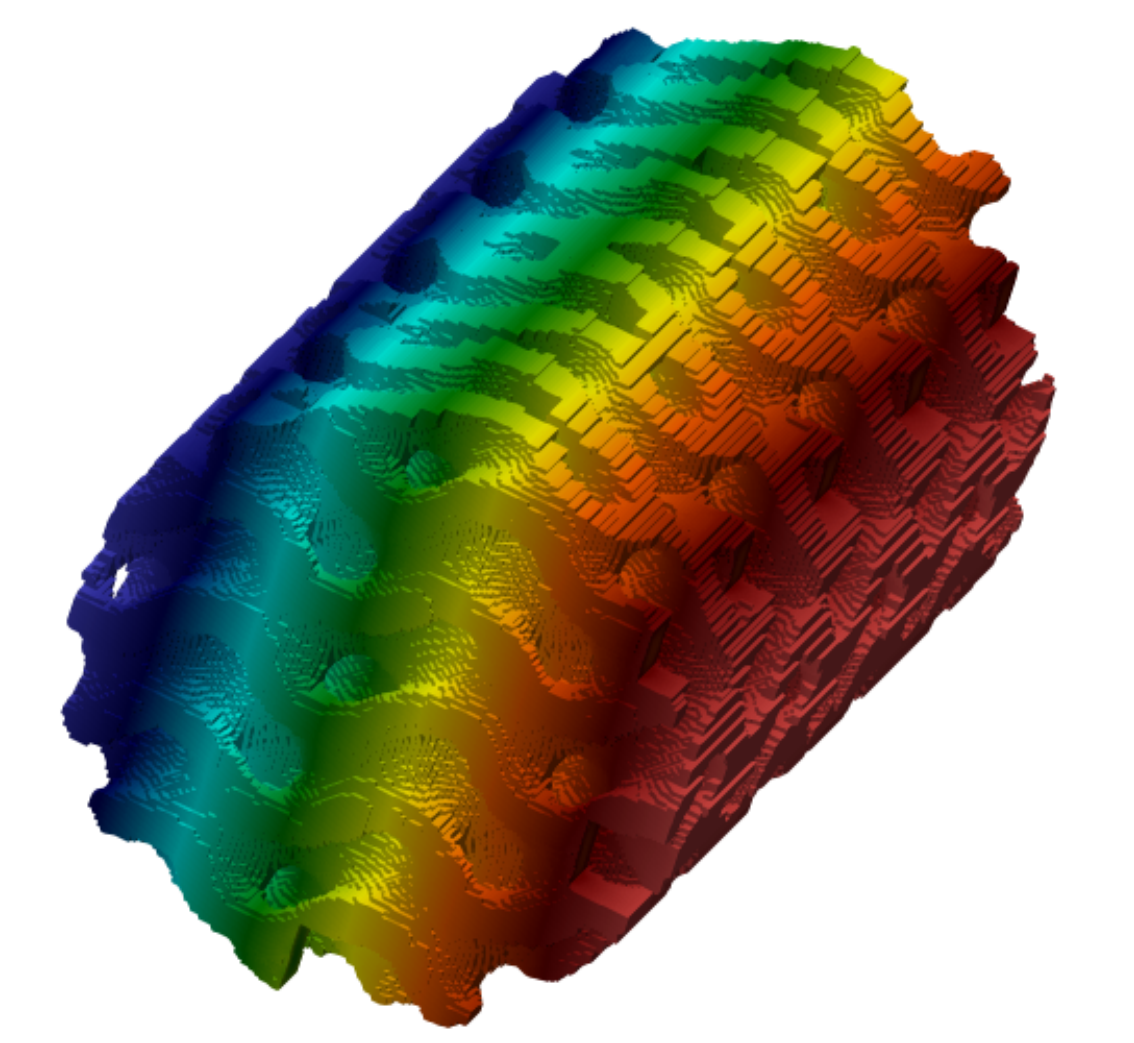}
        \caption{$x$-direction displacement, $U_x$.}
        \label{fig:gyroid_ux}
    \end{subfigure}
    \begin{subfigure}[!b]{0.28\linewidth}
        \centering
        \includegraphics[trim=400 100 400 100,clip,width=\linewidth]{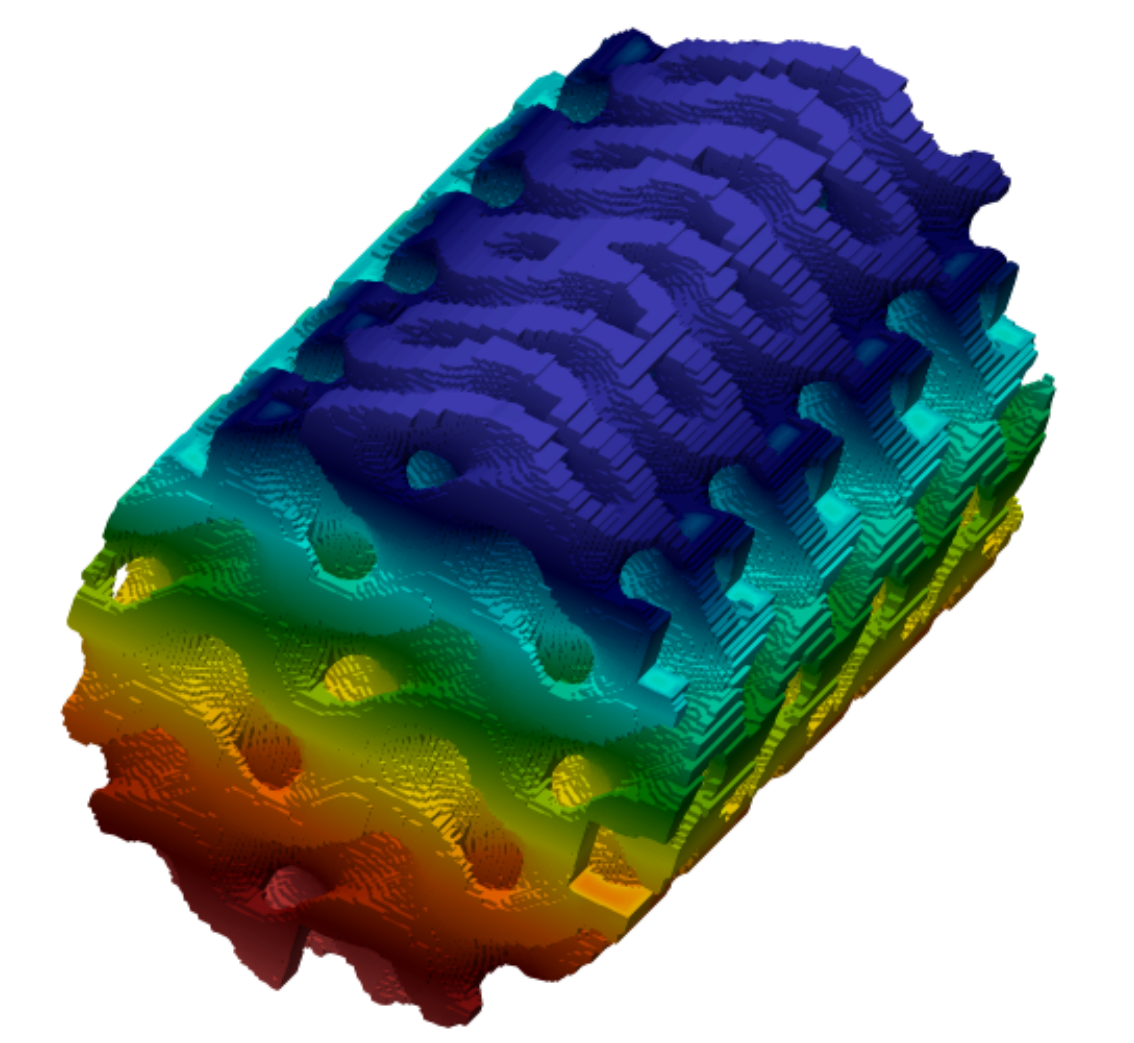}
        \caption{$y$-direction displacement, $U_y$.}
        \label{fig:gyroid_uy}
    \end{subfigure}
    \begin{subfigure}[!b]{0.37\linewidth}
        \centering
        \includegraphics[trim=400 100 400 110,clip,width=.72\linewidth]{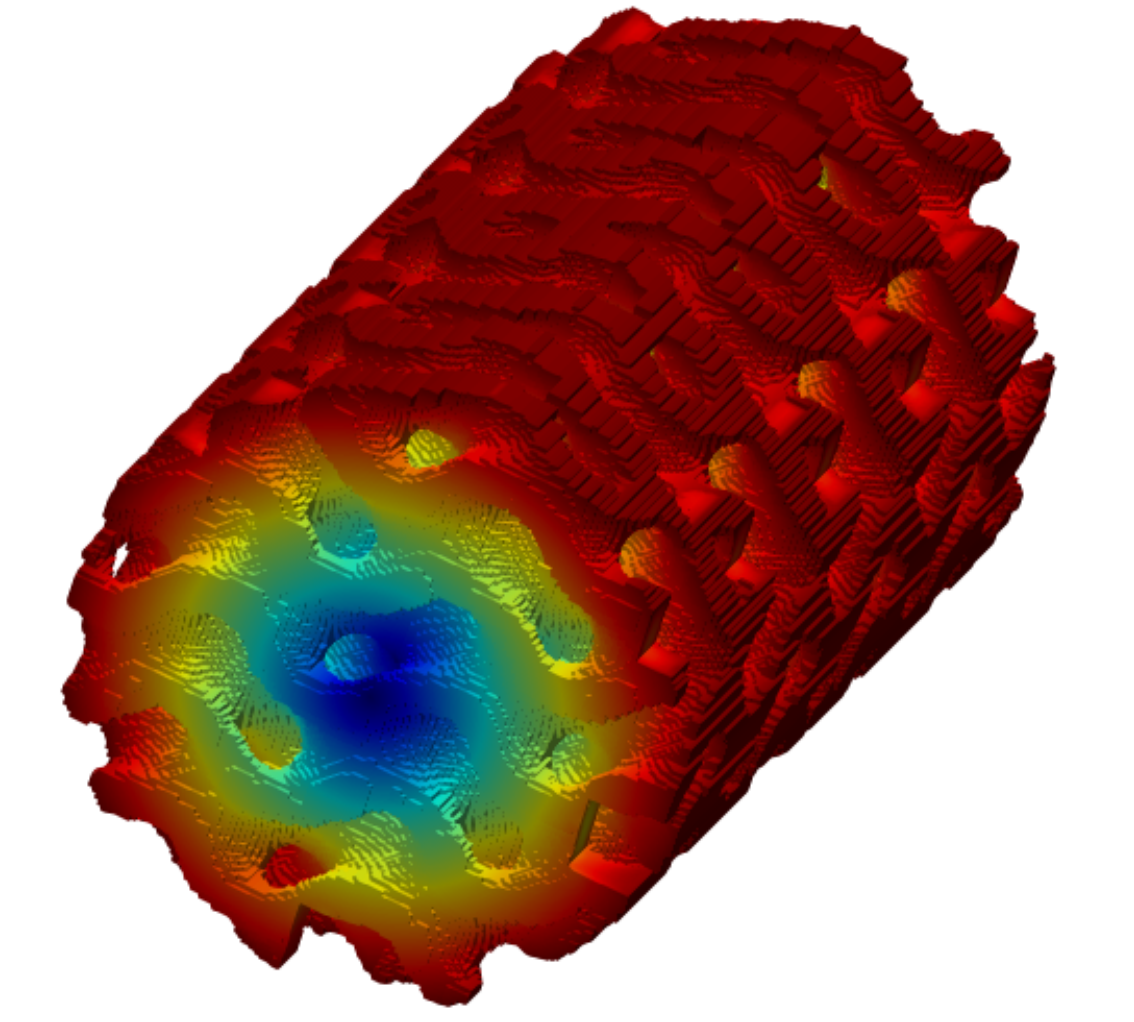}
        \caption{\small Radial displacement magnitude, $\sqrt{u_x^2 + u_y^2 + u_z^2}$.}
        \label{fig:gyroid_ur}
    \end{subfigure}
    \caption{Displacement fields (a) $U_x$ and (b) $U_y$ under a radially increasing displacement boundary condition. Displacements are minimal near the center and increase outward. (c) Radial displacement magnitude for the Gyroid, confirming consistency with the applied boundary condition.}
    \label{fig:gyroid_combined}
\end{figure*}

\begin{figure}[t!]
    \centering
    \begin{subfigure}[b]{0.45\linewidth}
        \centering
        \includegraphics[trim=400 100 400 10,clip,width=\linewidth]{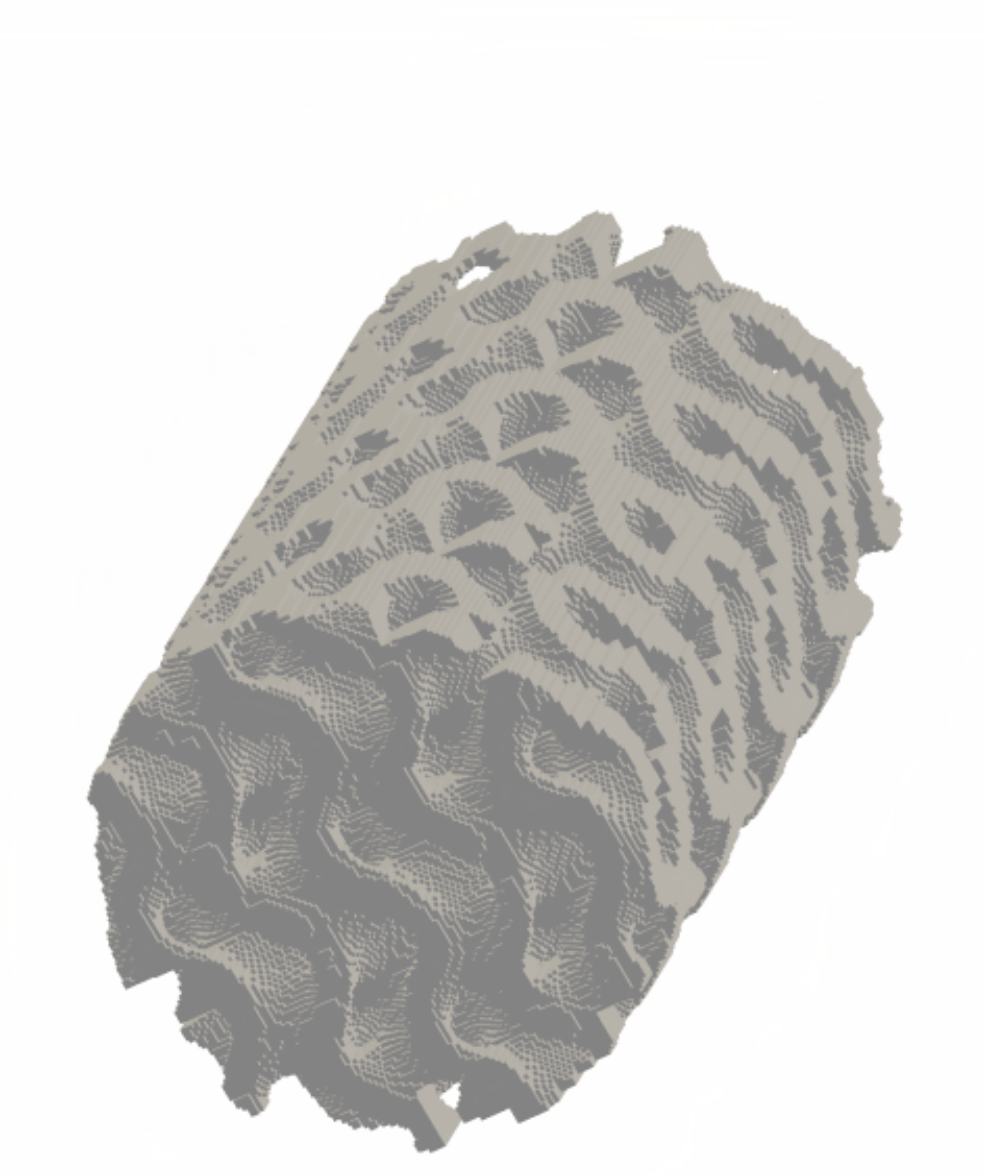}
        \caption{}
        \label{fig:gyroid_octree}
    \end{subfigure}
    \hspace{0.05\linewidth}
    \begin{subfigure}[b]{0.45\linewidth}
        \centering
        \includegraphics[trim=300 0 300 0,clip,width=\linewidth]{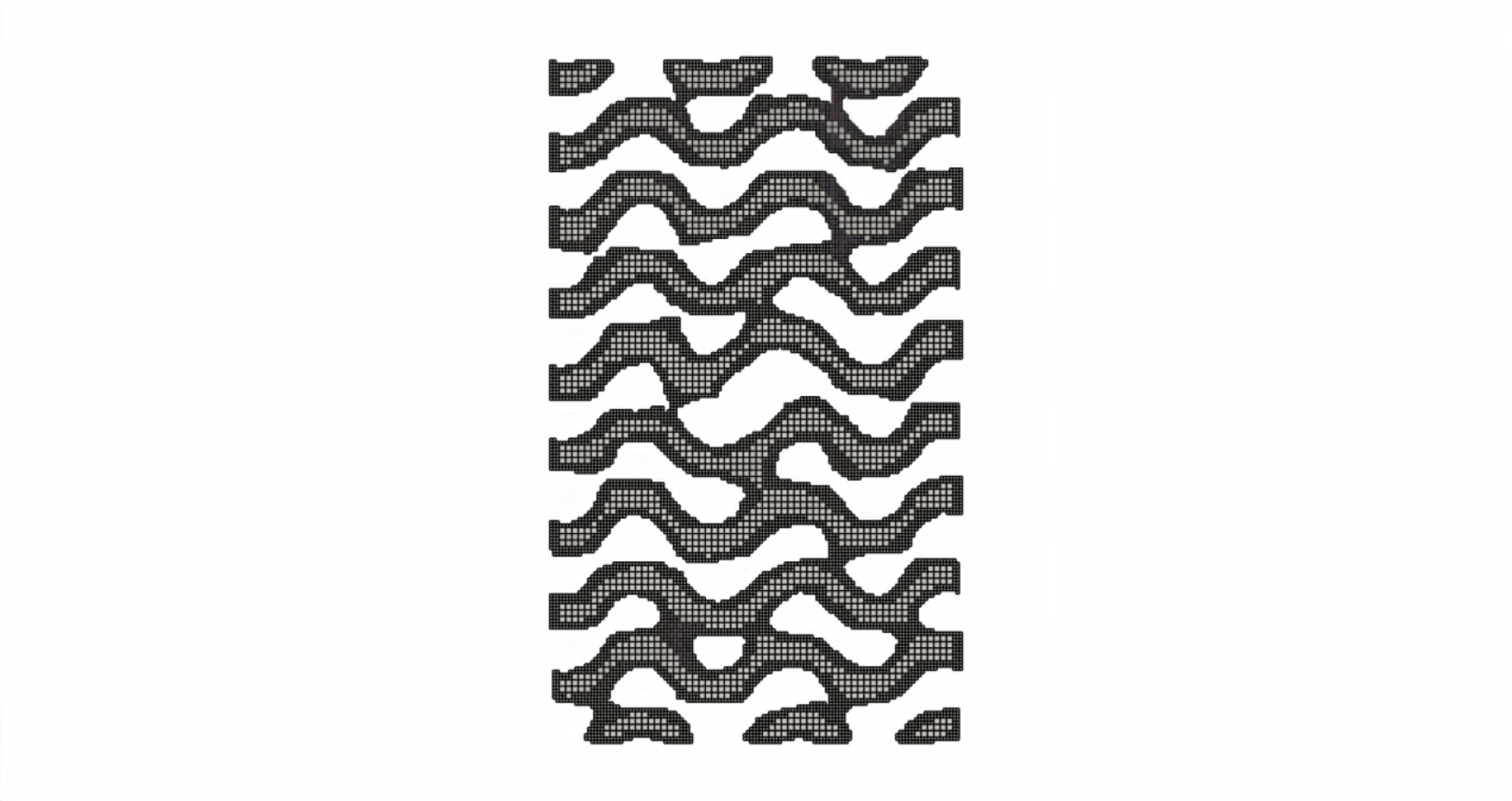}
        \caption{}
        \label{fig:gyroid_slice}
    \end{subfigure}
    \caption{(a) Octree representation of complicated Gyroid shape obtained with base refinement of level 6 and boundary level refinement of level 8. (b) A slice of the intricate Gyroid with adaptive refinement to capture the boundary,}
    \label{fig:gyroid}
\end{figure}

The Eiffel Tower model is subjected to plane strain condition with young's modulus of $10^{11}$ and poisson ratio of 0.33 with self-weight acting on the body. The simulation setup needs to solve 1.8 million degrees of freedom. \figref{fig:EiffelDisp} displays (a) the displacement field in the $x$-direction ($U_x$) and (b) the displacement in the $y$-direction ($U_y$). Both displacement fields follow a sinusoidal variation as per the applied boundary conditions. The results indicate a higher displacement magnitude in the $x$-direction, with the model being constrained in the $z$-direction, effectively capturing the expected deformation behavior.

\subsection{Gyroid Model}

\figref{fig:gyroid_octree} illustrates the octree-based INR representation of a complex Gyroid structure~\citep{schoen1970infinite}. The base level of refinement is set to 6, while the boundaries are refined to level 8 to accurately capture intricate geometric details. \figref{fig:gyroid_slice} shows a slice through the Gyroid, demonstrating the effectiveness of adaptive refinement in maintaining geometric fidelity. The complex internal structure of the gyroid is very difficult to capture, and getting a boundary-fitted mesh is a difficult undertaking, but the combination of INR and octree-based refinement enables the automatic capture of such intricate details with minimal manual intervention.

To perform the analysis, we use poisson ratio of 0.3 and youngs modulus of $7 \times 10^{10}$ with following boundary condition enforced using SBM:
\begin{subequations}
\begin{align}
u_x &= 0.01 \cdot R \cdot \frac{x}{R + 10^{-6}}, \label{eq:ux_rad} \\
u_y &= 0.01 \cdot R \cdot \frac{y}{R + 10^{-6}}, \label{eq:uy_rad} \\
u_z &= 0 \label{eq:uz_rad}
\end{align}
\end{subequations}

The intricacy of gyroid geometry yields almost 3 million degrees of freedom at this setup to be solved, which provides a clear indication of the computational scale involved in the simulation.\figref{fig:gyroid_ux} and \figref{fig:gyroid_uy} presents (a) the displacement field in the $x$-direction ($U_x$) and (b) the displacement in the $y$-direction ($U_y$) respectively under a radially increasing displacement boundary condition. Both $U_x$ and $U_y$ are constrained to zero at the center and increase outward, following the imposed boundary conditions. Finally, the bottom of \figref{fig:gyroid_combined} shows the displacement magnitude $\sqrt{u_x^2 + u_y^2 + u_z^2}$ for the Gyroid model, which conforms well to the expected radially increasing boundary condition. This result further supports the robustness of the INR-SBM framework in accurately representing and simulating complex geometries.

These results collectively demonstrate the effectiveness of the proposed INR-SBM approach in handling intricate geometries across different complexities, offering an efficient and accurate method to perform direct simulation with Implicit Neural Representation (INR).

\section{Conclusions}
\label{Sec:Conclusions}
This work introduces a novel computational framework that integrates Implicit Neural Representations (INRs) with the Shifted Boundary Method (SBM). Unlike traditional mesh-based methods, which require extensive preprocessing and suffer from discretization errors, INRs provide a smooth, adaptive, and compact way to encode complex geometries as continuous functions. By leveraging SBM, which eliminates the need for explicit mesh generation by imposing boundary conditions on surrogate boundaries, the proposed approach enables seamless incorporation of INR-based geometries into finite element simulations. The framework can be applied to INRs generated from diverse sources, including point clouds, image-based reconstructions, and AI-generated models. Results demonstrate improved computational efficiency, reduced manual intervention, and enhanced adaptability, highlighting the method’s potential impact across various scientific and engineering domains. Our approach provides a foundational step toward seamlessly integrating AI-generated geometries into high-fidelity simulations, paving the way for faster, more flexible, and data-driven mechanics workflows without the pain of intermediate mesh generation.

While the proposed INR-SBM framework significantly streamlines geometric processing and simulation workflows, several avenues for future exploration remain. One key direction is extending the framework to support fully nonlinear physics, such as large deformation mechanics and coupled multi-physics simulations, where implicit geometric representations could further simplify problem setup. Additionally, improving the training and inference efficiency of INRs for real-time or large-scale applications would enhance their practicality in computational engineering. Another promising area involves integrating adaptive refinement strategies that leverage neural representations to dynamically adjust simulation resolution based on local geometric complexity. Lastly, incorporating uncertainty quantification techniques within the INR-SBM framework could enable robust error estimation and confidence assessment, making the method more suitable for high-stakes applications in biomedical and aerospace engineering.

\bibliographystyle{elsarticle-num-names}
\bibliography{Refs}

\begin{thebibliography}{35}
\expandafter\ifx\csname natexlab\endcsname\relax\def\natexlab#1{#1}\fi
\providecommand{\url}[1]{\texttt{#1}}
\providecommand{\href}[2]{#2}
\providecommand{\path}[1]{#1}
\providecommand{\DOIprefix}{doi:}
\providecommand{\ArXivprefix}{arXiv:}
\providecommand{\URLprefix}{URL: }
\providecommand{\Pubmedprefix}{pmid:}
\providecommand{\doi}[1]{\href{http://dx.doi.org/#1}{\path{#1}}}
\providecommand{\Pubmed}[1]{\href{pmid:#1}{\path{#1}}}
\providecommand{\bibinfo}[2]{#2}
\ifx\xfnm\relax \def\xfnm[#1]{\unskip,\space#1}\fi
\bibitem[{McHenry and Bajcsy(2008)}]{mchenry2008overview}
\bibinfo{author}{K.~McHenry}, \bibinfo{author}{P.~Bajcsy}, \bibinfo{title}{An overview of 3D data content, file formats and viewers}, \bibinfo{type}{Technical Report} \bibinfo{number}{1205}, National Center for Supercomputing Applications (NCSA), \bibinfo{year}{2008}. \bibinfo{note}{Technical Report, 22 pages}.
\bibitem[{Chiba et~al.(1998)Chiba, Nishigaki, Yamashita, Takizawa, and Fujishiro}]{CHIBA1998145}
\bibinfo{author}{N.~Chiba}, \bibinfo{author}{I.~Nishigaki}, \bibinfo{author}{Y.~Yamashita}, \bibinfo{author}{C.~Takizawa}, \bibinfo{author}{K.~Fujishiro},
\newblock \bibinfo{title}{A flexible automatic hexahedral mesh generation by boundary-fit method},
\newblock \bibinfo{journal}{Computer Methods in Applied Mechanics and Engineering} \bibinfo{volume}{161} (\bibinfo{year}{1998}) \bibinfo{pages}{145--154}. \URLprefix \url{https://www.sciencedirect.com/science/article/pii/S0045782597003046}. \DOIprefix\doi{https://doi.org/10.1016/S0045-7825(97)00304-6}.
\bibitem[{Park et~al.(2019)Park, Florence, Straub, Newcombe, and Lovegrove}]{park2019deepsdf}
\bibinfo{author}{J.~J. Park}, \bibinfo{author}{P.~Florence}, \bibinfo{author}{J.~Straub}, \bibinfo{author}{R.~Newcombe}, \bibinfo{author}{S.~Lovegrove},
\newblock \bibinfo{title}{Deepsdf: Learning continuous signed distance functions for shape representation},
\newblock in: \bibinfo{booktitle}{Proceedings of the IEEE/CVF conference on computer vision and pattern recognition}, \bibinfo{year}{2019}, pp. \bibinfo{pages}{165--174}.
\bibitem[{Gropp et~al.(2020)Gropp, Yariv, Haim, Atzmon, and Lipman}]{gropp2020implicit}
\bibinfo{author}{A.~Gropp}, \bibinfo{author}{L.~Yariv}, \bibinfo{author}{N.~Haim}, \bibinfo{author}{M.~Atzmon}, \bibinfo{author}{Y.~Lipman},
\newblock \bibinfo{title}{Implicit geometric regularization for learning shapes},
\newblock \bibinfo{journal}{arXiv preprint arXiv:2002.10099}  (\bibinfo{year}{2020}).
\bibitem[{Mescheder et~al.(2019)Mescheder, Oechsle, Niemeyer, Nowozin, and Geiger}]{mescheder2019occupancy}
\bibinfo{author}{L.~Mescheder}, \bibinfo{author}{M.~Oechsle}, \bibinfo{author}{M.~Niemeyer}, \bibinfo{author}{S.~Nowozin}, \bibinfo{author}{A.~Geiger},
\newblock \bibinfo{title}{Occupancy networks: Learning 3d reconstruction in function space},
\newblock in: \bibinfo{booktitle}{Proceedings of the IEEE/CVF conference on computer vision and pattern recognition}, \bibinfo{year}{2019}, pp. \bibinfo{pages}{4460--4470}.
\bibitem[{Chen and Zhang(2019)}]{chen2019learning}
\bibinfo{author}{Z.~Chen}, \bibinfo{author}{H.~Zhang},
\newblock \bibinfo{title}{Learning implicit fields for generative shape modeling},
\newblock in: \bibinfo{booktitle}{Proceedings of the IEEE/CVF conference on computer vision and pattern recognition}, \bibinfo{year}{2019}, pp. \bibinfo{pages}{5939--5948}.
\bibitem[{Main and Scovazzi(2018{\natexlab{a}})}]{main2018shifted1}
\bibinfo{author}{A.~Main}, \bibinfo{author}{G.~Scovazzi},
\newblock \bibinfo{title}{The shifted boundary method for embedded domain computations. part i: Poisson and stokes problems},
\newblock \bibinfo{journal}{Journal of Computational Physics} \bibinfo{volume}{372} (\bibinfo{year}{2018}{\natexlab{a}}) \bibinfo{pages}{972--995}.
\bibitem[{Main and Scovazzi(2018{\natexlab{b}})}]{main2018shifted2}
\bibinfo{author}{A.~Main}, \bibinfo{author}{G.~Scovazzi},
\newblock \bibinfo{title}{The shifted boundary method for embedded domain computations. part ii: Linear advection--diffusion and incompressible navier--stokes equations},
\newblock \bibinfo{journal}{Journal of Computational Physics} \bibinfo{volume}{372} (\bibinfo{year}{2018}{\natexlab{b}}) \bibinfo{pages}{996--1026}.
\bibitem[{Hsu et~al.(2016)Hsu, Wang, Xu, Herrema, and Krishnamurthy}]{hsu2016direct}
\bibinfo{author}{M.-C. Hsu}, \bibinfo{author}{C.~Wang}, \bibinfo{author}{F.~Xu}, \bibinfo{author}{A.~J. Herrema}, \bibinfo{author}{A.~Krishnamurthy},
\newblock \bibinfo{title}{Direct immersogeometric fluid flow analysis using b-rep cad models},
\newblock \bibinfo{journal}{Computer Aided Geometric Design} \bibinfo{volume}{43} (\bibinfo{year}{2016}) \bibinfo{pages}{143--158}.
\bibitem[{Burman et~al.(2015)Burman, Claus, Hansbo, Larson, and Massing}]{burman2015cutfem}
\bibinfo{author}{E.~Burman}, \bibinfo{author}{S.~Claus}, \bibinfo{author}{P.~Hansbo}, \bibinfo{author}{M.~G. Larson}, \bibinfo{author}{A.~Massing},
\newblock \bibinfo{title}{Cutfem: discretizing geometry and partial differential equations},
\newblock \bibinfo{journal}{International Journal for Numerical Methods in Engineering} \bibinfo{volume}{104} (\bibinfo{year}{2015}) \bibinfo{pages}{472--501}.
\bibitem[{Burman and Hansbo(2012)}]{burman2012fictitious}
\bibinfo{author}{E.~Burman}, \bibinfo{author}{P.~Hansbo},
\newblock \bibinfo{title}{Fictitious domain finite element methods using cut elements: Ii. a stabilized nitsche method},
\newblock \bibinfo{journal}{Applied Numerical Mathematics} \bibinfo{volume}{62} (\bibinfo{year}{2012}) \bibinfo{pages}{328--341}.
\bibitem[{Saurabh et~al.(2021)Saurabh, Gao, Fernando, Xu, Khanwale, Khara, Hsu, Krishnamurthy, Sundar, and Ganapathysubramanian}]{saurabh2021industrial}
\bibinfo{author}{K.~Saurabh}, \bibinfo{author}{B.~Gao}, \bibinfo{author}{M.~Fernando}, \bibinfo{author}{S.~Xu}, \bibinfo{author}{M.~A. Khanwale}, \bibinfo{author}{B.~Khara}, \bibinfo{author}{M.-C. Hsu}, \bibinfo{author}{A.~Krishnamurthy}, \bibinfo{author}{H.~Sundar}, \bibinfo{author}{B.~Ganapathysubramanian},
\newblock \bibinfo{title}{Industrial scale large eddy simulations with adaptive octree meshes using immersogeometric analysis},
\newblock \bibinfo{journal}{Computers \& Mathematics with Applications} \bibinfo{volume}{97} (\bibinfo{year}{2021}) \bibinfo{pages}{28--44}.
\bibitem[{Colom{\'e}s et~al.(2021)Colom{\'e}s, Main, Nouveau, and Scovazzi}]{colomes2021weighted}
\bibinfo{author}{O.~Colom{\'e}s}, \bibinfo{author}{A.~Main}, \bibinfo{author}{L.~Nouveau}, \bibinfo{author}{G.~Scovazzi},
\newblock \bibinfo{title}{A weighted shifted boundary method for free surface flow problems},
\newblock \bibinfo{journal}{Journal of Computational Physics} \bibinfo{volume}{424} (\bibinfo{year}{2021}) \bibinfo{pages}{109837}.
\bibitem[{Chou et~al.(2023)Chou, Bahat, and Heide}]{chou2023diffusion}
\bibinfo{author}{G.~Chou}, \bibinfo{author}{Y.~Bahat}, \bibinfo{author}{F.~Heide},
\newblock \bibinfo{title}{Diffusion-sdf: Conditional generative modeling of signed distance functions},
\newblock in: \bibinfo{booktitle}{Proceedings of the IEEE/CVF international conference on computer vision}, \bibinfo{year}{2023}, pp. \bibinfo{pages}{2262--2272}.
\bibitem[{Huang and Lim(2020)}]{huang2020simulation}
\bibinfo{author}{T.~Huang}, \bibinfo{author}{H.-C. Lim},
\newblock \bibinfo{title}{Simulation of lid-driven cavity flow with internal circular obstacles},
\newblock \bibinfo{journal}{Applied Sciences} \bibinfo{volume}{10} (\bibinfo{year}{2020}) \bibinfo{pages}{4583}.
\bibitem[{Karatzas et~al.(2020)Karatzas, Stabile, Nouveau, Scovazzi, and Rozza}]{karatzas2020reduced}
\bibinfo{author}{E.~N. Karatzas}, \bibinfo{author}{G.~Stabile}, \bibinfo{author}{L.~Nouveau}, \bibinfo{author}{G.~Scovazzi}, \bibinfo{author}{G.~Rozza},
\newblock \bibinfo{title}{A reduced-order shifted boundary method for parametrized incompressible navier--stokes equations},
\newblock \bibinfo{journal}{Computer Methods in Applied Mechanics and Engineering} \bibinfo{volume}{370} (\bibinfo{year}{2020}) \bibinfo{pages}{113273}.
\bibitem[{Atallah et~al.(2021)Atallah, Canuto, and Scovazzi}]{atallah2021shifted}
\bibinfo{author}{N.~M. Atallah}, \bibinfo{author}{C.~Canuto}, \bibinfo{author}{G.~Scovazzi},
\newblock \bibinfo{title}{The shifted boundary method for solid mechanics},
\newblock \bibinfo{journal}{International Journal for Numerical Methods in Engineering} \bibinfo{volume}{122} (\bibinfo{year}{2021}) \bibinfo{pages}{5935--5970}.
\bibitem[{Yang et~al.(2024{\natexlab{a}})Yang, Saurabh, Scovazzi, Canuto, Krishnamurthy, and Ganapathysubramanian}]{yang2024optimal}
\bibinfo{author}{C.-H. Yang}, \bibinfo{author}{K.~Saurabh}, \bibinfo{author}{G.~Scovazzi}, \bibinfo{author}{C.~Canuto}, \bibinfo{author}{A.~Krishnamurthy}, \bibinfo{author}{B.~Ganapathysubramanian},
\newblock \bibinfo{title}{Optimal surrogate boundary selection and scalability studies for the shifted boundary method on octree meshes},
\newblock \bibinfo{journal}{Computer Methods in Applied Mechanics and Engineering} \bibinfo{volume}{419} (\bibinfo{year}{2024}{\natexlab{a}}) \bibinfo{pages}{116686}.
\bibitem[{Yang et~al.(2024{\natexlab{b}})Yang, Scovazzi, Krishnamurthy, and Ganapathysubramanian}]{yang2024simulating}
\bibinfo{author}{C.-H. Yang}, \bibinfo{author}{G.~Scovazzi}, \bibinfo{author}{A.~Krishnamurthy}, \bibinfo{author}{B.~Ganapathysubramanian},
\newblock \bibinfo{title}{Simulating incompressible flows over complex geometries using the shifted boundary method with incomplete adaptive octree meshes},
\newblock \bibinfo{journal}{arXiv preprint arXiv:2411.00272}  (\bibinfo{year}{2024}{\natexlab{b}}).
\bibitem[{Saurabh et~al.(2021)Saurabh, Ishii, Fernando, Gao, Tan, Hsu, Krishnamurthy, Sundar, and Ganapathysubramanian}]{saurabh2021scalable}
\bibinfo{author}{K.~Saurabh}, \bibinfo{author}{M.~Ishii}, \bibinfo{author}{M.~Fernando}, \bibinfo{author}{B.~Gao}, \bibinfo{author}{K.~Tan}, \bibinfo{author}{M.-C. Hsu}, \bibinfo{author}{A.~Krishnamurthy}, \bibinfo{author}{H.~Sundar}, \bibinfo{author}{B.~Ganapathysubramanian},
\newblock \bibinfo{title}{Scalable adaptive pde solvers in arbitrary domains},
\newblock in: \bibinfo{booktitle}{Proceedings of the International Conference for high performance computing, networking, storage and analysis}, \bibinfo{year}{2021}, pp. \bibinfo{pages}{1--15}.
\bibitem[{Jignasu et~al.(2024)Jignasu, Herron, Jiang, Sarkar, Hegde, Ganapathysubramanian, Balu, and Krishnamurthy}]{jignasu2024stitch}
\bibinfo{author}{A.~Jignasu}, \bibinfo{author}{E.~Herron}, \bibinfo{author}{Z.~Jiang}, \bibinfo{author}{S.~Sarkar}, \bibinfo{author}{C.~Hegde}, \bibinfo{author}{B.~Ganapathysubramanian}, \bibinfo{author}{A.~Balu}, \bibinfo{author}{A.~Krishnamurthy},
\newblock \bibinfo{title}{Stitch: Surface reconstruction using implicit neural representations with topology constraints and persistent homology},
\newblock \bibinfo{journal}{arXiv preprint arXiv:2412.18696}  (\bibinfo{year}{2024}).
\bibitem[{Liu et~al.(2019)Liu, Saito, Chen, and Li}]{liu2019learning}
\bibinfo{author}{S.~Liu}, \bibinfo{author}{S.~Saito}, \bibinfo{author}{W.~Chen}, \bibinfo{author}{H.~Li},
\newblock \bibinfo{title}{Learning to infer implicit surfaces without 3d supervision},
\newblock \bibinfo{journal}{Advances in Neural Information Processing Systems} \bibinfo{volume}{32} (\bibinfo{year}{2019}).
\bibitem[{Ben-Shabat et~al.(2022)Ben-Shabat, Koneputugodage, and Gould}]{ben2022digs}
\bibinfo{author}{Y.~Ben-Shabat}, \bibinfo{author}{C.~H. Koneputugodage}, \bibinfo{author}{S.~Gould},
\newblock \bibinfo{title}{Digs: Divergence guided shape implicit neural representation for unoriented point clouds},
\newblock in: \bibinfo{booktitle}{Proceedings of the IEEE/CVF Conference on Computer Vision and Pattern Recognition}, \bibinfo{year}{2022}, pp. \bibinfo{pages}{19323--19332}.
\bibitem[{Atzmon and Lipman(2020)}]{atzmon2020sal}
\bibinfo{author}{M.~Atzmon}, \bibinfo{author}{Y.~Lipman},
\newblock \bibinfo{title}{Sal: Sign agnostic learning of shapes from raw data},
\newblock in: \bibinfo{booktitle}{Proceedings of the IEEE/CVF conference on computer vision and pattern recognition}, \bibinfo{year}{2020}, pp. \bibinfo{pages}{2565--2574}.
\bibitem[{Wang et~al.(2021)Wang, Liu, Liu, Theobalt, Komura, and Wang}]{wang2021neus}
\bibinfo{author}{P.~Wang}, \bibinfo{author}{L.~Liu}, \bibinfo{author}{Y.~Liu}, \bibinfo{author}{C.~Theobalt}, \bibinfo{author}{T.~Komura}, \bibinfo{author}{W.~Wang},
\newblock \bibinfo{title}{Neus: Learning neural implicit surfaces by volume rendering for multi-view reconstruction},
\newblock \bibinfo{journal}{arXiv preprint arXiv:2106.10689}  (\bibinfo{year}{2021}).
\bibitem[{Niemeyer et~al.(2020)Niemeyer, Mescheder, Oechsle, and Geiger}]{niemeyer2020differentiable}
\bibinfo{author}{M.~Niemeyer}, \bibinfo{author}{L.~Mescheder}, \bibinfo{author}{M.~Oechsle}, \bibinfo{author}{A.~Geiger},
\newblock \bibinfo{title}{Differentiable volumetric rendering: Learning implicit 3d representations without 3d supervision},
\newblock in: \bibinfo{booktitle}{Proceedings of the IEEE/CVF conference on computer vision and pattern recognition}, \bibinfo{year}{2020}, pp. \bibinfo{pages}{3504--3515}.
\bibitem[{Ho et~al.(2020)Ho, Jain, and Abbeel}]{ho2020denoising}
\bibinfo{author}{J.~Ho}, \bibinfo{author}{A.~Jain}, \bibinfo{author}{P.~Abbeel},
\newblock \bibinfo{title}{Denoising diffusion probabilistic models},
\newblock \bibinfo{journal}{Advances in neural information processing systems} \bibinfo{volume}{33} (\bibinfo{year}{2020}) \bibinfo{pages}{6840--6851}.
\bibitem[{Rombach et~al.(2022)Rombach, Blattmann, Lorenz, Esser, and Ommer}]{rombach2022high}
\bibinfo{author}{R.~Rombach}, \bibinfo{author}{A.~Blattmann}, \bibinfo{author}{D.~Lorenz}, \bibinfo{author}{P.~Esser}, \bibinfo{author}{B.~Ommer},
\newblock \bibinfo{title}{High-resolution image synthesis with latent diffusion models},
\newblock in: \bibinfo{booktitle}{Proceedings of the IEEE/CVF conference on computer vision and pattern recognition}, \bibinfo{year}{2022}, pp. \bibinfo{pages}{10684--10695}.
\bibitem[{Erko{\c{c}} et~al.(2023)Erko{\c{c}}, Ma, Shan, Nie{\ss}ner, and Dai}]{erkocc2023hyperdiffusion}
\bibinfo{author}{Z.~Erko{\c{c}}}, \bibinfo{author}{F.~Ma}, \bibinfo{author}{Q.~Shan}, \bibinfo{author}{M.~Nie{\ss}ner}, \bibinfo{author}{A.~Dai},
\newblock \bibinfo{title}{Hyperdiffusion: Generating implicit neural fields with weight-space diffusion},
\newblock in: \bibinfo{booktitle}{Proceedings of the IEEE/CVF international conference on computer vision}, \bibinfo{year}{2023}, pp. \bibinfo{pages}{14300--14310}.
\bibitem[{Sitzmann et~al.(2020)Sitzmann, Martel, Bergman, Lindell, and Wetzstein}]{sitzmann2020implicit}
\bibinfo{author}{V.~Sitzmann}, \bibinfo{author}{J.~Martel}, \bibinfo{author}{A.~Bergman}, \bibinfo{author}{D.~Lindell}, \bibinfo{author}{G.~Wetzstein},
\newblock \bibinfo{title}{Implicit neural representations with periodic activation functions},
\newblock \bibinfo{journal}{Advances in neural information processing systems} \bibinfo{volume}{33} (\bibinfo{year}{2020}) \bibinfo{pages}{7462--7473}.
\bibitem[{Jacobson et~al.(2018)Jacobson, Panozzo et~al.}]{libigl}
\bibinfo{author}{A.~Jacobson}, \bibinfo{author}{D.~Panozzo}, et~al., \bibinfo{title}{{libigl}: A simple {C++} geometry processing library}, \bibinfo{year}{2018}. \bibinfo{note}{Https://libigl.github.io/}.
\bibitem[{Schillinger et~al.(2012)Schillinger, Ruess, Zander, Bazilevs, D{\"u}ster, and Rank}]{schillinger2012small}
\bibinfo{author}{D.~Schillinger}, \bibinfo{author}{M.~Ruess}, \bibinfo{author}{N.~Zander}, \bibinfo{author}{Y.~Bazilevs}, \bibinfo{author}{A.~D{\"u}ster}, \bibinfo{author}{E.~Rank},
\newblock \bibinfo{title}{Small and large deformation analysis with the p-and b-spline versions of the finite cell method},
\newblock \bibinfo{journal}{Computational Mechanics} \bibinfo{volume}{50} (\bibinfo{year}{2012}) \bibinfo{pages}{445--478}.
\bibitem[{Atallah et~al.(2020)Atallah, Canuto, and Scovazzi}]{atallah2020second}
\bibinfo{author}{N.~M. Atallah}, \bibinfo{author}{C.~Canuto}, \bibinfo{author}{G.~Scovazzi},
\newblock \bibinfo{title}{The second-generation shifted boundary method and its numerical analysis},
\newblock \bibinfo{journal}{Computer Methods in Applied Mechanics and Engineering} \bibinfo{volume}{372} (\bibinfo{year}{2020}) \bibinfo{pages}{113341}.
\bibitem[{Turk and Levoy(1994)}]{turk1994zippered}
\bibinfo{author}{G.~Turk}, \bibinfo{author}{M.~Levoy},
\newblock \bibinfo{title}{Zippered polygon meshes from range images},
\newblock in: \bibinfo{booktitle}{Proceedings of the 21st annual conference on Computer graphics and interactive techniques}, \bibinfo{year}{1994}, pp. \bibinfo{pages}{311--318}.
\bibitem[{Schoen(1970)}]{schoen1970infinite}
\bibinfo{author}{A.~H. Schoen}, \bibinfo{title}{Infinite periodic minimal surfaces without self-intersections}, volume \bibinfo{volume}{5541}, \bibinfo{publisher}{National Aeronautics and Space Administration}, \bibinfo{year}{1970}.

\end{thebibliography}

\end{document}